\documentclass[final,3p,times]{elsarticle}

\usepackage{amssymb,mathrsfs}
\usepackage{graphicx,multirow,caption,subcaption,multicol,setspace,amsmath,wrapfig}
\usepackage[linesnumbered,ruled]{algorithm2e}
\usepackage{algorithmic}
\usepackage{adjustbox}
\usepackage{booktabs,caption}
\usepackage[flushleft]{threeparttable}
\usepackage{lineno}
\usepackage{enumitem}
\setlist{nosep}
\usepackage{rotating,caption}
\usepackage[table]{xcolor}
\usepackage{soul,xcolor}
\usepackage{amsfonts}
\usepackage{makecell,eqnarray}
\biboptions{numbers,sort&compress}
\usepackage{amsthm}
\usepackage{bm}
\usepackage{wrapfig}
\theoremstyle{definition}
\newtheorem{definition}{Definition}[]

\usepackage{mathtools}
\DeclarePairedDelimiter\ceil{\lceil}{\rceil}

\graphicspath{ {Images/} }
\makeatletter
\newcommand{\algorithmfootnote}[2][\footnotesize]{%
  \let\old@algocf@finish\@algocf@finish
  \def\@algocf@finish{\old@algocf@finish
    \leavevmode\rlap{\begin{minipage}{\linewidth}
    #1#2
    \end{minipage}}%
  }%
}

\begin{document}
\begin{frontmatter}

\title{Orthogonal Floating Search Algorithms: From The Perspective of Nonlinear System Identification}

\author[label5]{Faizal Hafiz\corref{cor1}}
\address[label5]{Department of Electrical \& Computer Engineering, The University of Auckland, Auckland, New Zealand}
\ead{faizalhafiz@ieee.org}
\cortext[cor1]{Corresponding author}

\author[label5]{Akshya Swain}

\author[label1]{Eduardo Mendes}
\address[label1]{Department of Electronics Engineering, Federal University of Minas Gerais, Belo Horizonte, Brazil}

\begin{abstract}

The present study proposes a new Orthogonal Floating Search framework for structure selection of nonlinear systems by adapting the existing floating search algorithms for feature selection. The proposed framework integrates the concept of orthogonal space and consequent Error-Reduction-Ratio (ERR) metric with the existing floating search algorithms. On the basis of this framework, three well-known feature selection algorithms have been adapted which include the classical Sequential Forward Floating Search (SFFS), Improved sequential Forward Floating Search (IFFS) and Oscillating Search (OS). This framework retains the simplicity of classical Orthogonal Forward Regression with ERR (OFR-ERR) and eliminates the nesting effect associated with OFR-ERR. The performance of the proposed framework has been demonstrated considering several benchmark non-linear systems. The results show that most of the existing feature selection methods can easily be tailored to identify the correct system structure of nonlinear systems.

\end{abstract}

\begin{keyword}
Forward model selection \sep Model structure selection \sep NARX model \sep Nonlinear system \sep System identification 
\end{keyword}

\end{frontmatter}

\section{Introduction}
\label{sec1:intro}

Over the last several decades, researchers from  various fields of science and engineering have developed several methods to construct mathematical models from measured input-output data ( popularly known as system identification) ~\cite{Billings:2013,Ljung:1999}. Since most of the practical systems are nonlinear, the first step in the process of fitting the models is the choice of the model amongst various available models (e.g., Volterra, Wiener, Polynomial/Rational) which can effectively capture the dynamics in input-output data. In this study, the focus is on the identification of nonlinear systems represented by polynomial nonlinear auto-regressive with exogenous inputs (NARX) models~\cite{Billings:2013}. Note that the identification of system is a two stage process which involves 1) Determination of the significant terms and 2) Estimation of the corresponding parameters. While the parameters can be estimated using least-squares based algorithms, the determination of the significant terms is a challenging task and often referred to as \textit{`structure selection'}.

A close examination of the structure selection problem in nonlinear system identification reveals that this problem has significant similarities with the \textit{feature selection} problem encountered in the pattern recognition. In essence, both problems belong to a broader class of combinatorial optimization problem which is referred here as the \textit{attribute selection} problem. To understand this further, consider a superset of `$n$' number of \textit{`attributes'}, denoted by: $\mathcal{X}_{model}=\{ x_1, x_2, \dots x_{n} \}$, where any attribute (say `$x_i$'), could represent either a \textit{feature} (in \textit{pattern recognition}) or a \textit{term} of the NARX model (in \textit{system identification}). The goal of the attribute selection is to determine a subset of \textit{significant} attributes, $\mathcal{X}^{\star} \subset \mathcal{X}_{model}$, by maximizing a suitable criterion function,`$J(\cdotp)$', as follows:
\begin{linenomath*}
\begin{align}
    \label{eq:mss}
    J(\mathcal{X}^{\star}) = \max \limits_{\mathcal{X} \subset \mathcal{X}_{model}}  J(\mathcal{X})
\end{align}
\end{linenomath*}
An exhaustive search to solve such combinatorial problem is often intractable even for a moderate number of attributes, `$n$', as it requires the evaluation of $2^{n}$ subsets.

Over the years several approaches have been proposed to address the structure selection problem of NARX models, \textit{e.g.}, see~\cite{Hong:Mitchell:2008,Billings:Chen:Korenberg:1989,Baldacchino:Kadirkamanathan:2012,Baldacchino:Kadirkamanathan:2013,Falsone:Piroddi:2015,Tang:Long:2019}. Among these, the Orthogonal Forward Regression (OFR)~\cite{Korenberg:Billings:1988,Chen:Billings:1989,Billings:Chen:Korenberg:1989} is, arguably, computationally the most effective. Perhaps, for this reason, it has been extensively applied in many applications~\cite{Chiras:Evans:2001,Billings:2013}. The central idea of this approach is to decouple the terms in the orthogonal space so that the contribution of each term to the output variance can individually be determined. From the search perspective, the OFR is a \textit{sequential greedy} approach where the term with the highest performance metric, referred to as \textit{Error-Reduction-Ratio} (ERR), is included in each step and it has proven to be very effective in various applications~\cite{Chiras:Evans:2001,Hong:Mitchell:2008}. However, it has been shown that under certain conditions the OFR may yield sub-optimal term subsets~\cite{Mao:Billings:1997,Piroddi:Spinelli:2003,Wei:Billings:2006,Guo:Billings:2015,Guo:Guo:2016}. The reason for such suboptimal performance is often attributed to the performance metric, ERR. For example, Mao and Billings~\cite{Mao:Billings:1997} have shown that for the given term, the value of ERR is dependent on the \textit{orthogonalization path}, \textit{i.e.}, the order in which the terms are orthogonalized. This may lead to the selection of spurious terms, when there is an information overlap among the non-orthogonal terms, especially in the earlier stages of the search~\cite{Guo:Billings:2015}.

To alleviate these issues, several approaches have been proposed to extend/modify the classical OFR algorithm. These can be categorized mainly into two approaches: 1) Two-step approach~\cite{Mao:Billings:1997,Li:Peng:2006,Guo:Billings:2015} and 2) Improved performance metric~\cite{Piroddi:Spinelli:2003,Wei:Billings:2006,Guo:Guo:2016}. 

In the two-step approach, OFR is applied first to identify the initial term subset which is further refined by a secondary search procedure. For example, in~\cite{Mao:Billings:1997}, the initial term subset is refined by applying Genetic Algorithm (GA) to identify the optimal \textit{orthogonalization path}. However, the explosive growth of the search space limits the application of this approach, \textit{e.g.}, for `$n$' number of terms, the total number of possible orthogonalization paths becomes `$n!$'. In comparison, if GA (or any suitable algorithm) is applied directly to select term subset the search space reduces from $n!$ to $2^n$~\cite{Hafiz:Swain:CEC:2018}. The `iterative OFR' (iOFR) algorithm~\cite{Guo:Billings:2015} uses each term in the initial term subset as a \textit{pivot} to identify new, and possibly better, term subsets in the secondary search. In~\cite{Li:Peng:2006}, a \textit{backward elimination} approach is proposed as the secondary search to remove the spurious terms contained in the initial term subset. Most of the two-step approach, e.g.~\cite{Li:Peng:2006}, are usually effective provided all the system/significant terms are identified in the initial term subset. 

The focus of the second approach is to replace the performance metric,  `ERR'~\cite{Piroddi:Spinelli:2003,Wei:Billings:2006,Guo:Guo:2016}. For this purpose, several performance metrics have been proposed  based on Simulation Error~\cite{Piroddi:Spinelli:2003}, Mutual Information~\cite{Wei:Billings:2006} and Ultra-orthogonal least squares~\cite{Guo:Guo:2016}. While these metrics improve the search performance of OFR in many cases, they often increase the computational complexity of OFR. 

It is worth to emphasize that the aforementioned approaches to improve OFR are based on the prevailing belief that the sub-optimal performance of OFR is due \textit{only} to its performance metric, ERR. However, in this study, it is argued that this may be in part due to the \textit{uni-directional} sequential search nature of OFR where terms are progressively added into the term subset until pre-specified ERR threshold is reached. This leads to the \textit{`nesting effect'}, \textit{i.e.}, once a term is included in the subset, it cannot be removed from the term subset. This imposes a hierarchy of terms, whereas in practice, all significant terms are equally important. Note that the nesting effect is not new and it has been well-known among the feature selection community since the early seventies~\cite{Michael:Lin:1973,Stearns:1976,Kittler:1978}. Over the years, state-of-the-art \textit{`floating'} feature selection algorithms have specifically been developed to address the \textit{nesting effect}~\cite{Pudil:1994,Somol:Pudil:1999,Somol:Pudil:2000,Nakariyakul:Casasent:2009}. Since there already exist a class of \textit{floating} feature selection algorithms and given that both structure selection and feature selection share a common search objective, the question arises: \textit{Can the feature selection algorithms be adapted to address the structure selection problem?}

The present study aims to address this apropos question. In particular, the proposed Orthogonal Floating Search Framework integrates the key properties of the classical Sequential Forward Floating Search (SFFS) and its variants such as Improved Forward Floating Search (IFFS) and Oscillating Search (OS) with the Error-Reduction-Ratio (ERR) to determine the structure of nonlinear systems. In each step of these algorithms, after adding a \textit{`significant'} term (discussed in Section~\ref{s:termsig}), the selected terms are re-examined to identify and remove the \textit{`least significant'} term (discussed in Section~\ref{s:termsig}). Thus, the spurious terms can eventually be replaced by the system/significant terms during the search process. The search performance is therefore significantly improved without the need for a second search procedure or a complicated performance measure. The efficacy of the proposed approach is demonstrated on several benchmark non-linear systems. The worst case scenario (\textit{non-persistent excitation}) and the identification of a discrete-time model of a continuous-time system are also considered. The results of this investigation convincingly demonstrate that the feature selection algorithms can indeed be tailored for the purpose of structure selection.

The rest of the article is organized as follows: The polynomial NARX model and the OFR approach are discussed briefly in Section~\ref{s:background}. Next, the proposed Orthogonal Floating Search Framework is discussed in detail in Section~\ref{s:proposedOFS}. The framework of this investigation is described in Section~\ref{s:IF}.  Finally, the results of this investigation are discussed at length in Section~\ref{s:res}.

\section{Preliminaries}
\label{s:background}

In the following, the polynomial NARX model and the orthogonal regression approach are briefly discussed.

\subsection{The Polynomial NARX Model}
\label{s:NARX}

The NARX model represents a non-linear system as a function of recursive lagged input and output terms as follows:
\begin{linenomath*}
\begin{align*}
y(k) & = F^{n_l} \ \{ \ y(k-1),\ldots,y(k-n_y),u(k-1),\ldots, u(k-n_u) \ \}+e(k)
\end{align*}
\end{linenomath*}
where $y(k)$ and $u(k)$ respectively represent the output and input at time intervals $k$, $n_y$ and $n_u$ are corresponding lags and $F^{n_l}\{ \cdotp \}$ is some nonlinear function of degree $n_l$. 

The \textit{total number of possible terms} or \textit{model size} ($n$) of the NARX model is given by,
\begin{linenomath*}
\begin{align}
\label{eq:Nt}
n & = \sum_{i=0}^{n_l} n_i, \ n_0=1 \ \textit{and \ } n_i = \frac{n_{i-1}(n_y+n_u+i-1)}{i}, \ i=1,\ldots, n_l 
\end{align}
\end{linenomath*}
This model is essentially linear-in-parameters and can be expressed as:
\begin{linenomath*}
\begin{align}
\label{eq:NARXmodel}
    y(k) & = \sum_{i=1}^{n} \theta_i x_i(k) + e(k), \ \ k = 1,2,\dots \mathcal{N}\\
    \text{where, } x_1(k) & = 1, \ \ \text{and \ } x_i(k)  = \prod_{j=1}^{p_y}y(k-n_{y_j})\prod_{k=1}^{q_u}u(k-n_{u_k}) \, \ i=2,\ldots, n, \nonumber
\end{align}
\end{linenomath*}
$p_y,q_u \geq 0$; $1\leq p_y+q_u \leq n_l$; $1 \leq n_{y_j}\leq n_y$;$1 \leq n_{u_k}\leq n_u$; $n_l$ is the degree of polynomial expansion; `$\mathcal{N}$' denotes the total number of data points.

\subsection{Orthogonal Regression}
\label{s:OFR}   

In this study, the orthogonalization is used to determine the significance of a particular term, which will be discussed in Section~\ref{s:termsig}. It is, therefore, pertinent to briefly discuss the concept of orthogonal regression introduced by Billings \textit{et al.}~\cite{Korenberg:Billings:1988,Chen:Billings:1989,Billings:Chen:Korenberg:1989}. In the orthogonal regression, as the name suggests, the original terms are replaced by the corresponding orthogonal terms. Given that the orthogonal terms are decoupled, the significance of each term can be determined independently in the orthogonal space. This could be achieved by a simple Gram-Schmidt orthogonalization:
\begin{linenomath*}
\begin{align*}
    w_1 = x_1, \ \ w_i = x_i - \sum \limits_{m=1}^{i-1} \frac{\langle w_m  x_i \rangle}{w_m^T w_m}
\end{align*}
\end{linenomath*}
Consequently, the NARX model (\ref{eq:NARXmodel}) can be represented in the orthogonal space as follows:
\begin{linenomath*}
\begin{align*}
    y(k) & = \sum_{i=1}^{n} w_i(k) g_i + e(k), \ \ k = 1,2,\dots \mathcal{N}
\end{align*}
\end{linenomath*}
Given that, in the orthogonal space, $w_i^Tw_j=0$ holds for $i\neq j$, the significance of the $i^{th}$ term, $x_i$, can be easily be determined using the Error-Reduction-Ratio (ERR) metric as follows:
\begin{linenomath*}
\begin{align}
\label{eq:err}
    ERR (x_i) = \frac{g_i^2 w_i^T w_i}{y^T y}, \ \ \text{where, \ } g_i = \frac{w_i^T y}{w_i^T w_i}, \   i \in [1,n]
\end{align}
\end{linenomath*}
\section{Proposed Orthogonal Floating Search Framework}
\label{s:proposedOFS}


This study aims to investigate whether the existing feature selection algorithms can be tailored to address the non-linear system identification problem. For this purpose, a new search framework is being proposed which integrates the term orthogonalization into `\textit{floating}' feature selection algorithms and therefore referred to as `\textit{Orthogonal Floating Search Framework}'. 

The term orthogonalization decouples the structure selection and the parameter estimation, which are crucial steps of the nonlinear system identification. It is, therefore, a vital first step to adapt the feature selection algorithms for system identification. Further, the following `\textit{floating}' feature selection algorithms have been adapted: \textit{Sequential Forward Floating Search} (SFFS)~\cite{Pudil:1994}, \textit{Improved Forward Floating Search} (IFFS)~\cite{Nakariyakul:Casasent:2009} and \textit{Oscillating Search} (OS)~\cite{Somol:Pudil:2000}. The motivation for the selection of these methods is two-fold: 1) The term orthogonalization can be integrated with relative ease into the floating search methods. 2) The step-wise subset building approach of the sequential floating methods (\textit{e.g.}, SFFS and IFFS) is similar to that of the Orthogonal Forward Regression (OFR-ERR)~\cite{Billings:Chen:Korenberg:1989}. Further, in OS, the depth of the search process can easily be controlled as per the prevailing requirements, as will be discussed in Section~\ref{s:o2s}.

The concept of feature significance is central to the floating search methods. In this study, we extend this concept in the context of system identification and integrate the term orthogonalization to determine the significance of a term/term subset. This is discussed in Section~\ref{s:termsig}. Further, each search method and associated adaption are discussed in the context of the system identification in Section~\ref{s:OSF}, Section~\ref{s:OIF} and Section~\ref{s:o2s}. Note that the adapted versions of SFFS, IFFS and OS are referred to respectively as Orthogonal Sequential Floating search (OSF), Orthogonal Improved sequential Floating search (OIF) and Orthogonal Oscillating Search (O$^2$S). 

\subsection{Term Significance}
\label{s:termsig}

Determination of feature significance is the core of the floating search methods~\cite{Pudil:1994,Somol:Pudil:1999,Somol:Pudil:2000,Nakariyakul:Casasent:2009}, where in each search step, the most significant feature/features are selected from the pool of available features and then the least significant feature/features are removed, from the selected feature subset. If a similar approach is to be applied to address the structure selection problem, it is essential to determine first the `\textit{significance}' of a term/term subset. For this purpose, in this study, we extend the definitions of feature significance developed by Pudil, Somol and co-workers~\cite{Pudil:1994,Somol:Pudil:1999,Somol:Pudil:2000} as follows:

Let $\mathcal{X}_k = \{ x_1, x_2, \dots x_k \}$ be the term subset selected by the search algorithm at particular search step from the superset of NARX terms, $\mathcal{X}_{model} = \{ x_1, x_2, \dots x_{n} \}$. Consequently, the subset of available terms is given by $\mathcal{X}_{model} \setminus \mathcal{X}_k $. Let $\mathcal{X}_o^i$ denote $i^{th}$ subset containing `$o$' number of terms. The criterion function corresponding to the term subset $\mathcal{X}_k$ is denoted by $J(\mathcal{X}_k)$ and is given by
\begin{align}
    \label{eq:J}
    J(\mathcal{X}_k) = \sum \limits_{i=1}^{k} ERR (x_i), \ \ \forall x_i \in \mathcal{X}_k
\end{align}
where, $ERR(x_i)$ is given by (\ref{eq:err}).

For these subsets, the significance of term/term subset is defined as follows:
\begin{definition}{} 
\label{def:1}
\textit{The most significant term with respect to $\mathcal{X}_k$ is denoted as `$x^{MS}$' and is defined as:}
\begin{align}
    x^{MS} = \{x_i \ | \ J(\mathcal{X}_k \cup x_i) = \max \limits_{i} J(\mathcal{X}_k \cup x_i), x_i \in \mathcal{X}_{model} \setminus \mathcal{X}_k \}, \ \forall i \in [1, n]
\end{align}
\end{definition}

\begin{definition}{} 
\label{def:2}
\textit{The least significant term in $\mathcal{X}_k$ is denoted as `$x^{LS}$' and is defined as:}
\begin{align}
    x^{LS} = \{x_i \ | \ J(\mathcal{X}_k \setminus x_i) = \max \limits_{i} J(\mathcal{X}_k \setminus x_i), x_i \in \mathcal{X}_k \}, \ \forall i \in [1, n] 
\end{align}
\end{definition}

\begin{definition}{} 
\label{def:3}
\textit{The most significant subset of `$o$' number of terms with respect to $\mathcal{X}_k$ is denoted as `$\mathcal{X}_o^{MS}$' and is defined as:}
\begin{align}
    \mathcal{X}_o^{MS} = \{ \mathcal{X}_o ^ i \ | \ J(\mathcal{X}_k \cup \mathcal{X}_o^i) = \max \limits_{i} J(\mathcal{X}_k \cup \mathcal{X}_o^i), \  \mathcal{X}_o^i \in \mathcal{X}_{model} \setminus \mathcal{X}_k \}
\end{align}
\end{definition}
\begin{definition}{} 
\label{def:4}
\textit{The least significant subset of `$o$' number of terms in $\mathcal{X}_k$ is denoted as `$\mathcal{X}_o^{LS}$' and is defined as:}
\begin{align}
    \mathcal{X}_o^{LS} = \{ \mathcal{X}_o ^ i \ | \ J(\mathcal{X}_k \setminus \mathcal{X}_o^i) = \max \limits_{i} J(\mathcal{X}_k \setminus \mathcal{X}_o^i), \  \mathcal{X}_o^i \in \mathcal{X}_k \}
\end{align}
\end{definition}
\begin{algorithm}[!t]
    \small
    \SetKwInOut{Input}{Input}
    \SetKwInOut{Output}{Output}
    \SetKwComment{Comment}{*/ \ \ \ }{}
    \Input{Input-output Data, $(u,y)$; Number of required terms, `$\xi$'}
    \Output{Identified Structure, $\mathcal{X}$ }
    \algorithmfootnote{Note that the OSF does not include the `\textit{Term Swapping}' procedure shown in Line~\ref{l:oif4}-\ref{l:oif5}.}
    \BlankLine
    Begin with the empty subset: $\mathcal{X} \leftarrow \varnothing$, $ k = 0$, $J(\mathcal{X}_i) =0,  \forall i \in [1,\xi]$\\
    \BlankLine
    \While{$k<\xi$}
    {
    \BlankLine
    \Comment*[h] {Sequential Forward Selection}\\
    Add the most significant term to $\mathcal{X}$, \textit{i.e.}, $\hat{\mathcal{X}} \leftarrow \{ \mathcal{X} \cup x^{MS}\}$ \nllabel{l:oif1}\\
    \BlankLine
    \uIf{$J(\hat{\mathcal{X}}) > J(\mathcal{X}_{k+1})$ \nllabel{l:oif10}}
    {$\mathcal{X} \leftarrow \hat{\mathcal{X}}$; $J(\mathcal{X}) \leftarrow J(\hat{\mathcal{X}})$ }
    \Else{$\mathcal{X} \leftarrow \mathcal{X}_{k+1}$; $J(\mathcal{X}) \leftarrow J(\mathcal{X}_{k+1})$}
    $k \leftarrow k+1$ \nllabel{l:oif11}\\
    \BlankLine
    \Comment*[h] {Backwards Elimination}\\
    Set the \textit{first exclusion flag}, $f_1=1$ \nllabel{l:oif21}\\
    \While{$k>2$ \nllabel{l:oif2}}
    {
    \BlankLine
    Identify the least significant term, $x^{LS}$, in $\mathcal{X}$\\
    \BlankLine
    \uIf{$ (f_1=1 \ \wedge \ x^{LS}=x^{MS})$ OR $J(\mathcal{X} \backslash x^{LS}) \leq J(\mathcal{X}_{k-1})$}
    {Stop Elimination.}
    \Else
    {Remove the least significant term, \textit{i.e.},\\
    $\mathcal{X} \leftarrow \mathcal{X} \backslash x^{LS}$; $J(\mathcal{X}) \leftarrow J(\mathcal{X} \backslash x^{LS})$; $k \leftarrow k-1$}
    $f_1=0$ } \nllabel{l:oif22}
    \BlankLine
    \Comment*[h] {Term Swapping : Included only in OIF }\\
    \For{$i=1$ to $k$ \nllabel{l:oif4}}
    {Remove the $i^{th}$ term, \textit{i.e.}, $\hat{\mathcal{X}}^i \leftarrow \mathcal{X} \backslash x_i$\\
     Add the most significant term to $\hat{\mathcal{X}}^i$, \textit{i.e.}, $\hat{\mathcal{X}}^i \leftarrow \{ \hat{\mathcal{X}}^i \cup x^{MS}\}$ 
    }
    Select the best swapping subset, \textit{i.e.}, $\hat{\mathcal{X}} = \{ \hat{\mathcal{X}}^i | J(\hat{\mathcal{X}}^i) = \max \limits_{i=1:k} J(\hat{\mathcal{X}}^i)  \}$\\
    \If{$J(\hat{\mathcal{X}}) > J(\mathcal{X}_k)$ \nllabel{l:oif50}}
    {$\mathcal{X} \leftarrow \hat{\mathcal{X}}$; $J(\mathcal{X}) \leftarrow J(\hat{\mathcal{X}})$\\
    \If{$k>2$}
    { Go To Step~\ref{l:oif2} for the Backward Elimination}
    }\nllabel{l:oif5}
    }
\caption{Orthogonal Floating Structure Selection.}
\label{al:oif}
\end{algorithm}
\subsection{Orthogonal Sequential Floating Search}
\label{s:OSF}

The principle of the generalized floating search was introduced by Pudil \textit{et al.} in~\cite{Pudil:1994} which is essentially a sequential approach with a distinct ability to `\textit{backtrack}', \textit{i.e.}, the \textit{selected}/\textit{discarded} term can be \textit{discarded}/\textit{selected} in later stages. It is worth to note that the floating search can proceed either in \textit{forward} (bottom-up) or \textit{backward} (top-down) direction. In this study, we focus on the forward or bottom-up approach. This is referred to as \textit{Orthogonal Sequential Floating} (OSF) search. The pseudo code for OSF is shown in Algorithm~\ref{al:oif}. Note that the OSF does not include `\textit{term swapping}' procedure (Line~\ref{l:oif4}-\ref{l:oif5}, Algorithm~\ref{al:oif}), in contrast to the improved floating search approach (which will be discussed in Section~\ref{s:OIF}).

The search process in OSF begins with an empty term subset and is continued till the term subset with the desired cardinality `$\xi$' is obtained. Each search step essentially involves two procedures: inclusion of the \textit{most significant} term followed by the removal of the \textit{least significant} term/terms through \textit{backtracking}. This procedure is discussed briefly in the following:

Let $\mathcal{X}_k$ denote a term subset with $k$ number of terms at a particular search step. First, following Definition~\ref{def:1}, the \textit{most significant} term ($x^{MS}$) with respect to $\mathcal{X}_k$, is identified from the pool of available terms. This term is included in the term subset provided it leads to a better criterion function, \textit{i.e.}, $J(\mathcal{X}_k \cup x^{MS}) > J(\mathcal{X}_{k+1})$. This procedure is outlined in Lines~\ref{l:oif1}-\ref{l:oif11}, Algorithm~\ref{al:oif}. After including a term, the focus is on removing least-significant terms present in the selected term subset through adaptive backtracking, as shown in Lines~\ref{l:oif21}-\ref{l:oif22}, Algorithm~\ref{al:oif}. It is worth to emphasize that the backtracking continues to remove the least significant term until an improvement in the criterion function is obtained, \textit{i.e.}, $J(\mathcal{X}_k \setminus x^{LS}) > J(\mathcal{X}_{k-1})$.

\begin{figure}[!t]
\parbox[t]{.42\textwidth}
{\null
  \centering
  \includegraphics[width=0.47\textwidth]{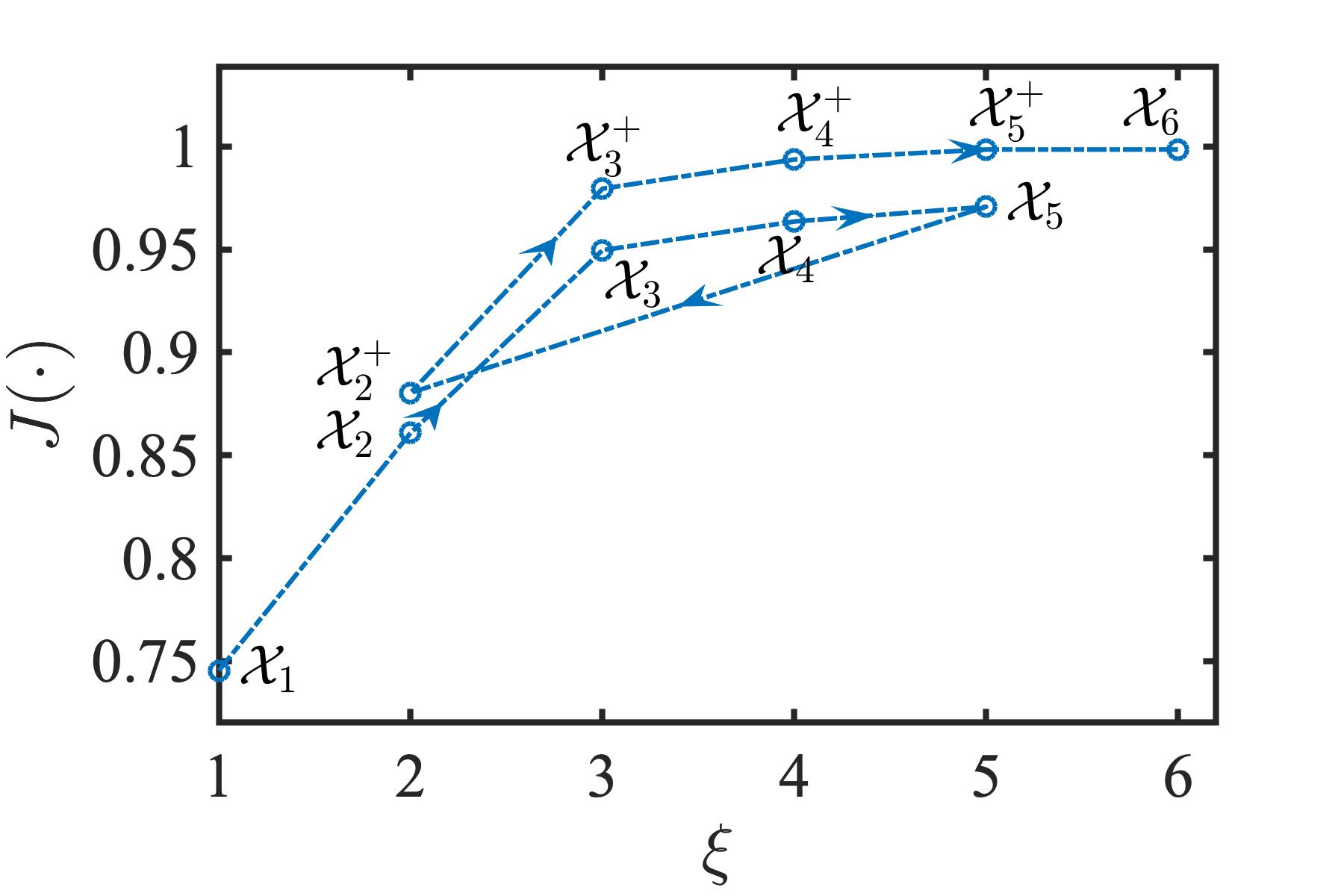}
  \captionof{figure}{Search Behavior of OSF for the numerical example with $\xi=6$. Note that the `backtracking' at Step 6 led to better subsets, \textit{e.g.}, $J(\mathcal{X}_2^+) > J(\mathcal{X}_2)$, $J(\mathcal{X}_3^+) > J(\mathcal{X}_3)$ and so on.}%
  \label{f:SB1}
}
\hfill
\parbox[t]{.57\textwidth}
{\null \centering
  \captionof{table}[t]{Search Steps Involved in OSF $^\dagger$}%
  \label{t:sbosf}%
  \vskip\abovecaptionskip
  \begin{adjustbox}{max width=0.49\textwidth}
  \begin{threeparttable}
    
    \begin{tabular}{cccc}
    \toprule
    \textbf{Step} & \boldmath$\xi_{step}$ & \textbf{Subset}$^\dagger$ & \boldmath$J(\cdotp)$ \\
    \midrule
    1     & 1     & $\mathcal{X}_1 = \{ x_1 \} $ & 0.7454  \\ [0.5ex]
    2     & 2     & $\mathcal{X}_2 = \{ x_1, \mathcal{T}_{1} \}$  & 0.8609  \\ [0.5ex]
    3     & 3     & $\mathcal{X}_3 = \{ x_1, x_3, \mathcal{T}_{1} \}$ & 0.9496  \\ [0.5ex]
    4     & 4     & $\mathcal{X}_4 = \{ x_1, x_3, x_{4}, \mathcal{T}_{1} \}$ & 0.9636  \\ [0.5ex]
    5     & 5     & $\mathcal{X}_5 = \{ x_1, x_3, x_{4}, x_{5}, \mathcal{T}_{1}\}$ & 0.9708  \\ [0.5ex]
    6     & 2     & $\mathcal{X}_{2}^+ = \{x_2, x_{5} \}$ & 0.8804   \\ [0.5ex]
    7     & 3     & $\mathcal{X}_3^+ = \{x_2, x_3, x_{5} \}$ & 0.9795   \\ [0.5ex]
    8     & 4     & $\mathcal{X}_4^+ = \{x_2, x_3, x_{5}, \mathcal{T}_{2} \}$ & 0.9937   \\ [0.5ex]
    9     & 5     & $\mathcal{X}_5^+ = \{x_1, x_2, x_3, x_{4}, x_{5} \}$ & 0.9986  \\ [0.5ex]
    10    & 6     & $\mathcal{X}_{6} = \{x_1, x_2, x_3, x_{4}, x_{5}, \mathcal{T}_{3} \}$ & 0.9986  \\ [0.5ex]
    \bottomrule
    \end{tabular}%

     \begin{tablenotes}
      \small
      \item $\dagger$ System Terms: $x_1 \rightarrow c$, $x_2 \rightarrow y(k-1)$, $x_3 \rightarrow u(k-2)$, $x_4 \rightarrow y(k-2)^2$, $x_5 \rightarrow u(k-1)^2$
      \smallskip
      \item Spurious Terms: $\mathcal{T}_1 \rightarrow y(k-1)u(k-1)^2$, $\mathcal{T}_2 \rightarrow y(k-1)y(k-2)^2$, $\mathcal{T}_3 \rightarrow u(k-1)u(k-2)^2$
    \end{tablenotes}
  \end{threeparttable}
 \end{adjustbox}
}
\end{figure} 

To gain an insight into the search dynamics of the proposed OSF, consider the following system.
\begin{align}
    \label{eq:numExample}
    \mathcal{S}_1 : y(k) = 0.5 + 0.5y(k-1) + 0.8u(k-2) - 0.05y(k-2)^2 + u(k-1)^2 + e(k)
\end{align}
For identification purposes, $1000$ input-output data-points are generated by exciting the system $\mathcal{S}_1$ with zero-mean uniform white noise sequence with unit variance, $u \sim WUN(0,1)$ and Gaussian white noise, $e \sim WGN(0,0.05)$. The model set of $165$ NARX terms is obtained with the following specifications: $[n_u, n_y, n_l] = [4,4,3]$.

The OSF is applied to identify six significant terms of $\mathcal{S}_1$, \textit{i.e.}, $\xi=6$. Note that the floating search methods require \textit{a priori} specification of subset size, `$\xi$'. A simple approach to select the appropriate subset size $\xi$ and the associated issues will be discussed in Section~\ref{s:XiSel}. The search behavior of OSF is explained as follows: In each search step, the term subset, its cardinality and the corresponding criterion function are recorded, which are shown in Table~\ref{t:sbosf} and Fig.~\ref{f:SB1}. For the sake of clarity, the system terms and spurious terms are respectively denoted by `$x$' and `$\mathcal{T}$'. Further, a better subset for the given cardinality is annotated by `$^+$', \textit{e.g.}, $\mathcal{X}_2^+$ indicates that $J(\mathcal{X}_2^+) > J(\mathcal{X}_2)$. 

The positive effects of backtracking are clearly evident in Fig.~\ref{f:SB1}. The backtracking occurs after the subset $\mathcal{X}_5$ is obtained. Note that the subsets obtained after backtracking ($\mathcal{X}_2^+, \mathcal{X}_3^+,\mathcal{X}_4^+$ and $\mathcal{X}_5^+$) yield improved criterion function in comparison to the previous subsets with the similar cardinality ($\mathcal{X}_2, \mathcal{X}_3, \mathcal{X}_4$ and $\mathcal{X}_5$). A closer examination of subsets given in Table~\ref{t:sbosf} reveals that the backtracking could remove the spurious term `$\mathcal{T}_1$' which is first included at Step 2 and remains in the selected subsets till Step 5. Further, the subset containing all the system terms, $\mathcal{X}_5^+ = \{ x_1, x_2, x_3, x_4, x_5 \}$, is obtained during backtracking at Step 6. This subset is eventually selected at Step 9 when the forward inclusion procedure failed to yield a better subset, \textit{i.e.}, the search is restored to previous best subset (in this case $\mathcal{X}_5^+$, as $J(\mathcal{X}_4^+ \cup x^{MS}) < J(\mathcal{X}_5^+)$; see Line~\ref{l:oif10}-\ref{l:oif11}, Algorithm~\ref{al:oif}).
\subsection{Orthogonal Improved Floating Search}
\label{s:OIF}

The next algorithm considered in this study is based on the improved floating search proposed in~\cite{Nakariyakul:Casasent:2009}. This algorithm adds a new procedure referred to as `\textit{swapping}' to replace a weak feature, besides retaining both the procedures of the floating search. The improved floating search is adapted for the structure selection by including the ERR metric associated with orthogonal least square algorithms of~\cite{Billings:Chen:Korenberg:1989}. This is referred to as Orthogonal Improved Floating (OIF) search and discussed briefly in the following.


The procedures involved in OIF are shown in Algorithm~\ref{al:oif}. As discussed earlier, OIF retains the `\textit{forward inclusion}' (Line~\ref{l:oif1}-\ref{l:oif11}, Algorithm~\ref{al:oif}) and `\textit{backtracking}' (Line~\ref{l:oif21}-\ref{l:oif22}, Algorithm~\ref{al:oif}) procedures of the OSF and the same discussions are valid here. In addition, the \textit{swapping} procedure is introduced after \textit{backtracking} as shown in Line~\ref{l:oif4}-\ref{l:oif5}. The objective of this procedure is to replace a weak/non-significant term which is explained as follows:

Assume that $\mathcal{X}_k = \{x_1, x_2, \dots x_k \} $ is the term subset obtained after \textit{backtracking}. To replace a non-significant term, several new candidate term subsets are generated from $\mathcal{X}_k$. The $i^{th}$ candidate term subset (denoted as $\hat{\mathcal{X}^i}$) is generated from $\mathcal{X}_k$ in the following two steps:
\begin{enumerate}
    \item First, the $i^{th}$ term is removed from $\mathcal{X}_k$, \textit{i.e.}, $\hat{\mathcal{X}^i} \leftarrow \{ \mathcal{X}_k \setminus x_i \}$
    \item  Next, the most significant term with respect to $\hat{\mathcal{X}^i}$ is included, \textit{i.e.}, $\hat{\mathcal{X}^i} \leftarrow \{ \hat{\mathcal{X}^i} \cup x^{MS} \}$
\end{enumerate}    
Following these steps, a total of `$k$' candidate term subsets are obtained from $\mathcal{X}_k$ and the subset with the highest criterion function, $J(\cdotp)$, is referred to as `\textit{swapping subset}'(denoted by $\hat{\mathcal{X}}$). If the swapping subset yields an improved criterion function, then it replaces the current term subset $\mathcal{X}_k$ and sent for backtracking, as outlined in Line~\ref{l:oif50}-\ref{l:oif5}, Algorithm~\ref{al:oif}. 

\begin{figure}[!t]
\parbox[t]{.49\textwidth}
{\null
  \centering
  \includegraphics[width=0.47\textwidth]{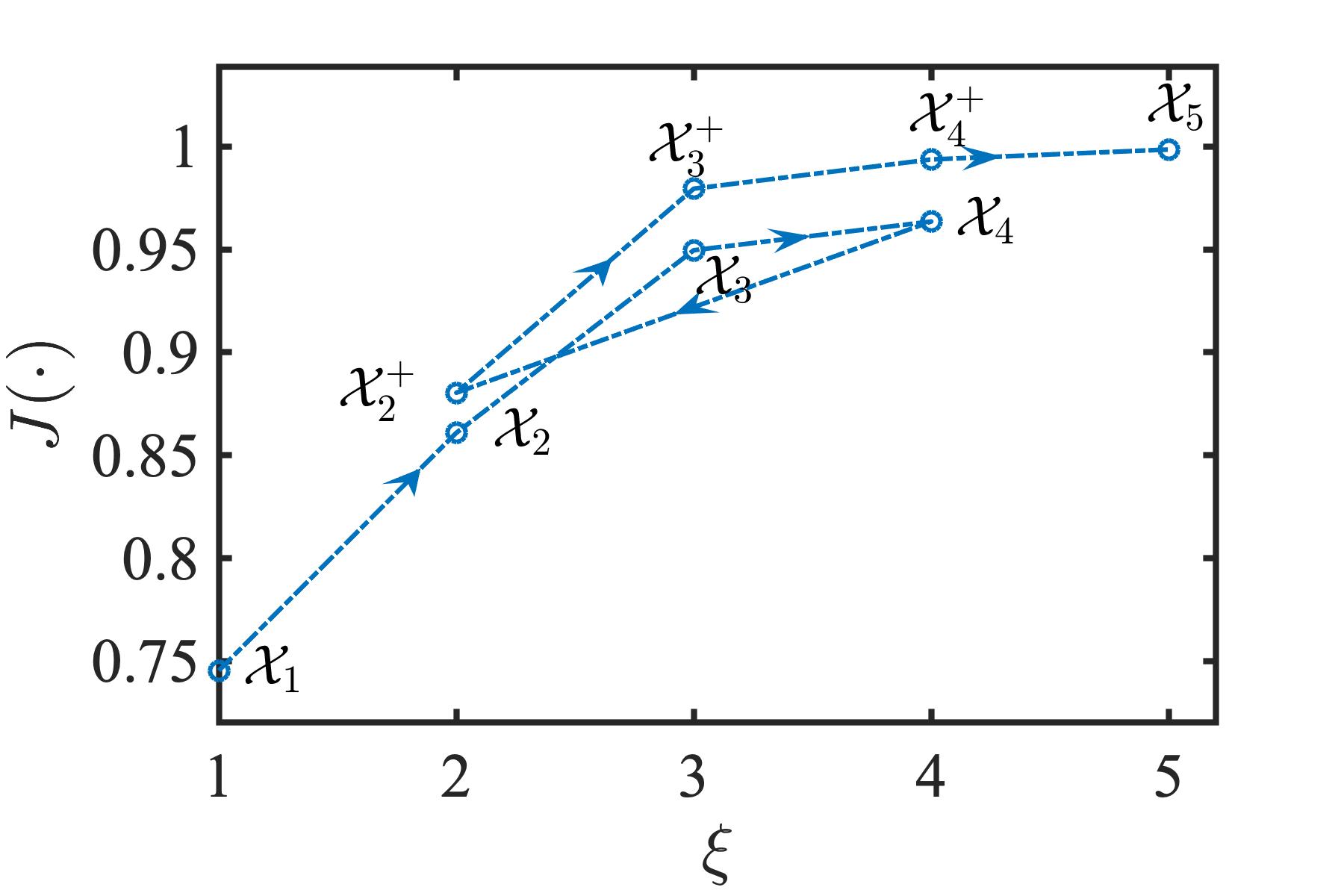}
  \captionof{figure}{Search Behavior of OIF for the numerical example with $\xi=5$. Note the backtracking after $\mathcal{X}_4$ leading to better subsets, $\mathcal{X}_2^+$, $\mathcal{X}_3^+$ and $\mathcal{X}_4^+$.}%
  \label{f:SB2}
}
\hfill
\parbox[t]{.5\textwidth}
{\null \centering
  \captionof{table}[t]{Search Steps Involved in OIF $^\dagger$}%
  \label{t:sboif}%
  \vskip\abovecaptionskip
  \begin{adjustbox}{max width=0.49\textwidth}
  \begin{threeparttable}
    
    \begin{tabular}{cccc}
    \toprule
    \textbf{Step} & \boldmath$\xi_{step}$ & \textbf{Subset} & \boldmath$J(\cdotp)$ \\
    \midrule
    1     & 1     & $\mathcal{X}_1 = \{ x_1 \} $ & 0.7454  \\ [0.5ex]
    2     & 2     & $\mathcal{X}_2 = \{ x_1, \mathcal{T}_{1} \}$  & 0.8609  \\ [0.5ex]
    3     & 3     & $\mathcal{X}_3 = \{ x_1, x_3, \mathcal{T}_{1} \}$ & 0.9496  \\ [0.5ex]
    4     & 4     & $\mathcal{X}_4 = \{ x_1, x_3, x_{4}, \mathcal{T}_{1} \}$ & 0.9636  \\ [0.5ex]
    5     & 2     & $\mathcal{X}_2^+ = \{x_2, x_{5} \}$ & 0.8804  \\ [0.5ex]
    6     & 3     & $\mathcal{X}_3^+ = \{x_2, x_3, x_{5} \}$ & 0.9795  \\ [0.5ex]
    7     & 4     & $\mathcal{X}_4^+ = \{x_2, x_3, x_{5}, \mathcal{T}_{2} \}$ & 0.9937  \\ [0.5ex]
    8     & 5     & $\mathcal{X}_5 = \{x_1, x_2, x_3, x_{4}, x_{5} \}$ & 0.9986  \\ [0.5ex]
    \bottomrule
    \end{tabular}%

     \begin{tablenotes}
      \small
      \item $\dagger$ System Terms: $x_1 \rightarrow c$, $x_2 \rightarrow y(k-1)$, $x_3 \rightarrow u(k-2)$, $x_4 \rightarrow y(k-2)^2$, $x_5 \rightarrow u(k-1)^2$
      \smallskip
      \item Spurious Terms: $\mathcal{T}_1 \rightarrow y(k-1)u(k-1)^2$, $\mathcal{T}_2 \rightarrow y(k-1)y(k-2)^2$
    \end{tablenotes}
  \end{threeparttable}
 \end{adjustbox}
}
\end{figure}

Thus, OIF takes a two-pronged approach to remove weak/non-significant terms: the \textit{backtracking} removes the weak term/terms and the \textit{swapping} replaces a weak term. Especially, the swapping procedure encourages the exploration of new term subsets which is likely to yield improved search performance in comparison to OSF. To investigate this further, the OIF is applied to identify the significant terms of the system $\mathcal{S}_1$ considered in (\ref{eq:numExample}) under the similar test conditions.  

The final subsets obtained in each step and the corresponding criterion function are shown in Table~\ref{t:sboif}. The variations in the criterion function are shown in Fig.~\ref{f:SB2}. Though, the overall search dynamics of both OSF (Fig.~\ref{f:SB1}) and OIF (Fig.~\ref{f:SB2}) may appear similar in nature, it is interesting to note that the backtracking in OIF occurs one step earlier (Step 5, Table~\ref{t:sboif}) in comparison to OSF (Step 6, Table~\ref{t:sbosf}). This could be explained by the swapping procedure of OIF which enables the inclusion of missing system term `$x_2$' one step earlier.

\begin{algorithm}[!t]
    \small
    \SetKwInOut{Input}{Input}
    \SetKwInOut{Output}{Output}
    \SetKwComment{Comment}{*/ \ \ \ }{}
    \Input{Input-output Data, $(u,y)$; Number of required terms, `$\xi$'}
    \Output{Identified Structure, $\mathcal{X}_\xi$ }
    \BlankLine
    \Comment*[h] {Initialize}\\
    \BlankLine
    Select the maximum search depth, $\mathcal{O}_{max}$\\
    \BlankLine
    Set the search depth, $o=1$ and `\textit{flags}', $f_1=0$, $f_2=0$\\
    $\mathcal{X}_\xi \leftarrow \varnothing$
    \BlankLine
    \For{i=1 to $\xi$ \nllabel{l:o2s10}}
     {Add the most significant term to $\mathcal{X}_\xi$, \textit{i.e.}, $\mathcal{X}_\xi \leftarrow \{ \mathcal{X}_\xi \cup x^{MS}\}$} \nllabel{l:o2s11}
    \BlankLine
    \While{$o < \mathcal{O}_{max}$}
    {
    \BlankLine
    \BlankLine
    \Comment*[h] {Down Swing}\\
    \BlankLine
    Remove `$o$' number of least significant terms from $\mathcal{X}_\xi$, \textit{i.e.}, $\mathcal{X}_{\xi-o} \leftarrow \mathcal{X}_\xi \backslash \mathcal{X}_o^{LS}$ \nllabel{l:ols20} \\
    \BlankLine
    Add `$o$' number of most significant terms to $\mathcal{X}_{\xi-o}$, \textit{i.e.}, $\hat{\mathcal{X}}_\xi \leftarrow \{ \mathcal{X}_{\xi-o} \cup \mathcal{X}_o^{MS} \} $\\
    \BlankLine
    \uIf{$J(\hat{\mathcal{X}}_\xi)>J(\mathcal{X}_\xi)$}
    {$\mathcal{X}_\xi \leftarrow \hat{\mathcal{X}}_\xi$; $J(\mathcal{X}_\xi) \leftarrow J(\hat{\mathcal{X}}_\xi)$; $f_1 \leftarrow 0$ \Comment*[h] {better subset is found}}
   \Else{$f_1 \leftarrow 1$ \Comment*[h] {down swing is unsuccessful}} \nllabel{l:ols21}
    \BlankLine
    \If{$f_1=1 \ \wedge \ f_2=1$}{$o=o+1$ \Comment*[h] {increase search depth}}
    \BlankLine
    \BlankLine
    \Comment*[h] {Up Swing}\\
    \BlankLine
    Add `$o$' number of most significant terms to $\mathcal{X}_\xi$, \textit{i.e.}, $\mathcal{X}_{\xi+o} \leftarrow \{ \mathcal{X}_\xi \cup \mathcal{X}_o^{MS} \} $ \nllabel{l:o2s30}\\
    \BlankLine
    Remove `$o$' number of least significant terms from  $\mathcal{X}_{\xi+o}$, \textit{i.e.}, $\hat{\mathcal{X}}_\xi \leftarrow \mathcal{X}_{\xi+o} \backslash \mathcal{X}_o^{LS}$ \\
    \BlankLine
    \uIf{$J(\hat{\mathcal{X}}_\xi)>J(\mathcal{X}_\xi)$}
    {$\mathcal{X}_\xi \leftarrow \hat{\mathcal{X}}_\xi$; $J(\mathcal{X}_\xi) \leftarrow J(\hat{\mathcal{X}}_\xi)$; $f_2 \leftarrow 0$ \Comment*[h] {better subset is found}}
   \Else{$f_2 \leftarrow 1$ \Comment*[h] {up swing is unsuccessful}}\nllabel{l:o2s31} 
    \BlankLine
    \If{$f_1=1 \ \wedge \ f_2=1$}{$o=o+1$ \Comment*[h] {increase search depth}}
    } 
\caption{Orthogonal Oscillating Search (O$^2$S) for Structure Selection }
\label{al:o2s}
\end{algorithm}
\subsection{Orthogonal Oscillating Search}
\label{s:o2s}

The third algorithm considered in this study is the oscillating search introduced by Somol and Pudil in~\cite{Somol:Pudil:2000} which is the natural successor of the floating search principle. It retains one of the key properties of floating search,\textit{i.e.}, the ability to \textit{backtrack}. Unlike the floating search principle, where the focus is on the individual feature, the oscillating search focuses on a subset of features. The basic search engine in this method is the `\textit{swing}' procedure where the subset under consideration is perturbed by successive \textit{addition}/\textit{removal} and \textit{removal}/\textit{addition} of multiple features. In this study, the oscillating search is also adapted for the structure selection by the introduction of `orthogonalization'  and referred to as Orthogonal Oscillating Search (O$^2$S). The steps involved in O$^2$S are outlined in Algorithm~\ref{al:o2s} and discussed briefly in the following:

Let `$\xi$' denote the required number of terms. The search is initialized with a subset, $\mathcal{X}_\xi$, containing $\xi$ number of terms. This initial subset is obtained by successively adding the most significant terms (See Definition~\ref{def:1}) as outlined Line~\ref{l:o2s10}-\ref{l:o2s11}, Algorithm~\ref{al:o2s}. Subsequently, this subset is perturbed throughout the search process by the `\textit{swing}' procedures. The rationale behind the swing procedure is that a better subset could be obtained by replacing multiple (say, `$o$') number of weak or non-significant terms present in the current subset with the similar number of relevant or significant terms. This could be achieved by the following two `swing' procedures: 1) Down Swing and 2) Up Swing. 

During the `\textit{down swing}', at first `$o$' number of least significant terms in $\mathcal{X}_\xi$ are removed which yields a subset of $(\xi - o)$ terms, $\mathcal{X}_{\xi-o}$. Subsequently, `$o$' number of most significant terms are added to $\mathcal{X}_{\xi-o}$ to yield a new, and possibly a better subset of $\xi$ terms, $\hat{\mathcal{X}_\xi}$. This new subset is retained provided the criterion function is improved. The down swing procedure is outlined in  Line~\ref{l:ols20}-\ref{l:ols21}, Algorithm~\ref{al:o2s}.

\begin{table}[!t]
  \centering
  \caption{Search Steps Involved in O$^2$S with $\xi=5$ and $\mathcal{O}_{max}=2$}
  \small
  \label{t:sbo2s}%
  \begin{adjustbox}{max width=0.8\textwidth}
  \begin{threeparttable}
    
    \begin{tabular}{cccccccc}
    \toprule
    \textbf{Step} & \makecell{\textbf{Search} \\ \textbf{Depth} \\ `\boldmath$o$'}     & \boldmath$\xi_{step}$ & \textbf{Subset}$^\dagger$ & \boldmath$J(\cdotp)$ & \boldmath$f_1$ & \boldmath$f_2$ & \textbf{Remark} \\
    \midrule

    1     & -     & 5     & $\mathcal{X}_5 =\{ x_1, x_3, x_4, x_5, \mathcal{T}_1 \}$ & 0.97079 & 0     & 0     & Initial Model \\[0.8ex]
    \midrule
    2     & \multirow{2}[2]{*}{1} & 4 & $\mathcal{X}_4 =\{x_1, x_3, x_4, \mathcal{T}_1 \}$ &  &       &       & \multirow{2}{*}{\makecell{Down Swing \\ Did Not \\ Improve}} \\ [1.5ex]
    3     &       & 5     & $\mathcal{X}_5 =\{ x_1, x_3, x_4, x_5, \mathcal{T}_1 \}$ & 0.97079 & 0     & 0     &  \\[1.5ex]
    \midrule
    4     & \multirow{2}[2]{*}{1} & 6 & $\mathcal{X}_6 =\{x_1, x_2, x_3, x_4, x_5, \mathcal{T}_2\}$ &  &       &       & \multirow{2}{*}{\makecell{Up Swing\\ Improved}} \\[0.8ex]
    5     &       & 5     & \boldmath{$\mathcal{X}_5^+ =\{x_1, x_2, x_3, x_4, x_5 \}$} & \boldmath{$0.99856$} & 1     & 0     &  \\[0.8ex]
    \midrule
    6     & \multirow{2}[2]{*}{1} & 4     & $\mathcal{X}_4 =\{ x_2, x_3, x_4, x_5 \}$ &  &       &       & \multirow{2}{*}{\makecell{Down Swing \\ Did Not \\ Improve}} \\[1.5ex]
    7     &       & 5     & $\mathcal{X}_5 =\{ x_1, x_2, x_3, x_4, x_5 \}$ & 0.99856 & 1     & 0     &  \\[1.5ex]
    \midrule
    8     & \multirow{2}[2]{*}{1} & 6     & $\mathcal{X}_6 =\{ x_1, x_2, x_3, x_4, x_5, \mathcal{T}_2 \}$ &  &       &       & \multirow{2}{*}{\makecell{Up Swing\\ Did Not \\ Improve}} \\[1.5ex]
    9     &       & 5     & $\mathcal{X}_5 =\{ x_1, x_2, x_3, x_4, x_5 \}$ & 0.99856 & 1     & 1     &  \\[1.5ex]
    \midrule
    10    & \multirow{2}[2]{*}{2} & 3     &  $\mathcal{X}_3 =\{ x_2, x_3, x_5 \}$ &  &       &       & \multirow{2}{*}{\makecell{Down Swing \\ Did Not \\ Improve}} \\[1.5ex]
    11    &       & 5     & $\mathcal{X}_5 =\{ x_2, x_3, x_5, \mathcal{T}_3, \mathcal{T}_4 \}$ & 0.99505 & 1     & 1     &  \\[1.5ex]

    \bottomrule
    \end{tabular}%

      \begin{tablenotes}
      \small
      \item $\dagger$ System Terms: $x_1 \rightarrow c$, $x_2 \rightarrow y(k-1)$, $x_3 \rightarrow u(k-2)$, $x_4 \rightarrow y(k-2)^2$, $x_5 \rightarrow u(k-1)^2$
      \medskip
      \item Spurious Terms: $\mathcal{T}_1 \rightarrow y(k-1)u(k-1)^2$, $\mathcal{T}_2 \rightarrow u(k-1) u(k-2)^2$, $\mathcal{T}_3 \rightarrow y(k-1)y(k-2)^2$, $\mathcal{T}_4 \rightarrow y(k-2)^3$
    \end{tablenotes}
    
  \end{threeparttable}
 \end{adjustbox}
\end{table}%

The Up-swing procedure, as the name suggests, first adds `$o$' number of most significant terms to $\mathcal{X}_\xi$ to yield a subset $\mathcal{X}_{\xi+o}$. Next, a new subset $\hat{\mathcal{X}_\xi}$ is obtained by removing the $o$ number of least significant terms from $\mathcal{X}_{\xi+o}$. The other aspects are similar to the down swing. This procedure is outlined in Line~\ref{l:o2s30}-\ref{l:o2s31}, Algorithm~\ref{al:o2s}.

It is clear that both swing procedures require identification of the following two subsets with respect to the term subset under consideration:  the subset of `$o$' most significant terms (denoted as $\mathcal{X}_o^{MS}$; see Definition~\ref{def:3}) and the subset containing `$o$' weak/non-significant terms (denoted as $\mathcal{X}_o^{LS}$; see Definition~\ref{def:4}). In this study, $\mathcal{X}_o^{MS}$ is identified using OSF search procedure described in Section~\ref{s:OSF}. The set of weak terms $\mathcal{X}_o^{LS}$ is identified using the \textit{top-down} or \textit{backward search} variant of OSF. 

Further, the `search depth' (denoted by `$o$') determines the number of terms which are either to be added or to be removed in a particular swing. The search depth is adaptive and its value `$o$' is dependent on the search dynamics. The search begins with $o=1$ and if two successive swings fail to identify a better term subset then it is incremented by $1$. The search depth is reset to $1$ whenever swing leads to an improved subset. The search is terminated when `$o$' reaches to a pre-specified maximum search depth, $\mathcal{O}_{max}$. Consequently, the depth of search can easily be controlled by varying $\mathcal{O}_{max}$. This is one of the distinct and important ability of O$^2$S which regulates the search efforts as per the prevailing requirement. 

\begin{figure}[!t]
\centering
\small
  \includegraphics[width=0.5\textwidth]{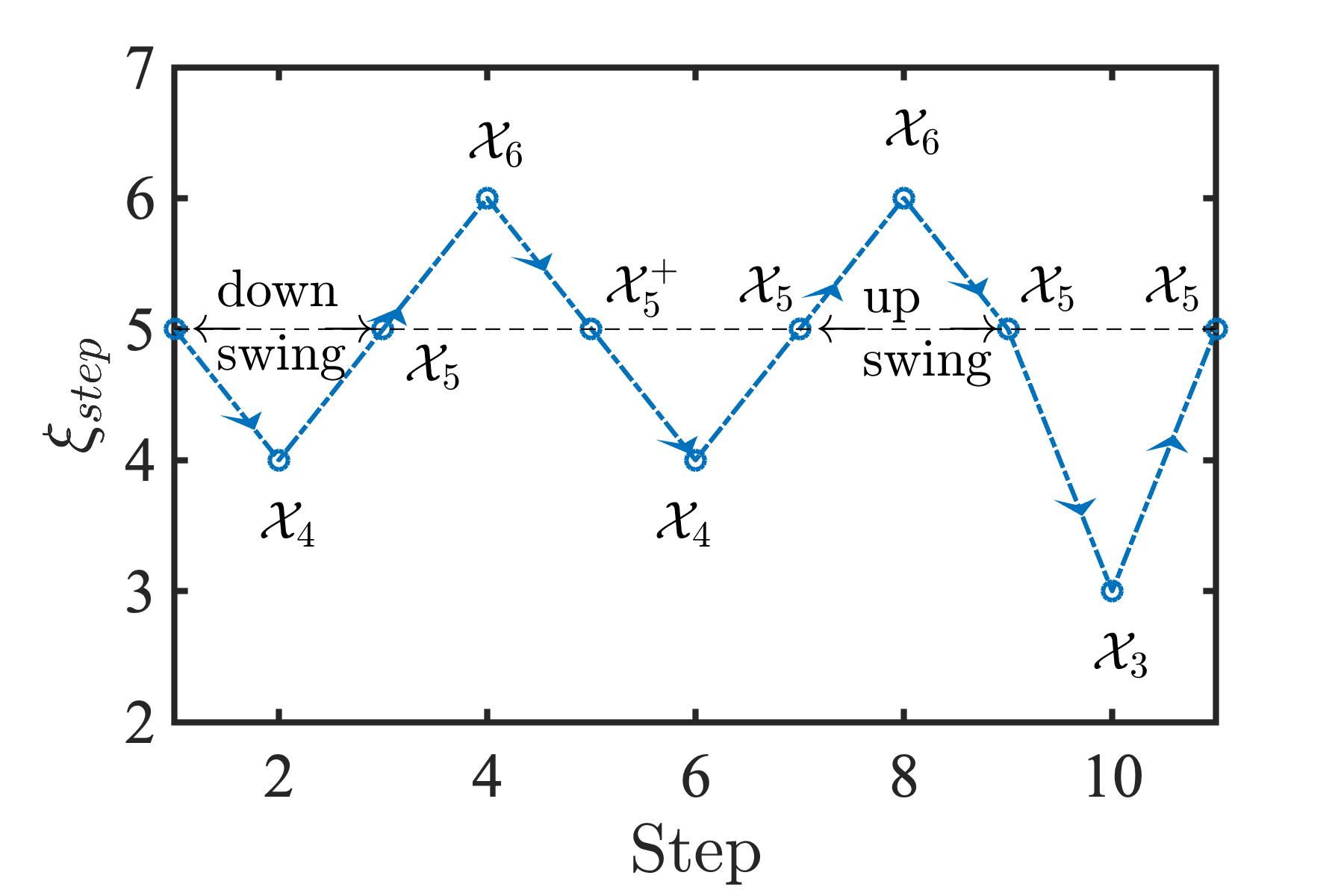}
\caption{The variation in subset cardinality during the O$^2$S Search}
\label{f:SBO2S}
\end{figure}

To analyze the search behavior of O$^2$S, it is applied to identify system terms of the system $\mathcal{S}_1$ considered in the numerical example~(\ref{eq:numExample}) with the following specifications: $\xi=5$ and $\mathcal{O}_{max}=2$. During this search for system terms, the term subset and the corresponding criterion function obtained in each step are recorded and shown here in Table~\ref{t:sbo2s}. The change in subset cardinality during the search process is shown in Fig.~\ref{f:SBO2S}. 

The search is initialized by forward inclusion of $5$ terms, as per Line~\ref{l:o2s10}-\ref{l:o2s11}, Algorithm~\ref{al:o2s}. This initial subset selects one spurious term, instead of the system term $x_2$, as shown in Step 1, Table~\ref{t:sbo2s}. The search begins by the down swing procedure which does not improve the subset, as seen in Step 3, Table~\ref{t:sbo2s}. However, in the subsequent up-swing, a better term subset, $\mathcal{X}_5^+$, is identified, \textit{i.e.}, $J(\mathcal{X}_5^+) > J(\mathcal{X}_5)$. Note that at this stage, all the system terms have been identified and the spurious term is discarded, as seen in Step 5, Table~\ref{t:sbo2s}. The subsequent down swing (Step 7) and up swing (Step 9) identifies the same subset with no improvement in the criterion function, $J(\cdotp)$. Consequently, the search depth is increased by `$1$', \textit{i.e.}, $o=2$. The search terminates at Step 11 as the down swing could not find a better subset and any further increase in search depth will lead to $o>\mathcal{O}_{max}$.

Note that the variation in subset cardinality during the aforementioned swing procedures are clearly visible in Fig.~\ref{f:SBO2S}. Especially, see the drop in subset cardinality at Step 10. The increase in search depth at this step requires the search for the term subset with $\{ \xi-o \} =3$ number of terms. 

\subsection{Selection of Subset Cardinality (Model Order Selection)}
\label{s:XiSel}

The floating search algorithms require the specification of `subset cardinality' or number of terms to be identified. Given that this is not known \textit{a priori}, in this study, the following procedure is followed to estimate the subset cardinality or `\textit{model order}'.

For the system under consideration, a search algorithm is applied to identify several term subsets of increasing cardinality in a predefined search interval, denoted by $[\xi_{min}, \xi_{max}]$. A set of subsets thus obtained can be used to estimate the model order using an appropriate Information Criterion (IC). The objective here is to locate a `\textit{plateau}' or `\textit{knee-point}' in the information criteria which would indicate an acceptable compromise for the \textit{bias-variance} dilemma. The term subsets corresponding to such \textit{knee-point} or \textit{plateau} can be selected as the system model. 

To further understand this procedure, consider the system $\mathcal{S}_1$ in numerical example~(\ref{eq:numExample}). For this system, OIF is applied to identify term subsets of increasing cardinality in the range of $[\xi_{min}, \xi_{max}]=[2,20]$. Thus, a family identified term subsets, denoted by `$\Omega$', is obtained as follows: 
\begin{align}
\label{eq:omega}
    \Omega = \{ \mathcal{X}_{\xi_{min}}, \mathcal{X}_{\xi_{min}+1}, \dots, \mathcal{X}_{\xi_{max}} \}
\end{align}

\begin{figure}[!t]
\centering
\small
  \includegraphics[width=0.48\textwidth]{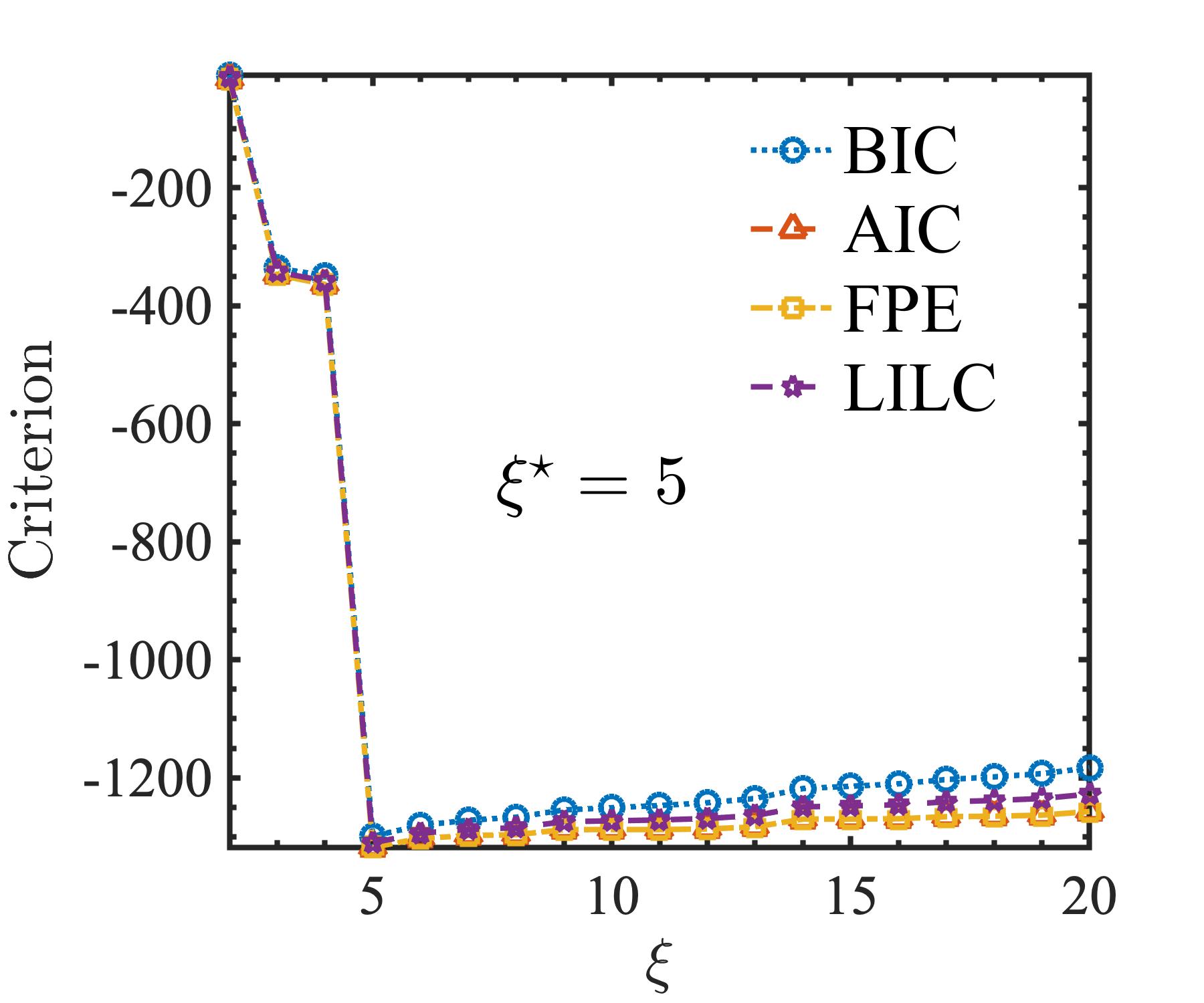}
\caption{The selection of subset cardinality or `model order'. $\xi^{\star}$ denotes the known cardinality of the system. Note that the knee-point for all criteria is obtained at $\xi=5$.}
\label{f:modelorder}
\end{figure}

Next, for each identified subset, $\mathcal{X}_\xi \in \Omega$, the various information criteria given in~\ref{s:appIC} are determined. Fig.~\ref{f:modelorder} shows the variation in information criteria as cardinality is varied from $\xi_{min}$ to $\xi_{max}$. It is observed that the `\textit{knee-point}' for all the criteria is obtained at $\xi=5$. This coincides with the actual cardinality of the system $\mathcal{S}_1$.

 In several comparative investigations~\cite{Stoica:Selen:2004a,Hafiz:Swain:CEC:2018}, the Bayesian Information Criterion (BIC) was found to be comparatively robust among the existing information criteria. Therefore, for the remainder of the study, the cardinality corresponding to the minimum BIC is selected as the model order. Although following this approach the correct model order could be obtained for all test system (as will be shown in Section~\ref{s:compEval}), in practice, we recommend the closer inspection of several term subsets surrounding the plateau or knee-point; as the existing research indicated that the information criteria tend to over-fit~\cite{Stoica:Selen:2004a}.

Since the existing research suggest that the dynamics of many nonlinear systems can be captured with $15$ or fewer significant terms~\cite{Chen:Billings:1989}, a pragmatic approach is followed in this study to fix the search interval of the cardinality as follows: $[\xi_{min}, \xi_{max}]$=$[2,20]$. This can be considered as a thumb-rule. However, some exception may exist in certain non-linear systems. A simple procedure to determine the cardinality interval in such a scenario is discussed in Section~\ref{s:commentModelOrder}.

\begin{algorithm}[!t]
    \small
    \SetKwInOut{Input}{Input}
    \SetKwInOut{Output}{Output}
    \SetKwComment{Comment}{*/ \ \ \ }{}
    \Input{Input-output Data, $(u,y)$}
    \Output{Identified Model, Structure and Coefficients}
    \BlankLine
    \Comment*[h] {Data Pre-processing}\\
    \BlankLine
    Generate set of model terms by specifying $n_u, n_y$ and $n_l$ of the NARX model \\
    \BlankLine
    \BlankLine
    \Comment*[h] {Search for the system structure}\\
    \BlankLine
    Select a orthogonal floating search algorithm: OSF, OIF (Algorithm~\ref{al:oif}) or O$^2$S (Algorithm~\ref{al:o2s})\\
    Select the search interval, $[\xi_{min}, \xi_{max}]$ \nllabel{line:1}\\
    $\Omega \leftarrow \varnothing$
    \BlankLine
    \For{each `k', $k\in[\xi_{min}, \xi_{max}]$} 
        { \BlankLine
           Identify subset of `$k$' significant terms,  $\mathcal{X}_k$\\
           $\Omega \leftarrow \{ \Omega \cup \mathcal{X}_k \} $
         } 
     \BlankLine
     Select the term-subset ($\mathcal{X}^{alg}$) with the minimum BIC, \textit{i.e.}, $\mathcal{X}^{alg} = \{ \mathcal{X}_i \ | \ BIC(\mathcal{X}_i) = \arg \min BIC(\mathcal{X}_i), \ \  \forall \ \mathcal{X}_i \in \Omega $ \}\nllabel{line:2}\\

\caption{Orthogonal Floating Structure Selection}
\label{al:genfloat}
\end{algorithm}

\section{Investigation Framework}
\label{s:IF}

In the following, the framework of this investigation is discussed. The overall procedure followed for the structure selection is shown in Algorithm~\ref{al:genfloat}. The test nonlinear system considered in this study are discussed in Section~\ref{s:Data}. Further, the possible search outcomes are discussed in Section~\ref{s:PM}, to evaluate the term subset found by the search algorithms qualitatively. 

\subsection{Test Non-linear Systems}
\label{s:Data}

In this study, a total of $6$ non-linear system have been selected from the existing investigations on structure selection~\cite{Mendes:1995,Mao:Billings:1997,Bonin:Pirrodi:2010,Piroddi:Spinelli:2003,Baldacchino:Kadirkamanathan:2013,Falsone:Piroddi:2015} and shown here in Table~\ref{t:sys}. The systems are excited by a white noise sequence having either uniform or Gaussian distribution as shown in Table~\ref{t:sys}. For identification purposes, a total of $1000$ input-output data points, $(u,y)$, are generated from each system. The structure selection is performed following the principle of \textit{cross-validation}; where $700$ data points are selected for the estimation purpose and the remaining data points are used for validation, \textit{i.e.}, $\mathcal{N}_v = 300$. For each system, the NARX model is generated by setting the input-output lags and the degree of non-linearity to: $[n_u,n_y,n_l]=[4,4,3]$. This gives the NARX model set, $X_{model}$, with a total of $165$ terms following~(\ref{eq:Nt}), \textit{i.e.}, $n=165$.
\begin{table}[!h]
  \centering
  \scriptsize
  \caption{Test Non-linear Systems}
  \label{t:sys} 
  \begin{adjustbox}{max width=\textwidth}
  \begin{threeparttable}
    \begin{tabular}{ccccccccc}
    \toprule
    
    \textbf{System} & \textbf{Known Structure} & \textbf{Input}(\boldmath$u$)$^\dagger$ & \textbf{Noise} (\boldmath$e$)$^\dagger$  \\
    \midrule

    $\mathcal{S}_1$    &  $y(k) = 0.5 + 0.5y(k-1) + 0.8u(k-2) + u(k-1)^2 - 0.05y(k-2)^2 + e(k)$     &  WUN$(0,1)$ & WGN$(0,0.05)$\\[2ex]
    $\mathcal{S}_2$    &  $y(k) = 0.5y(k-1) + 0.3u(k-1) + 0.3u(k-1)y(k-1) + 0.5u(k-1)^2 + e(k)$     &  WUN$(0,1)$ & WGN$(0,0.002)$\\ [2ex]
    $\mathcal{S}_3$    &  $y(k) = 0.8y(k-1) + 0.4u(k-1) + 0.4u(k-1)^2 + 0.4u(k-1)^3 + e(k) $         &  WGN$(0,1)$ & WGN$(0,0.33^2)$\\[4ex]
    
    $\mathcal{S}_4$    &  $y(k) =$  \makecell{$ 0.1586 y(k-1) + 0.6777 u(k-1) + 0.3037 y(k-2)^2$\\[0.7ex]
                                   $ -0.2566 y(k-2) u(k-1)^2 - 0.0339 u(k-3)^3 + e(k)$} &  WUN$(0,1)$ & WGN$(0,0.002)$\\[6ex]
                                   
    $\mathcal{S}_5$    &  $y(k) = $  \makecell{$0.7 y(k-1)u(k-1) - 0.5 y(k-2)$\\[0.7ex]
    $+ 0.6 u(k-2)^2 - 0.7 y(k-2)u(k-2)^2 + e(k)$} &  WUN$(-1,1)$ & WGN$(0,0.004)$\\[6ex]
                                   
    $\mathcal{S}_6$    &  $y(k) =$  \makecell{$0.2 y(k-1)^3 + 0.7 y(k-1)u(k-1) + 0.6 u(k-2)^2$\\[0.7ex]
                    $- 0.7 y(k-2)u(k-2)^2 - 0.5 y(k-2) + e(k)$} &  WUN$(-1,1)$ & WGN$(0,0.004)$\\[6ex]
    
    \bottomrule
    \end{tabular}%
    \begin{tablenotes}
      \scriptsize
       \item $\dagger$ WUN$(a,b)$ denotes white uniform noise sequence in the interval $[a,b]$; WGN$(\mu,\sigma)$ denotes white Gaussian noise sequence with the mean `$\mu$' and the variance `$\sigma$'.  
    \end{tablenotes}
  \end{threeparttable}
  \end{adjustbox}
\end{table}%

\subsection{Search Outcomes}
\label{s:PM}

For comparative evaluation purposes, the term subset found by each search algorithm is \textit{qualitatively} evaluated. For this purpose, the following term sets are defined with reference to the NARX model given by (\ref{eq:NARXmodel}),
\begin{itemize}
    \item $\mathcal{X}_{model}$ : the set containing all terms of the NARX model
    \item $\mathcal{X}^{\star}$ : the optimum term subset or the set of system terms, $\mathcal{X}^{\star} \subset \mathcal{X}_{model}$
    \item $\mathcal{X}^{alg}$ : subset of terms identified by the search algorithm, $\mathcal{X}^{alg} \subset \mathcal{X}_{model}$ 
    \item $\mathcal{X}_{spur}$ : set of \textit{spurious} terms which are selected by the search algorithm, but are not present in the actual system, \textit{i.e.}, $\mathcal{X}_{spur} = \mathcal{X}^{alg} \setminus \mathcal{X}^{\star}$
    \item $\varnothing$ : the null set
\end{itemize}
\smallskip

On the basis of these definitions, one of the following four search outcome can be expected: 

\begin{enumerate}
\item \textbf{Identification of the Correct Structure} (\textit{\textbf{Exact Fitting}}) :\\
\textit{In this scenario the identified model contains all the system terms and does not include any spurious terms, i.e.}, $\mathcal{X}^{alg}=\mathcal{X}^{\star}$ and $\mathcal{X}_{spur} = \varnothing$
\smallskip
\smallskip
\item \textit{\textbf{Over Fitting}} :\\
\textit{The identified model contains all the system terms; however spurious terms are also selected}, \textit{i.e.}, $\mathcal{X}^{alg} \supset \mathcal{X}^{\star}$ and $\mathcal{X}_{spur} \neq \varnothing$
\smallskip
\smallskip
\item \textit{\textbf{Under Fitting-1}} :\\
\textit{The algorithm fails to identify all the system terms; though it does not include any spurious terms}, \textit{i.e.}, $\mathcal{X}^{alg} \subset \mathcal{X}^{\star}$ and $\mathcal{X}_{spur} = \varnothing$
\smallskip
\smallskip
\item \textit{\textbf{Under Fitting-2}} :\\
\textit{The algorithm fails to identify all the system terms; however spurious terms are selected}, \textit{i.e.}, $\mathcal{X}^{alg} \not \supset \mathcal{X}^{\star}$ and $\mathcal{X}_{spur} \neq \varnothing$
\smallskip
\end{enumerate}

Thus, \textit{qualitatively}, the search is \textit{successful} when all the system/significant terms are identified, \textit{i.e.}, when the outcome is either \textit{Exact-Fitting} or \textit{Over-Fitting}. Note that at this stage, the structure identified by the algorithms are unaltered. The inclusion of few spurious terms can be tolerated provided that all significant terms are included (\textit{i.e.} Over-Fitting scenario) as the spurious terms can easily be identified and removed through a simple \textit{null-hypothesis} test on the corresponding coefficients.

\section{Results}
\label{s:res}

The goal of this study is to investigate the suitability of existing feature selection algorithms for the task of non-linear system identification. For this purpose, 3-well known floating search algorithms have been adapted in the proposed orthogonal floating search framework: OSF, OIF and O$^2$S, which are discussed in Section~\ref{s:proposedOFS}. The adapted algorithms are initially applied to identify the significant terms of $6$ test non-linear systems described in Section~\ref{s:Data}. The search performance of the algorithms is compared on these systems from various perspectives in Section~\ref{s:compEval}. The issues related to selection of cardinality interval $[\xi_{min}, \xi_{max}]$ are discussed in Section~\ref{s:commentModelOrder}.

Further, it is worth to note that while the proposed orthogonal floating search can work with any suitable criterion function, in this study, the Error-Reduction-Ratio (ERR) has been selected for this purpose due to its relative simplicity. However, ERR is often criticized for its `\textit{local}' nature and blamed for the unsatisfactory search outcome. This issue is investigated via a numerical example in Section~\ref{s:cERR}. 

Finally, the proposed orthogonal floating algorithms are applied to identify a discrete model of a continuous time system. The identified discrete model is validated through generalized frequency response functions. This part of the investigation is discussed in Section~\ref{s:Duff}.

\begin{figure*}[!t]
\centering
\small
\begin{subfigure}{.33\textwidth}
  \centering
  \includegraphics[width=\textwidth]{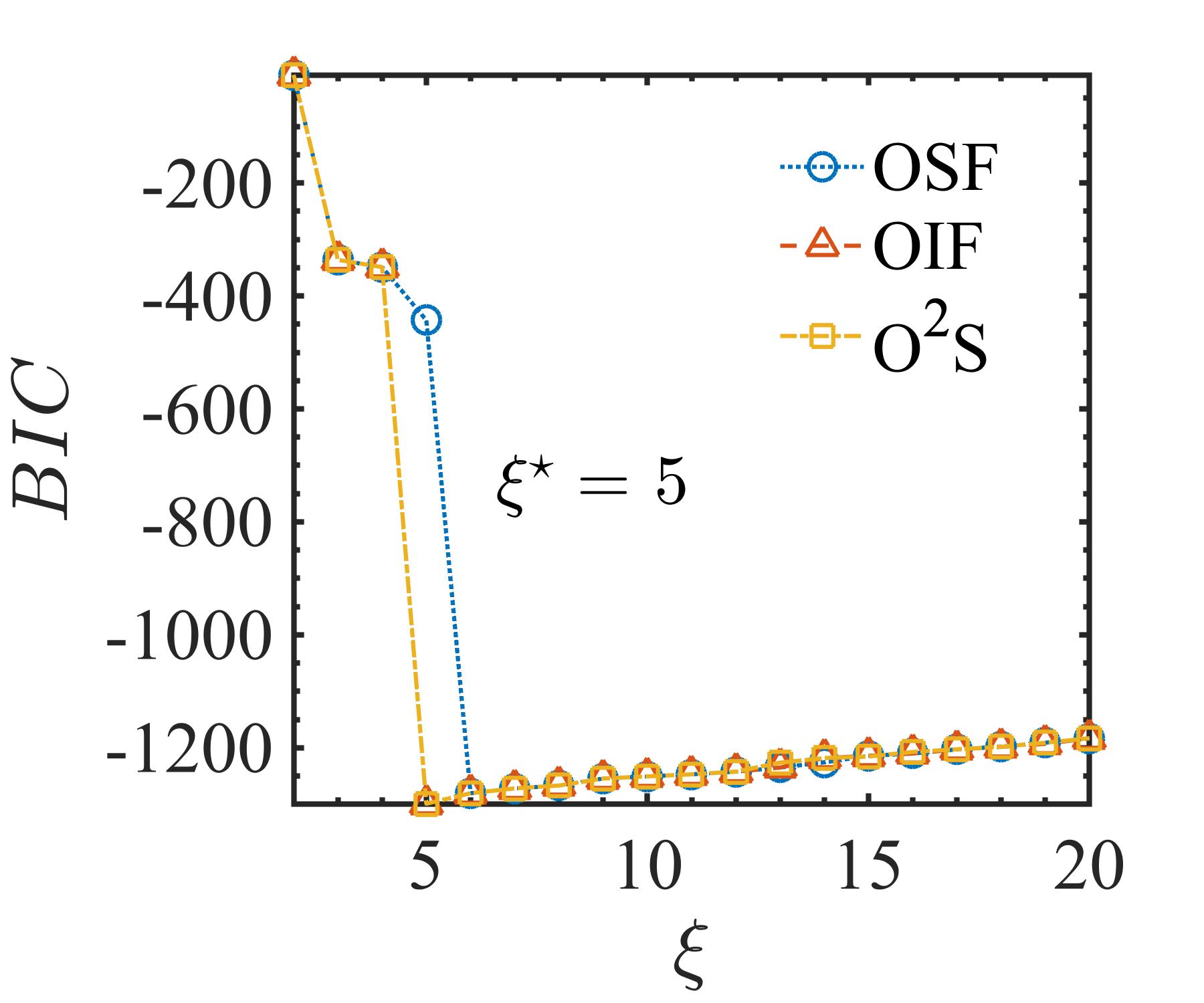}
  \caption{$\mathcal{S}_1$}
  \label{f:S1}
\end{subfigure}
\hfill
\begin{subfigure}{.33\textwidth}
  \centering
  \includegraphics[width=\textwidth]{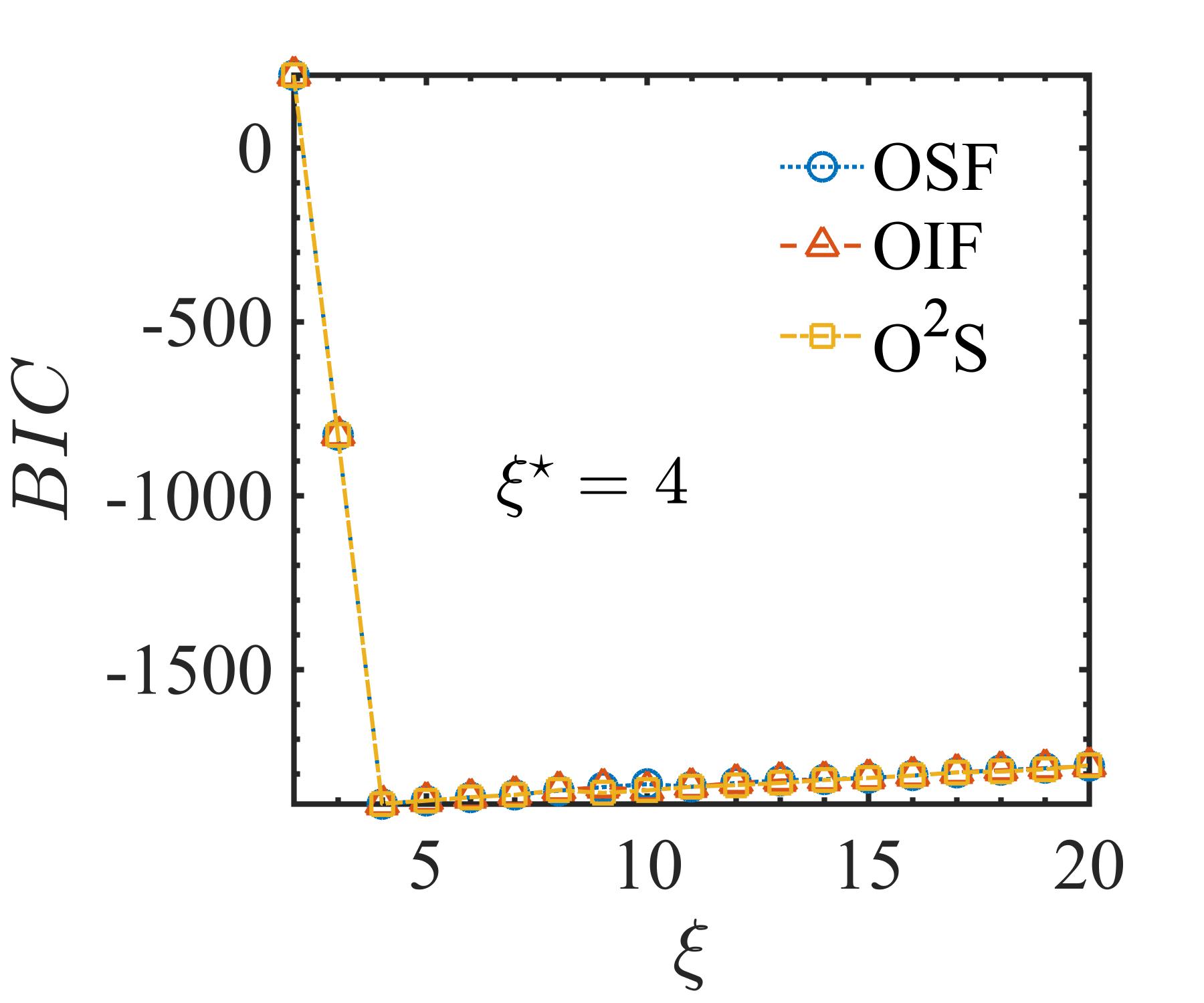}
  \caption{$\mathcal{S}_2$}
  \label{f:S2}
\end{subfigure}%
\begin{subfigure}{.33\textwidth}
  \centering
  \includegraphics[width=\textwidth]{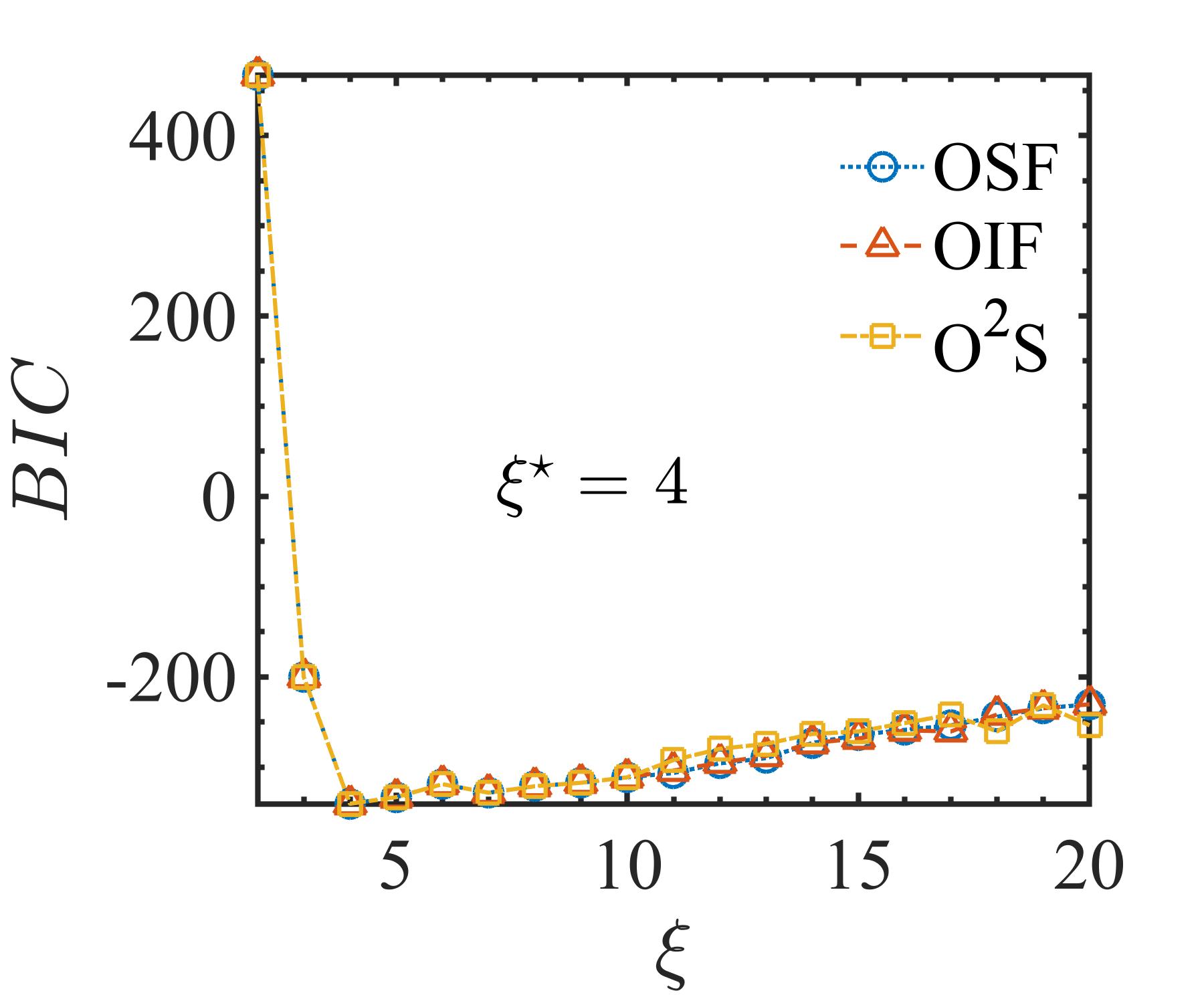}
  \caption{$\mathcal{S}_3$}
  \label{f:S3}
\end{subfigure}%
\hfill
\begin{subfigure}{.33\textwidth}
  \centering
  \includegraphics[width=\textwidth]{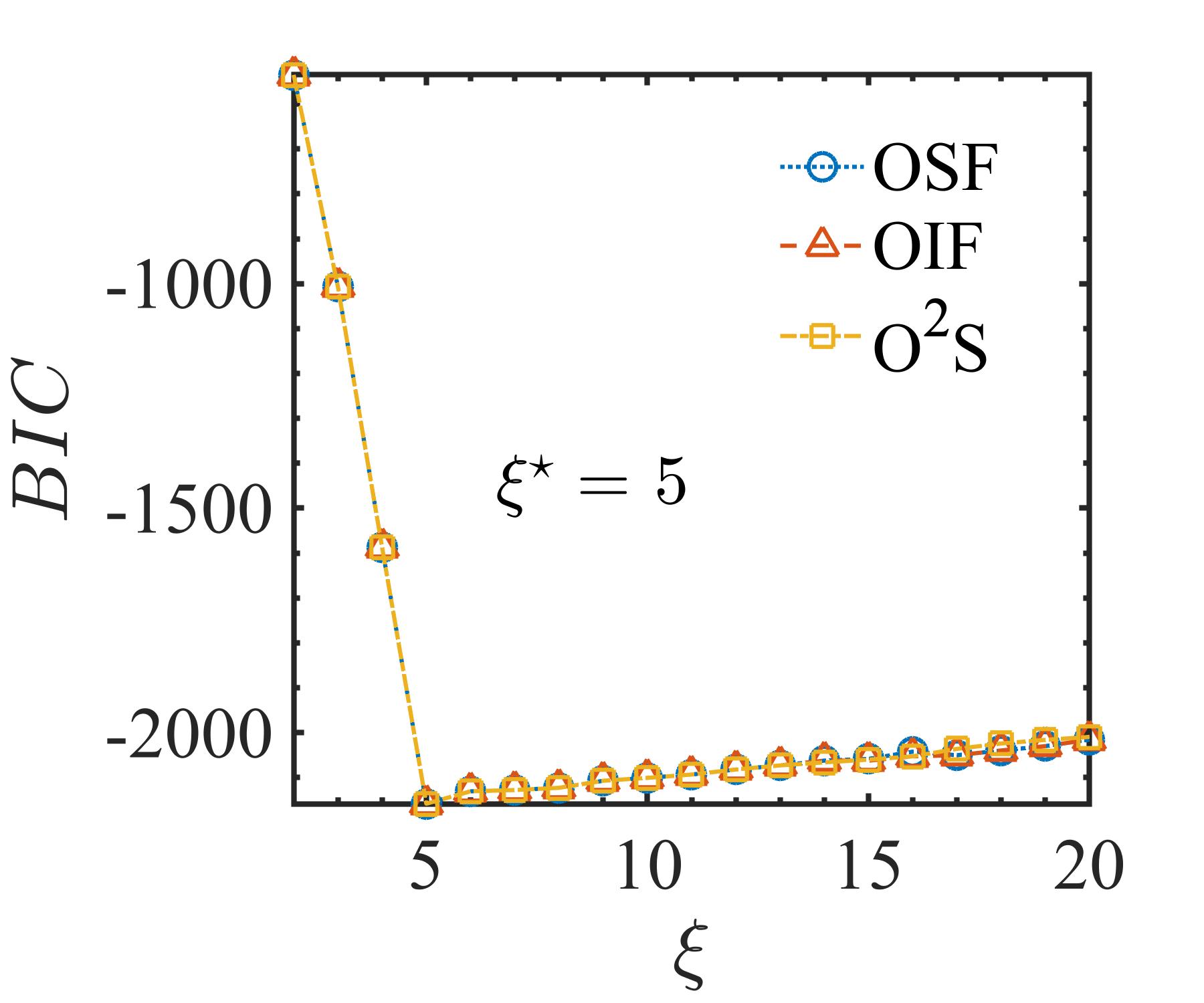}
  \caption{$\mathcal{S}_4$}
  \label{f:S4}
\end{subfigure}
\begin{subfigure}{.33\textwidth}
  \centering
  \includegraphics[width=\textwidth]{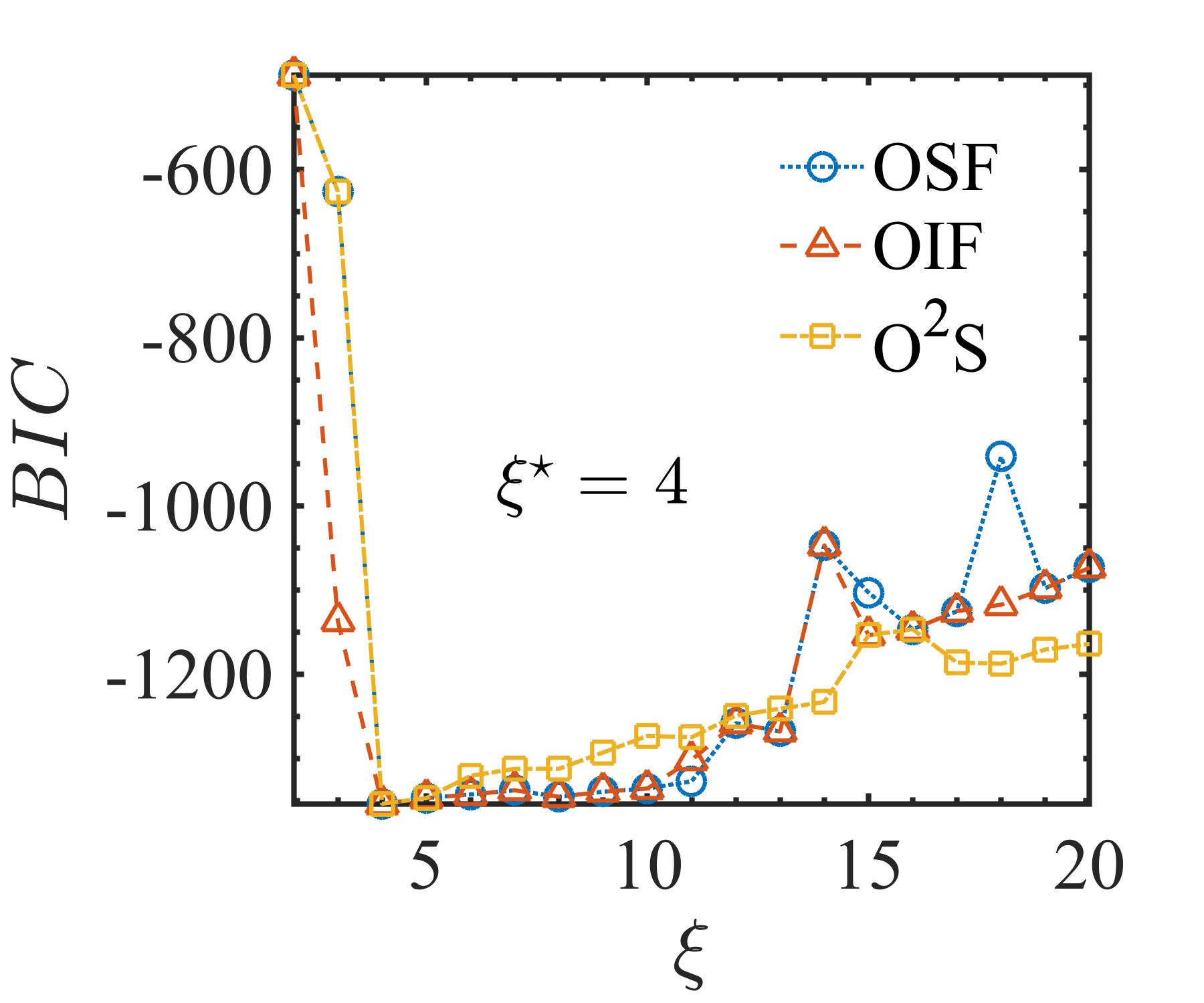}
  \caption{$\mathcal{S}_5$}
  \label{f:S5}
\end{subfigure}%
\hfill
\begin{subfigure}{.33\textwidth}
  \centering
  \includegraphics[width=\textwidth]{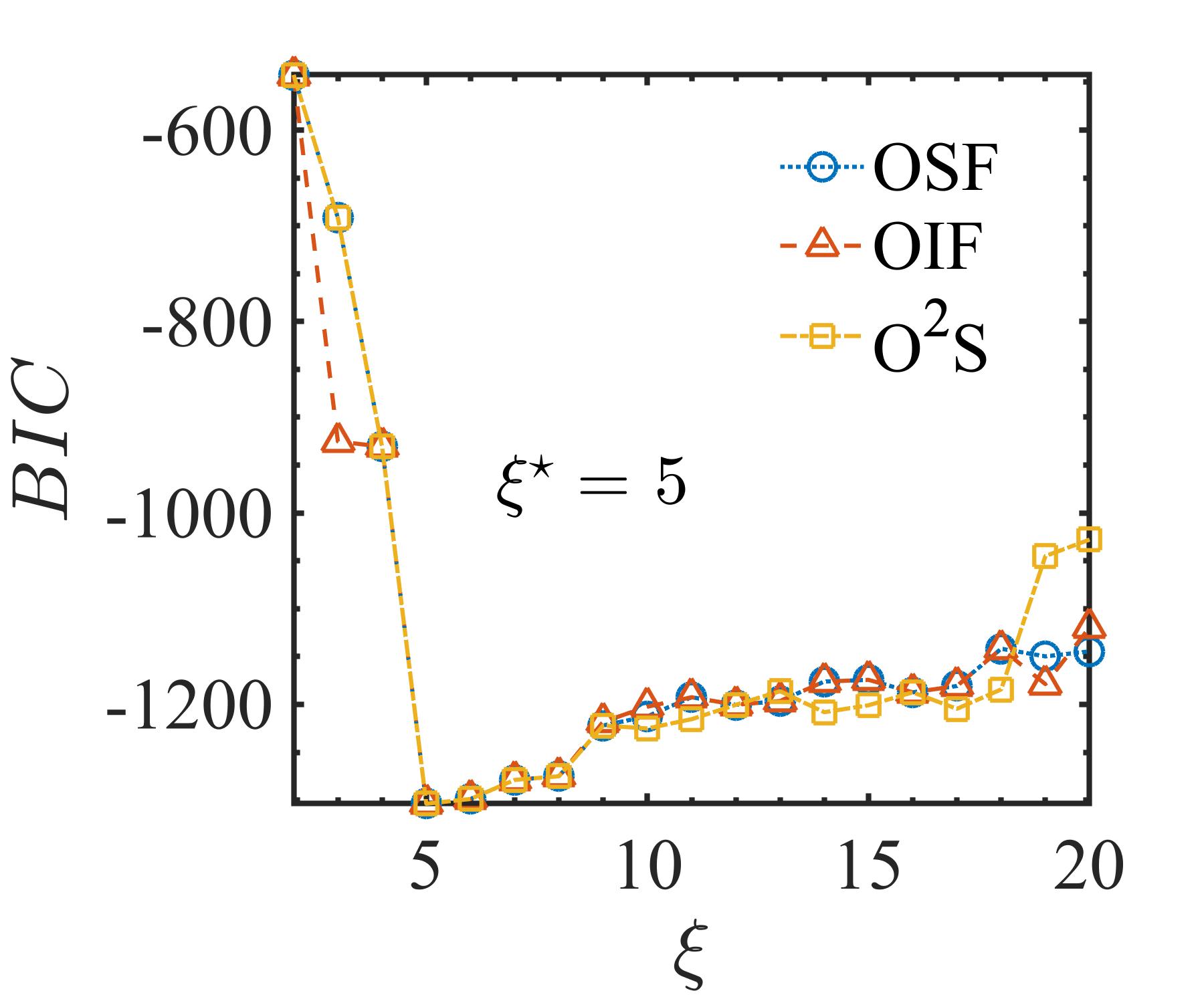}
  \caption{$\mathcal{S}_6$}
  \label{f:S6}
\end{subfigure}

\caption{Model Order Selection. `$\xi^{\star}$' denotes the known cardinality of the system.}
\label{f:mos}
\end{figure*}

\begin{table}[!t]
  \centering
  \scriptsize
  \caption{Search Outcomes$^\dagger$}
  \label{t:SO} 
  \begin{adjustbox}{max width=0.95\textwidth}
  \begin{threeparttable}
    \begin{tabular}{ccccccccc}
    
    \toprule
    \multirow{2}{*}{\textbf{System}} & \multirow{2}{*}{\boldmath$\xi^{\star}$} & \multicolumn{2}{c}{\textbf{OSF}} & \multicolumn{2}{c}{\textbf{OIF}} & \multicolumn{2}{c}{\boldmath\textbf{O}$^2$\textbf{S}} \\
    \cmidrule{3-8}          &       & \boldmath{$\xi_{alg}$}  & \textbf{Outcome} & \boldmath{$\xi_{alg}$}  & \textbf{Outcome} & \boldmath{$\xi_{alg}$}  & \textbf{Outcome} \\
    \midrule
    $\mathcal{S}_1$    & 5     & 6     & Over-Fitting & 5     & Exact-Fitting & 5     & Exact-Fitting \\ [0.5ex]
    $\mathcal{S}_2$    & 4     & 4     & Exact-Fitting & 4     & Exact-Fitting & 4     & Exact-Fitting \\ [0.5ex]
    $\mathcal{S}_3$    & 4     & 4     & Exact-Fitting & 4     & Exact-Fitting & 4     & Exact-Fitting \\ [0.5ex]
    $\mathcal{S}_4$    & 5     & 5     & Exact-Fitting & 5     & Exact-Fitting & 5     & Exact-Fitting \\ [0.5ex]
    $\mathcal{S}_5$    & 4     & 4     & Exact-Fitting & 4     & Exact-Fitting & 4     & Exact-Fitting \\ [0.5ex]
    $\mathcal{S}_6$    & 5     & 5     & Exact-Fitting & 5     & Exact-Fitting & 5     & Exact-Fitting \\ [0.5ex]
    \bottomrule
    
    \end{tabular}%
    \begin{tablenotes}
      \scriptsize
       \item $\dagger$ See Section~\ref{s:PM} for the definitions of search outcomes 
    \end{tablenotes}
  \end{threeparttable}
  \end{adjustbox}
\end{table}%
\begin{figure*}[!t]
\centering
\small
\begin{subfigure}{.32\textwidth}
  \centering
  \includegraphics[width=\textwidth]{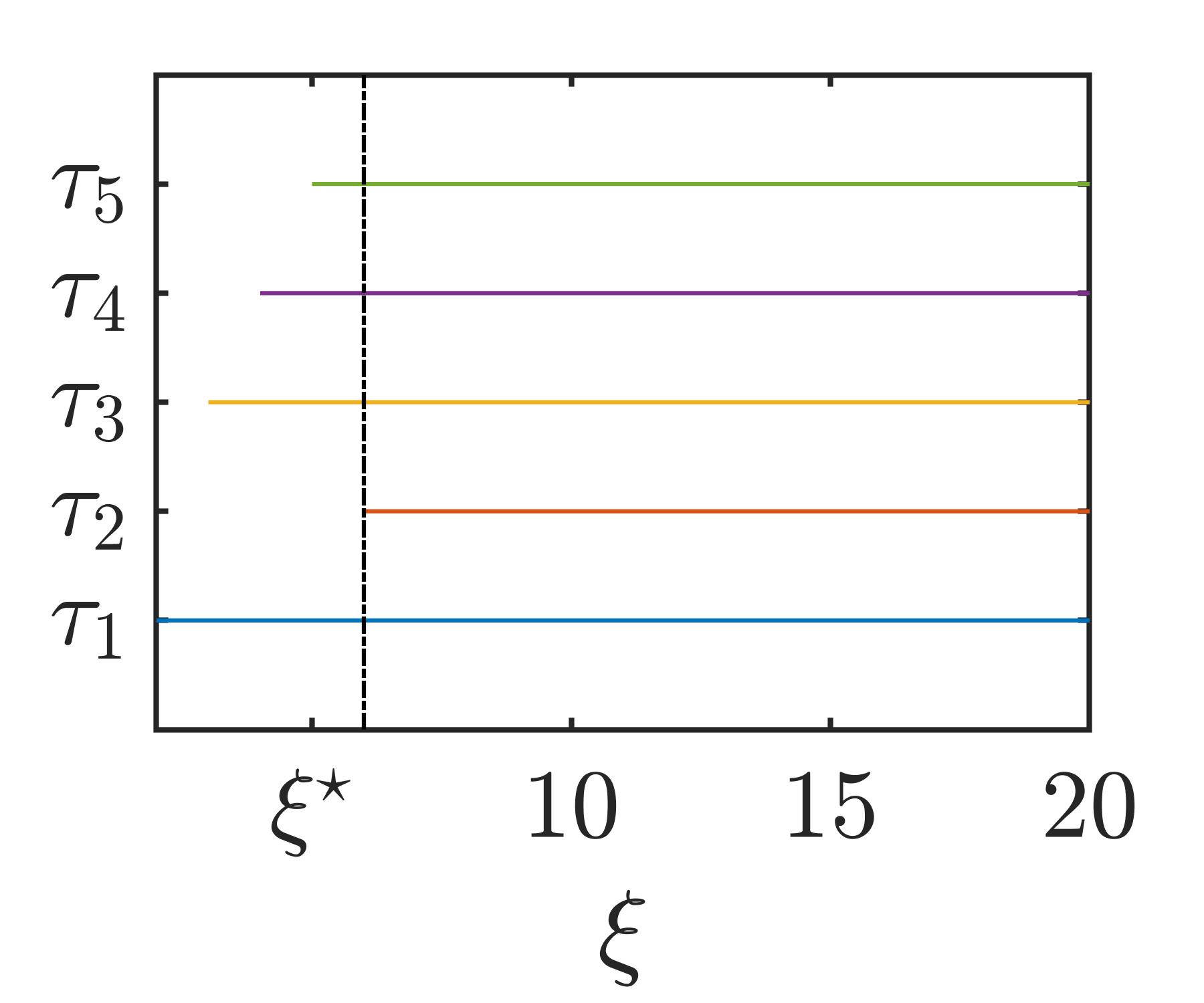}
  \caption{$\mathcal{S}_1+OSF$}
  \label{f:tss1osf}
\end{subfigure}%
\hfill
\begin{subfigure}{.32\textwidth}
  \centering
  \includegraphics[width=\textwidth]{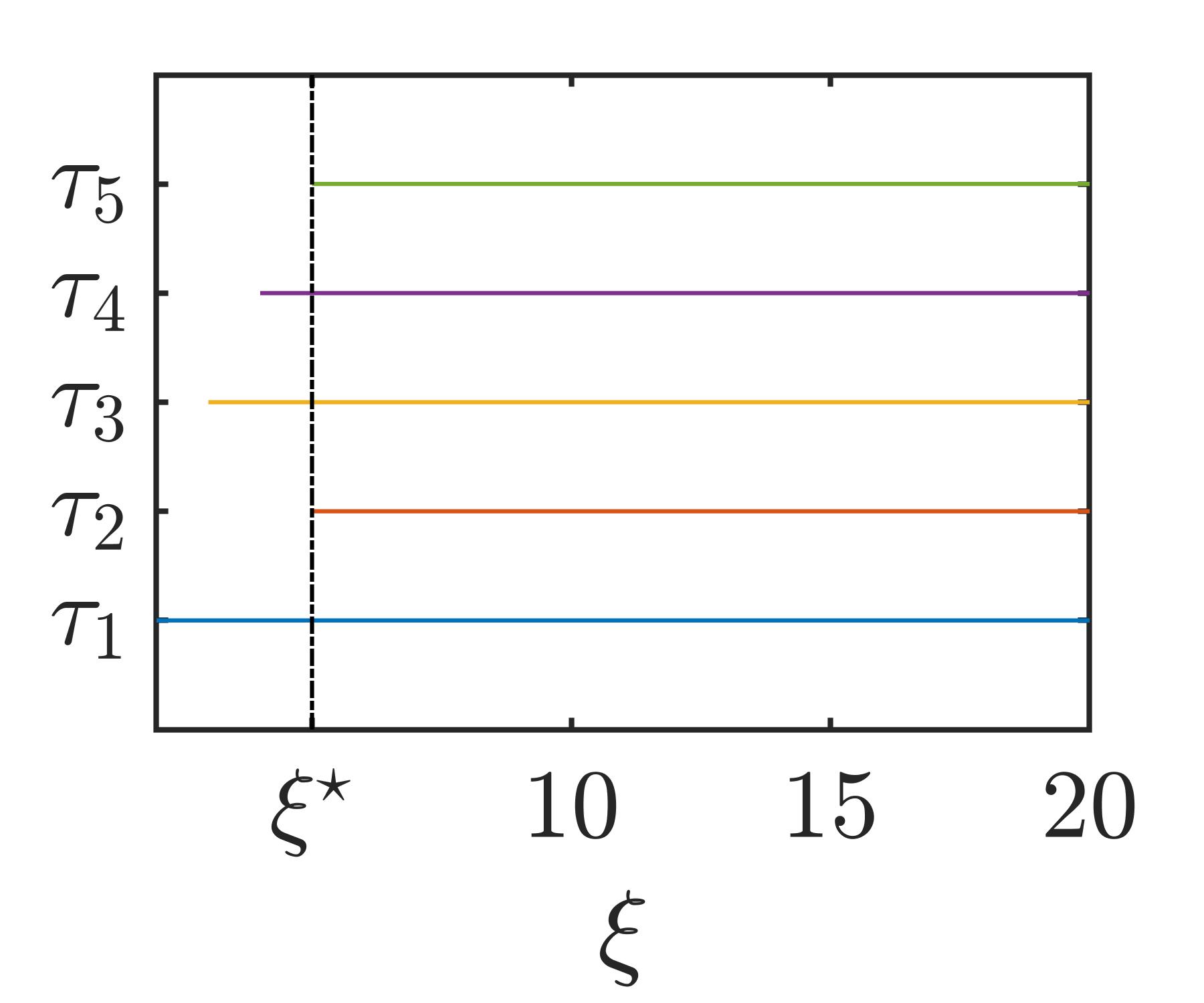}
  \caption{$\mathcal{S}_1+OIF$}
  \label{f:tss1oif}
\end{subfigure}
\hfill
\begin{subfigure}{.32\textwidth}
  \centering
  \includegraphics[width=\textwidth]{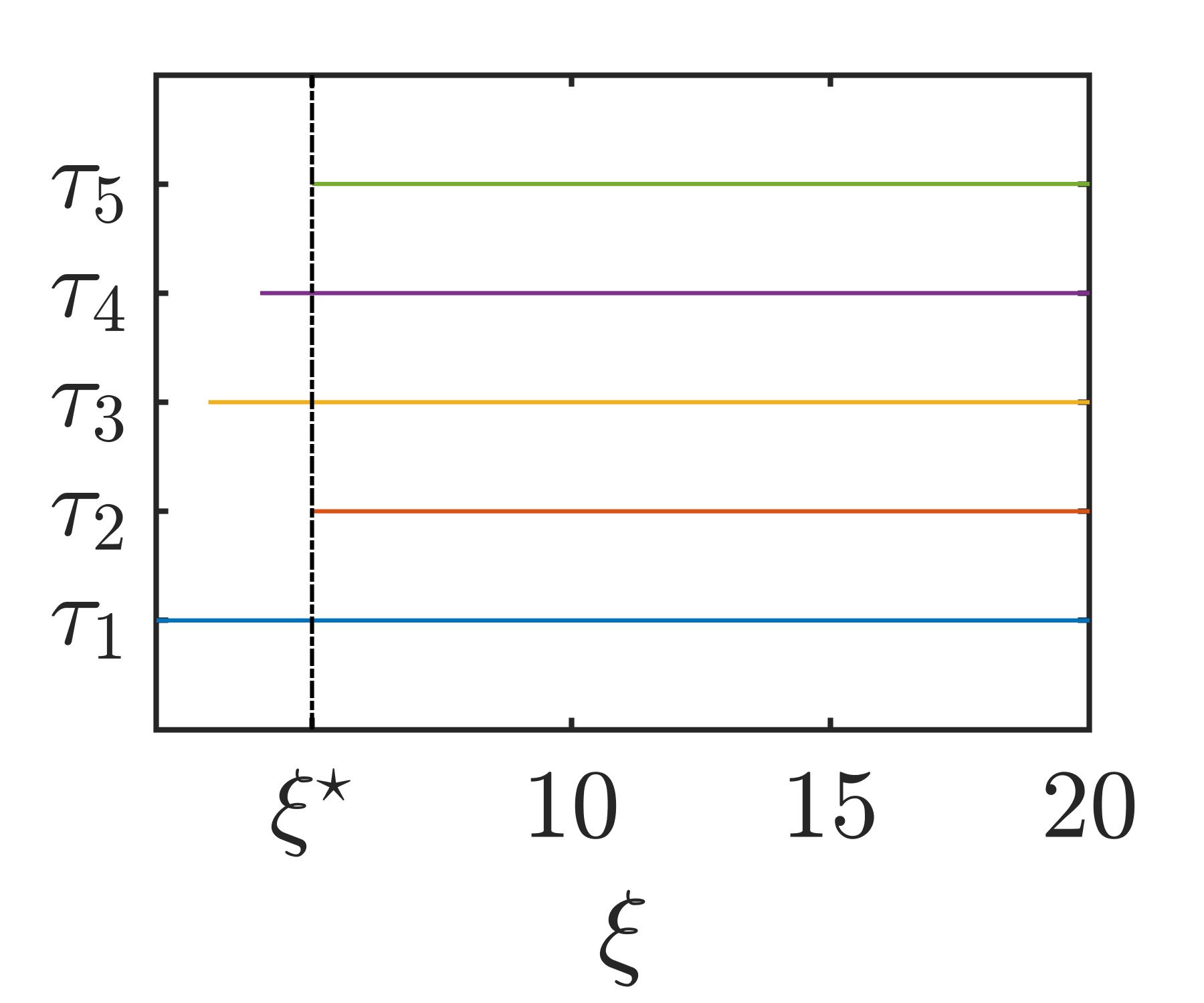}
  \caption{$\mathcal{S}_1+O^2S$}
  \label{f:tss1o2s}
\end{subfigure}
\hfill
\begin{subfigure}{.32\textwidth}
  \centering
  \includegraphics[width=\textwidth]{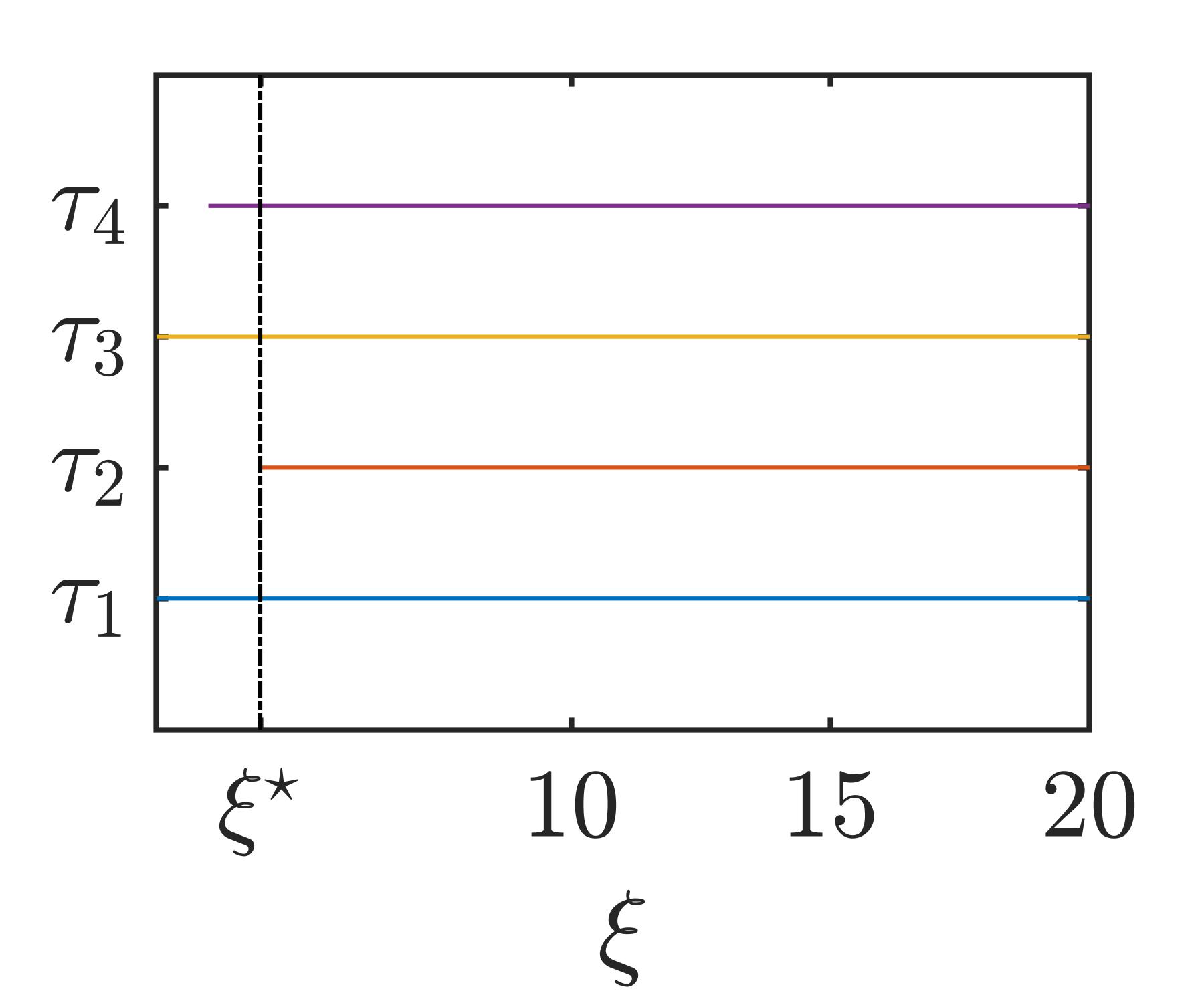}
  \caption{$\mathcal{S}_2+OSF$}
  \label{f:tss2osf}
\end{subfigure}%
\hfill
\begin{subfigure}{.32\textwidth}
  \centering
  \includegraphics[width=\textwidth]{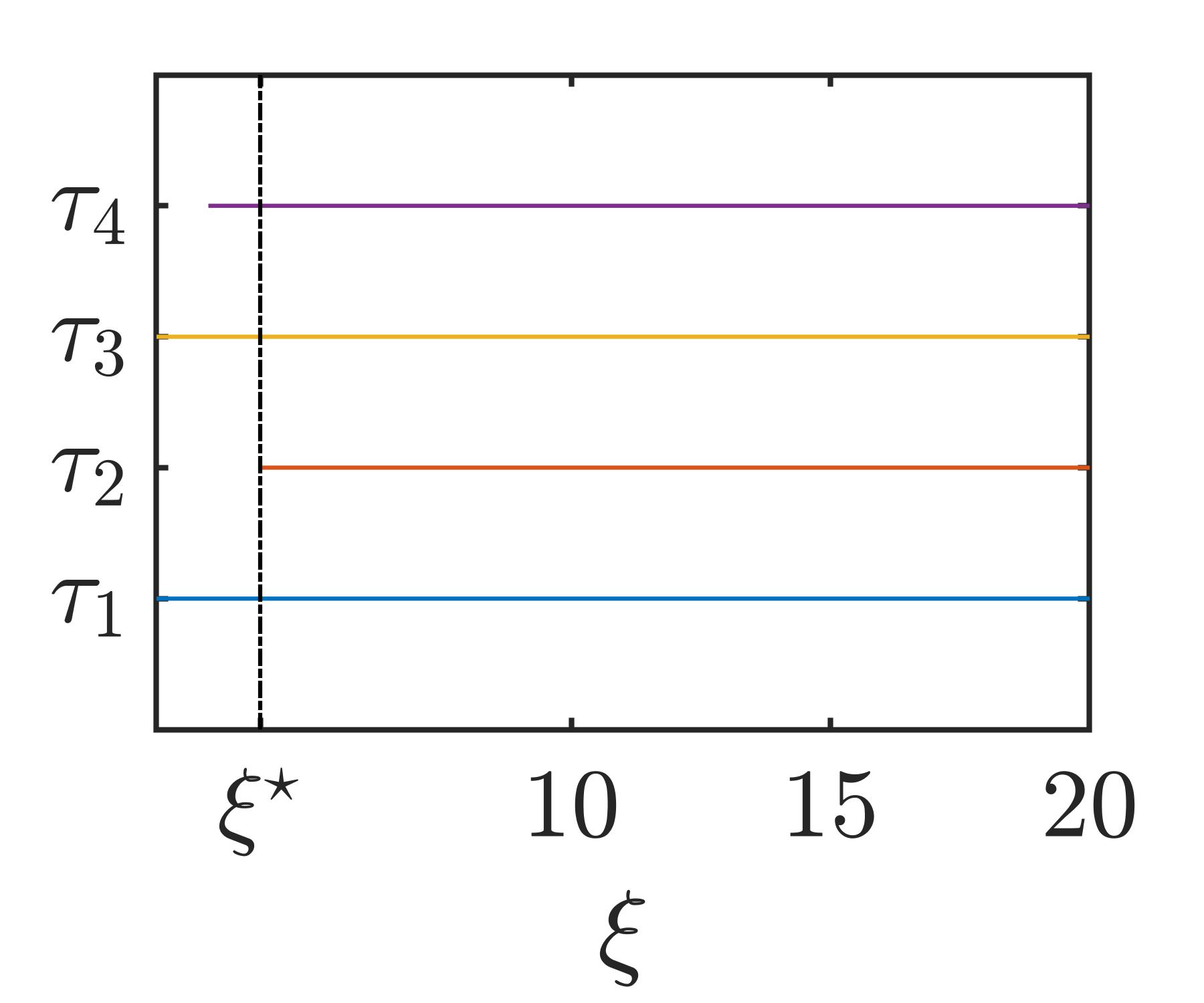}
  \caption{$\mathcal{S}_2+OIF$}
  \label{f:tss2oif}
\end{subfigure}
\hfill
\begin{subfigure}{.32\textwidth}
  \centering
  \includegraphics[width=\textwidth]{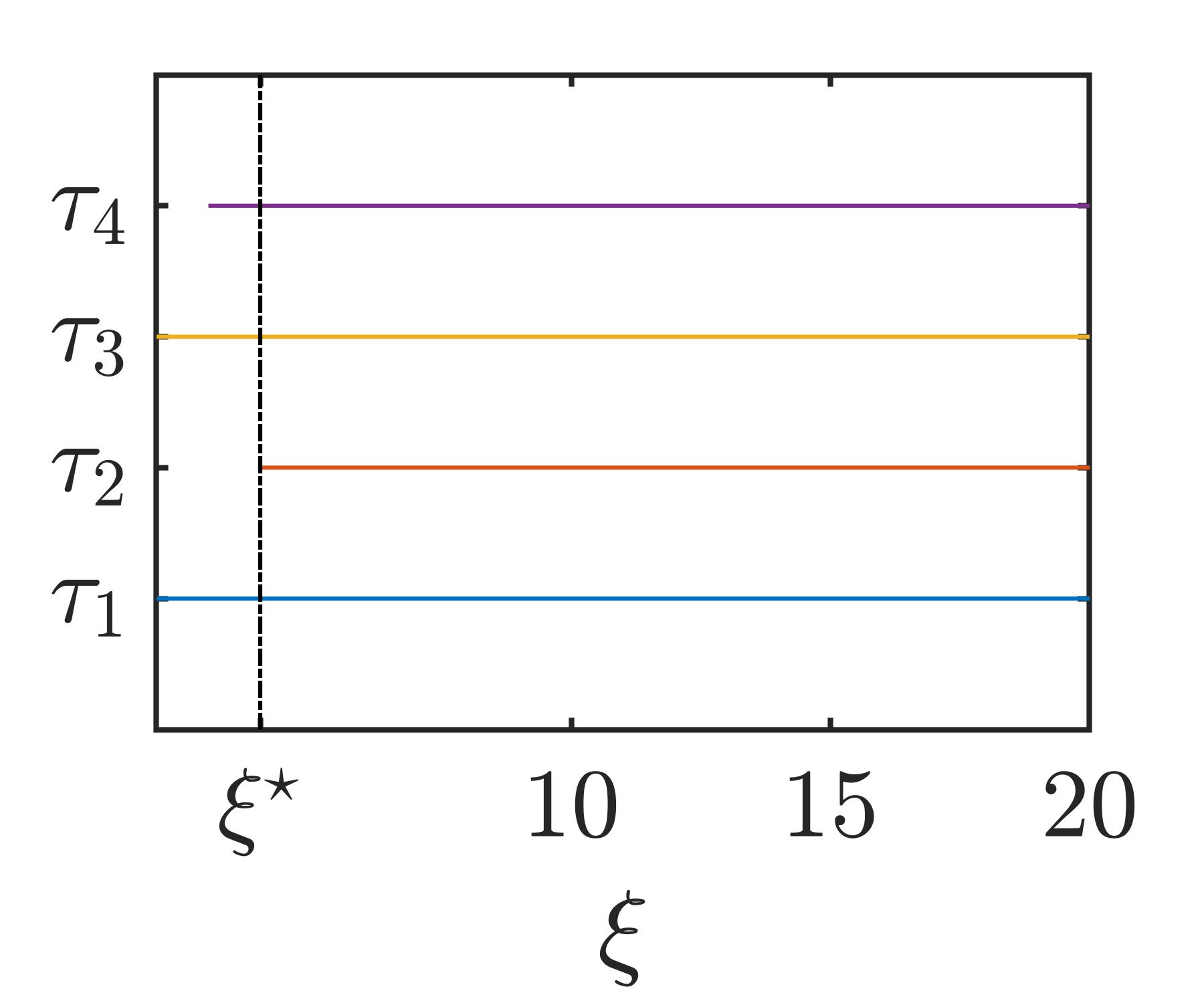}
  \caption{$\mathcal{S}_2+O^2S$}
  \label{f:tss2o2s}
\end{subfigure}%
\caption{Term Selection Behavior-I. $\xi^{\star}$ denotes actual cardinality of the system. The `dotted' vertical line indicate the value $\xi$ when all the system terms are identified.}
\label{f:TS1}
\end{figure*}
\begin{figure*}[!h]
\centering
\small
\begin{subfigure}{.32\textwidth}
  \centering
  \includegraphics[width=\textwidth]{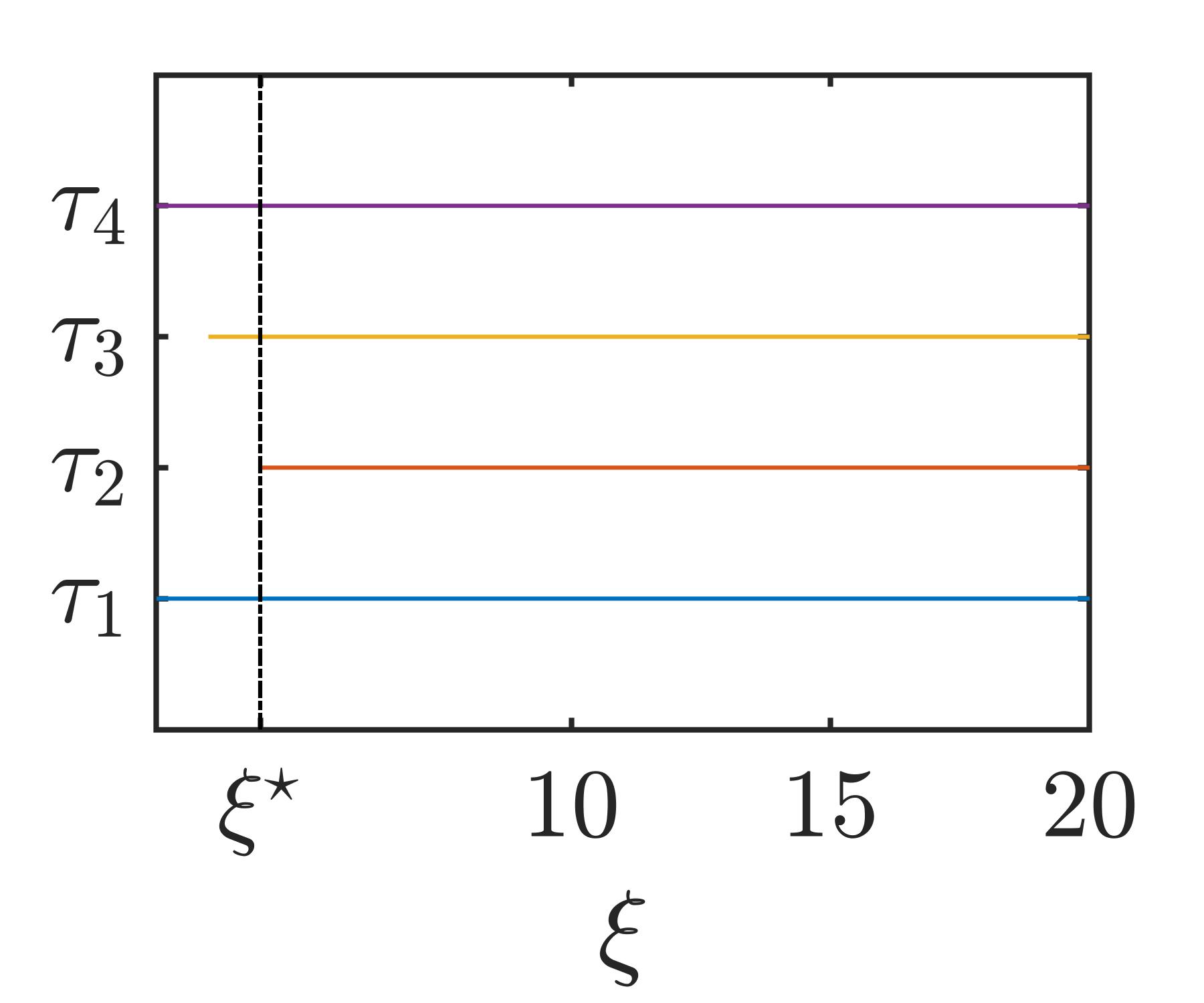}
  \caption{$\mathcal{S}_3+OSF$}
  \label{f:tss3osf}
\end{subfigure}%
\hfill
\begin{subfigure}{.32\textwidth}
  \centering
  \includegraphics[width=\textwidth]{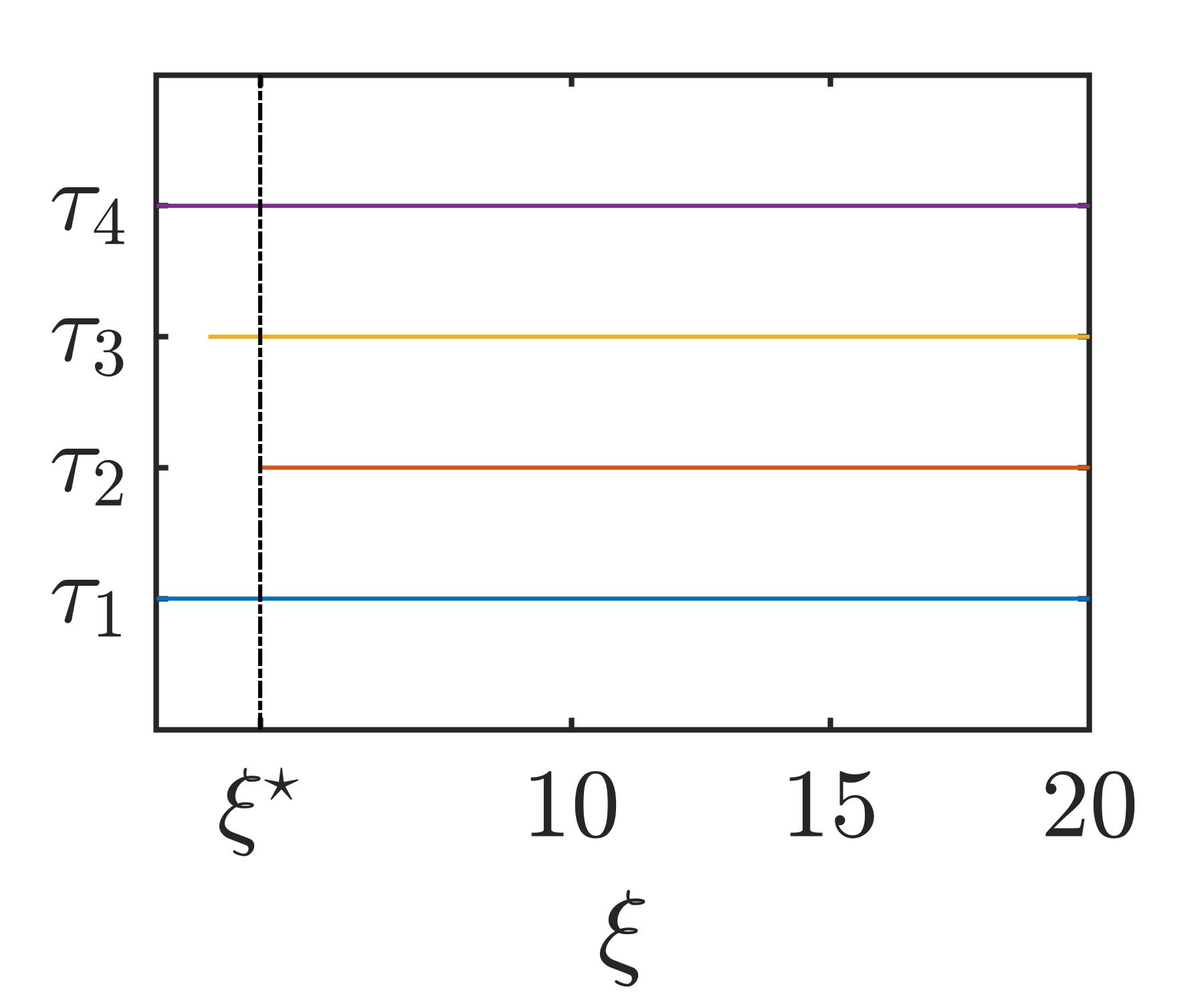}
  \caption{$\mathcal{S}_3+OIF$}
  \label{f:tss3oif}
\end{subfigure}
\hfill
\begin{subfigure}{.32\textwidth}
  \centering
  \includegraphics[width=\textwidth]{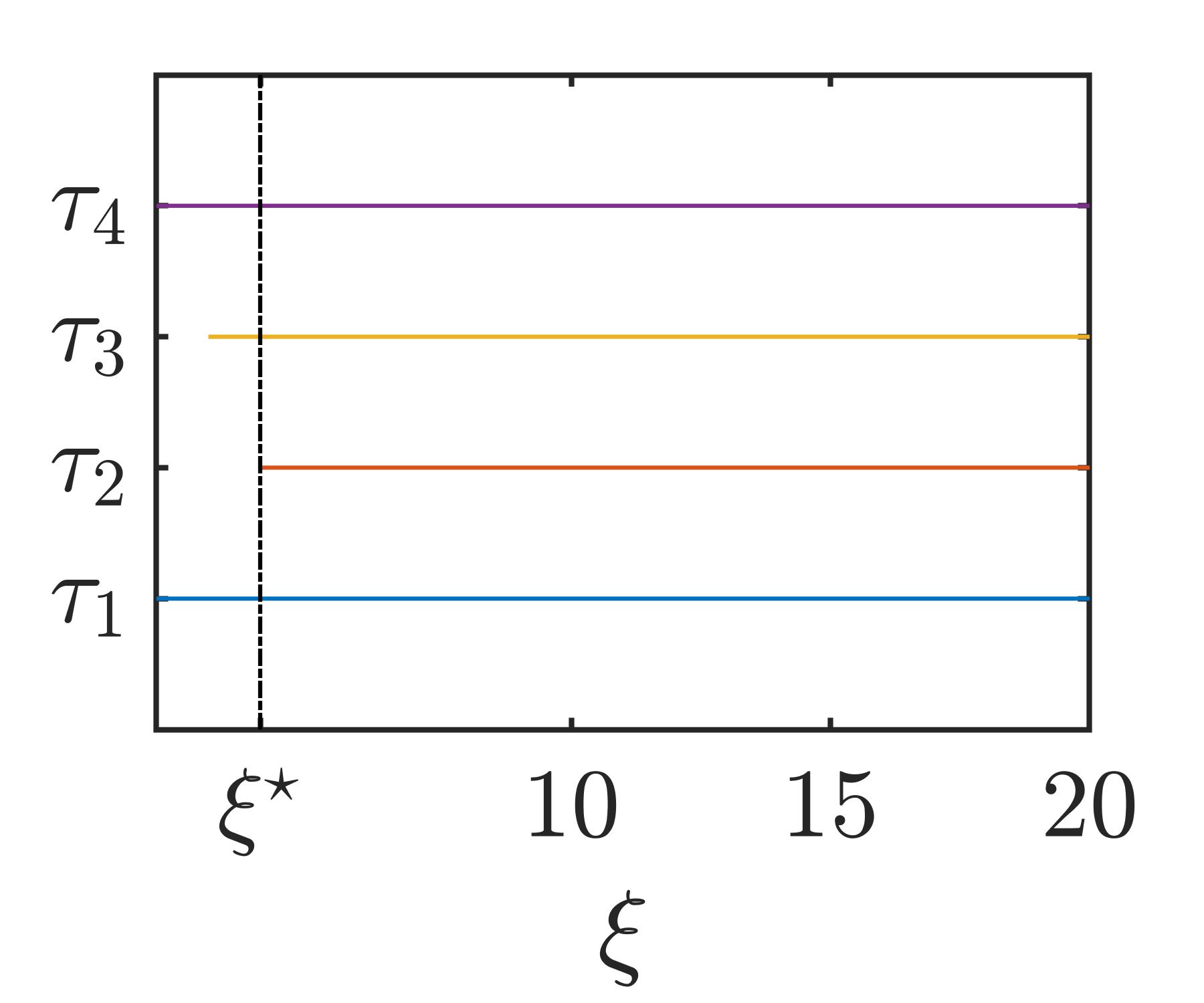}
  \caption{$\mathcal{S}_3+O^2S$}
  \label{f:tss3o2s}
\end{subfigure}%
\hfill
\begin{subfigure}{.32\textwidth}
  \centering
  \includegraphics[width=\textwidth]{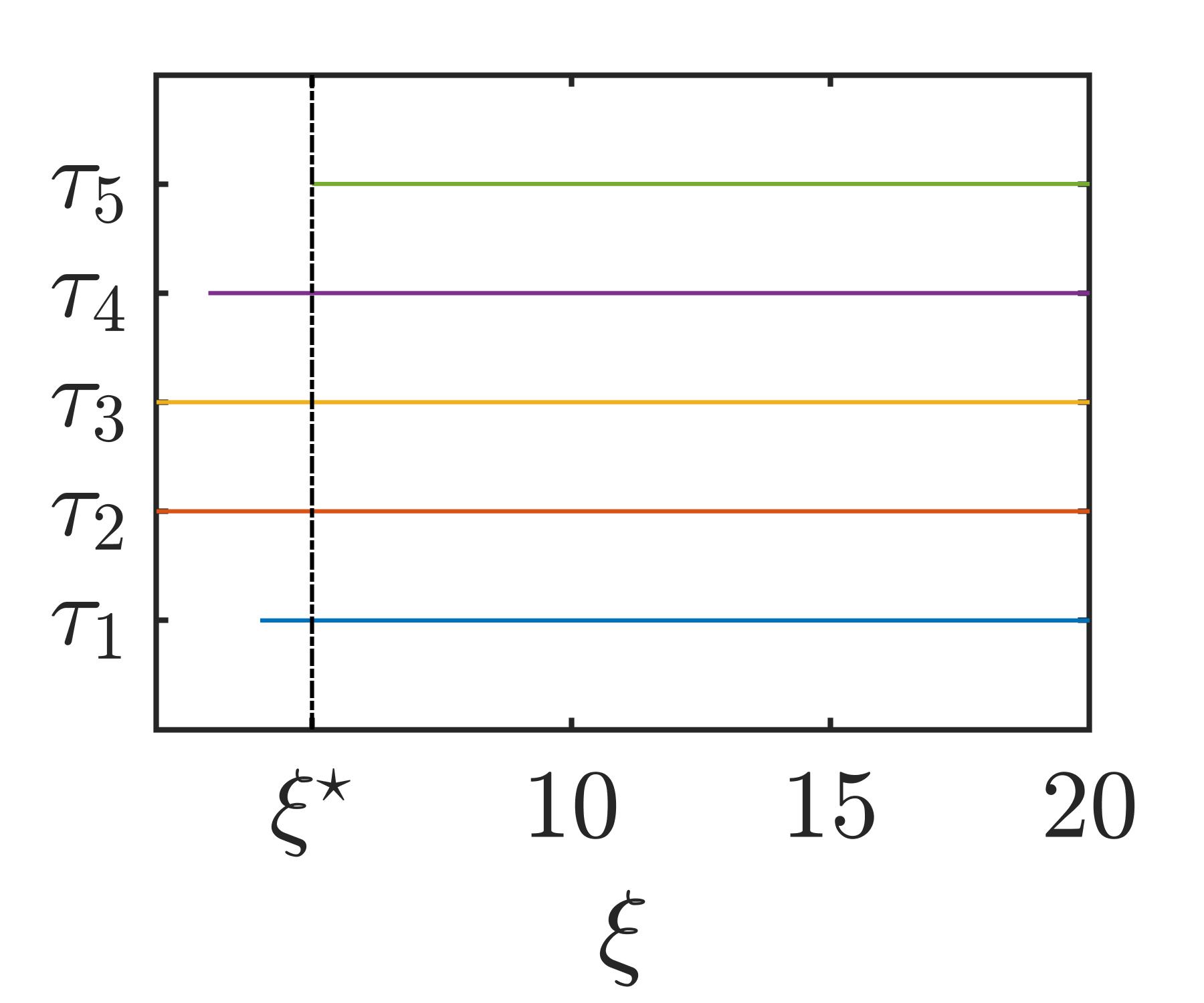}
  \caption{$\mathcal{S}_4+OSF$}
  \label{f:tss4osf}
\end{subfigure}%
\hfill
\begin{subfigure}{.32\textwidth}
  \centering
  \includegraphics[width=\textwidth]{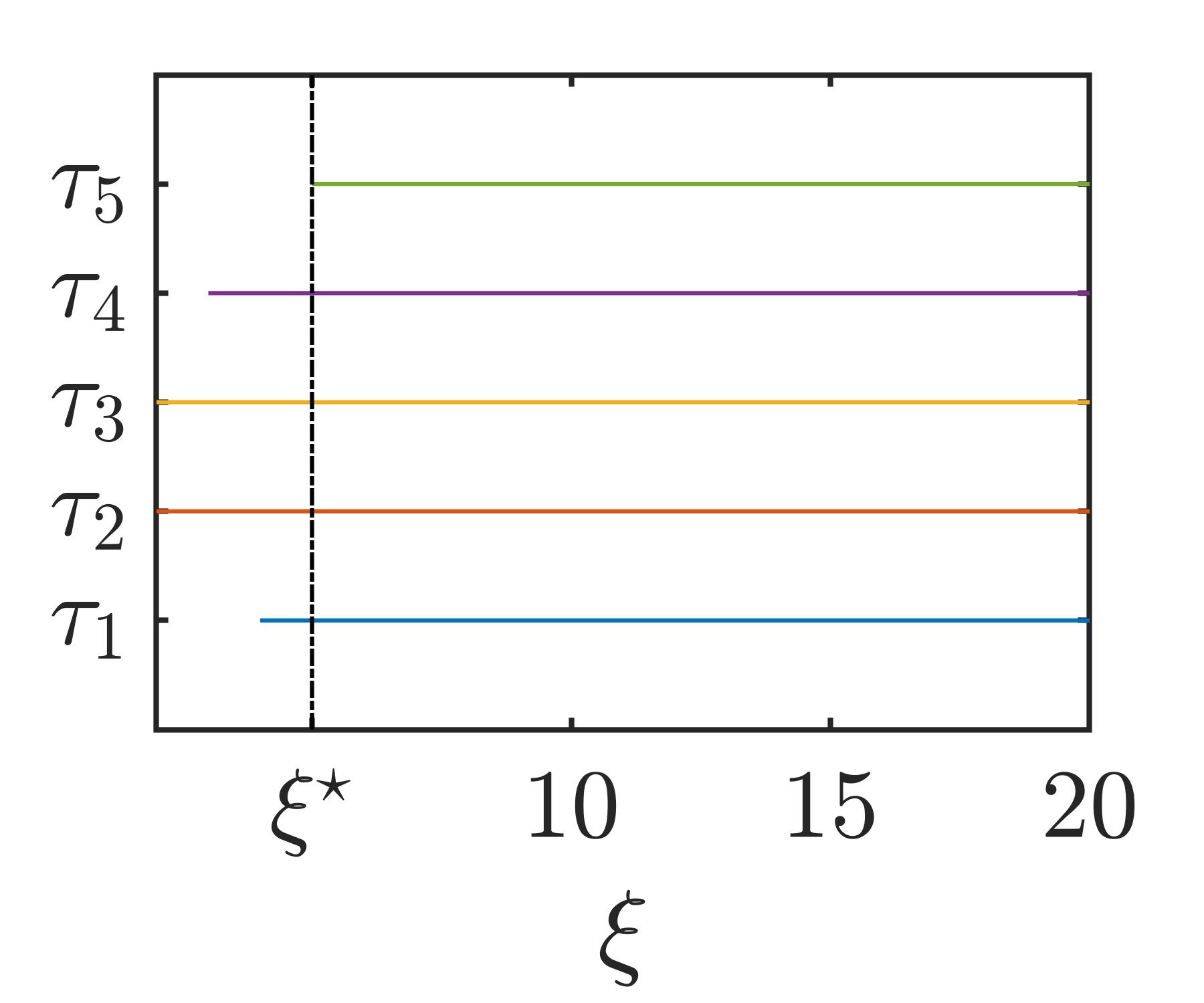}
  \caption{$\mathcal{S}_4+OIF$}
  \label{f:tss4oif}
\end{subfigure}
\hfill
\begin{subfigure}{.32\textwidth}
  \centering
  \includegraphics[width=\textwidth]{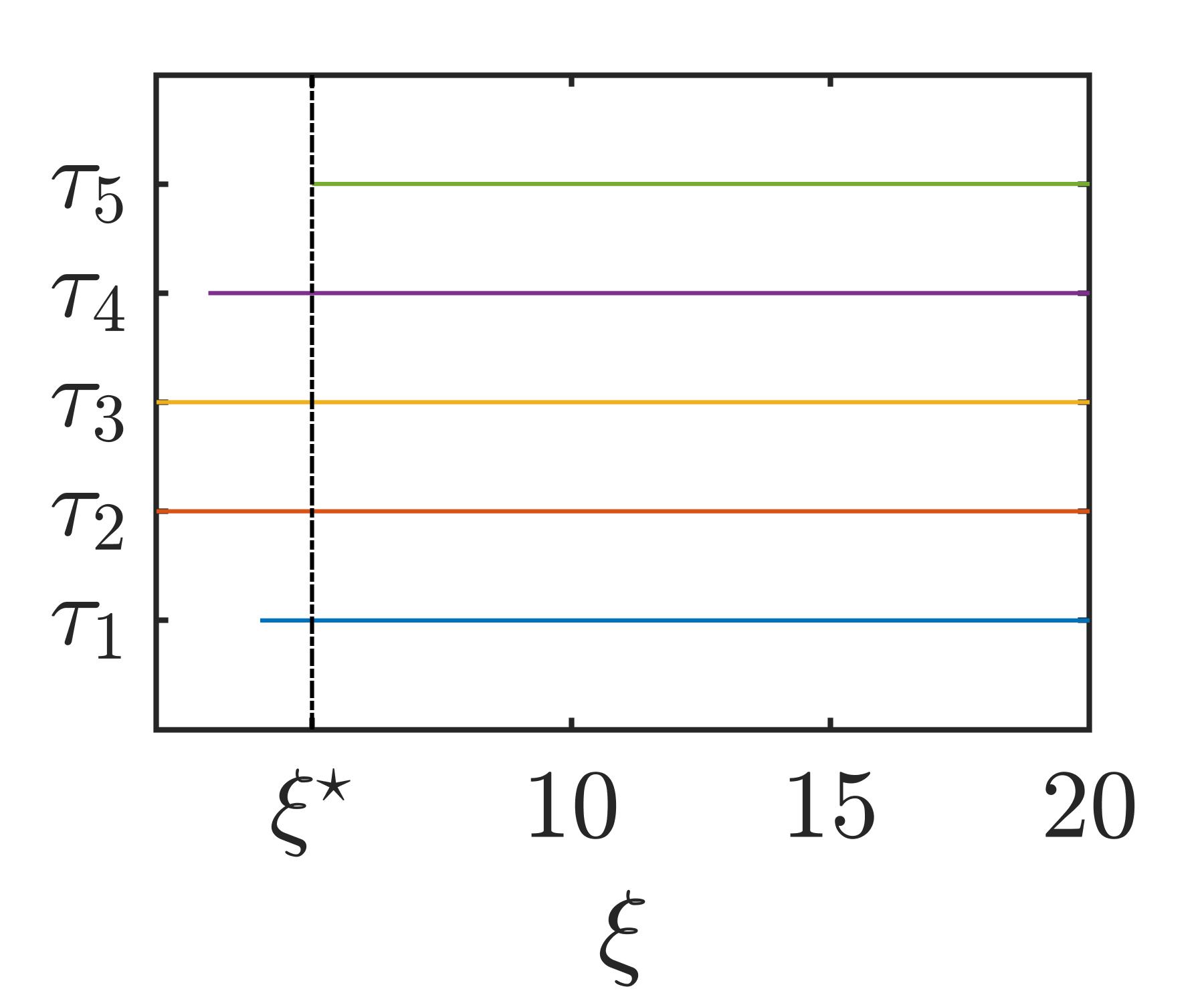}
  \caption{$\mathcal{S}_4+O^2S$}
  \label{f:tss4o2s}
\end{subfigure}%
\caption{Term Selection Behavior-II. $\xi^{\star}$ denotes actual cardinality of the system. The `dotted' vertical line indicate the value $\xi$ when all the system terms are identified.}
\label{f:TS2}
\end{figure*}
\begin{figure*}[!h]
\centering
\small
\begin{subfigure}{.32\textwidth}
  \centering
  \includegraphics[width=\textwidth]{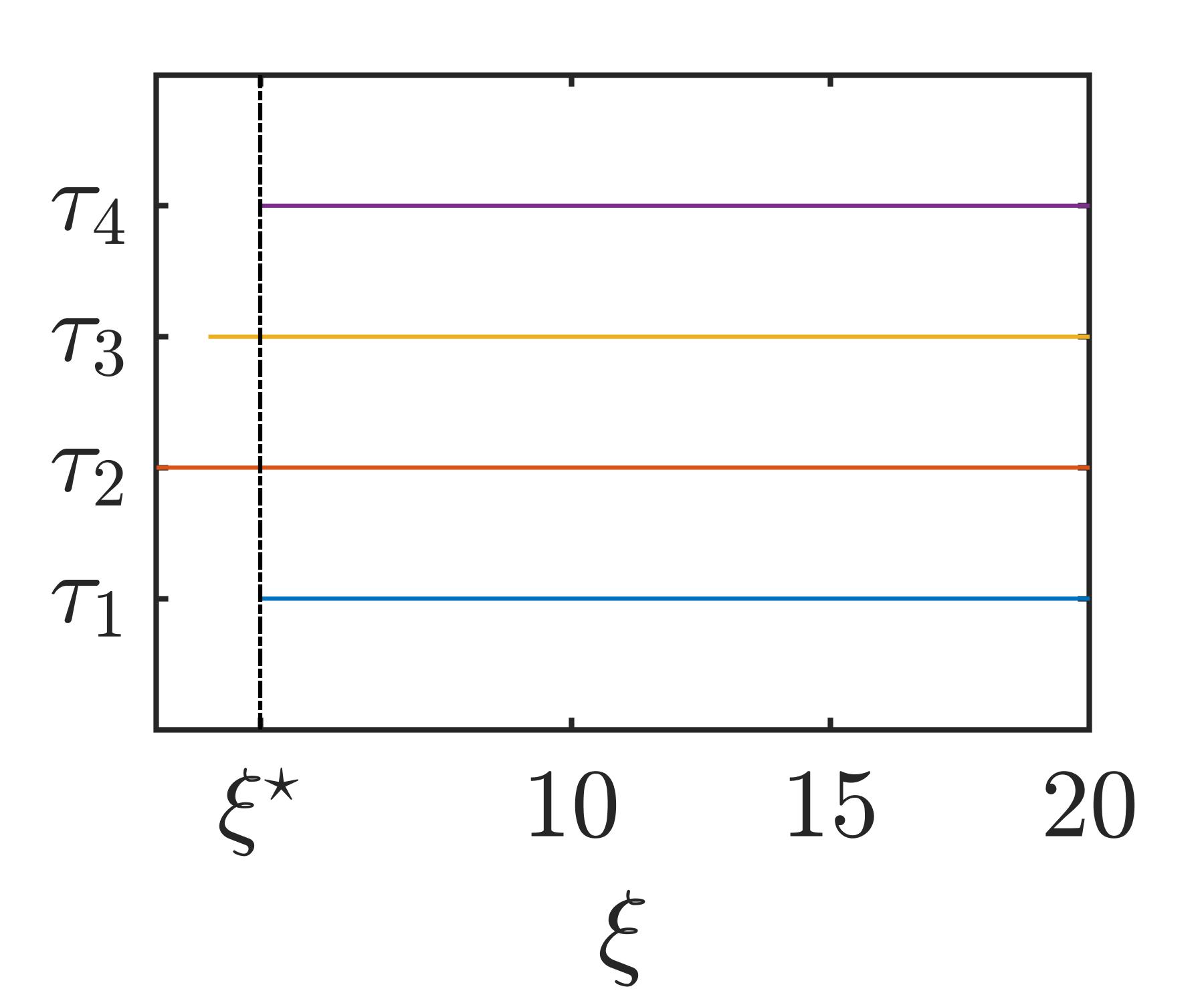}
  \caption{$\mathcal{S}_5+OSF$}
  \label{f:tss5osf}
\end{subfigure}%
\hfill
\begin{subfigure}{.32\textwidth}
  \centering
  \includegraphics[width=\textwidth]{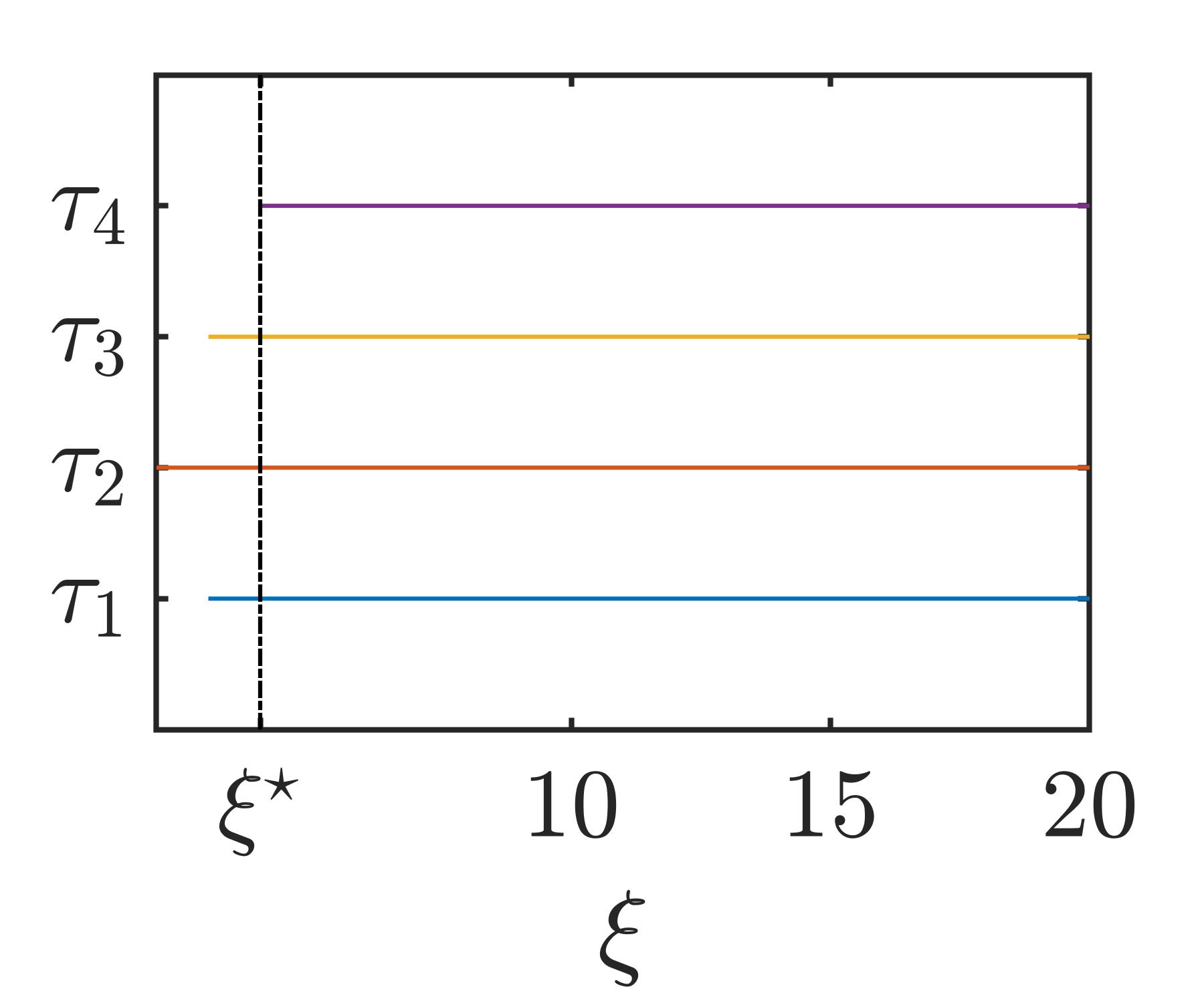}
  \caption{$\mathcal{S}_5+OIF$}
  \label{f:tss5oif}
\end{subfigure}
\hfill
\begin{subfigure}{.32\textwidth}
  \centering
  \includegraphics[width=\textwidth]{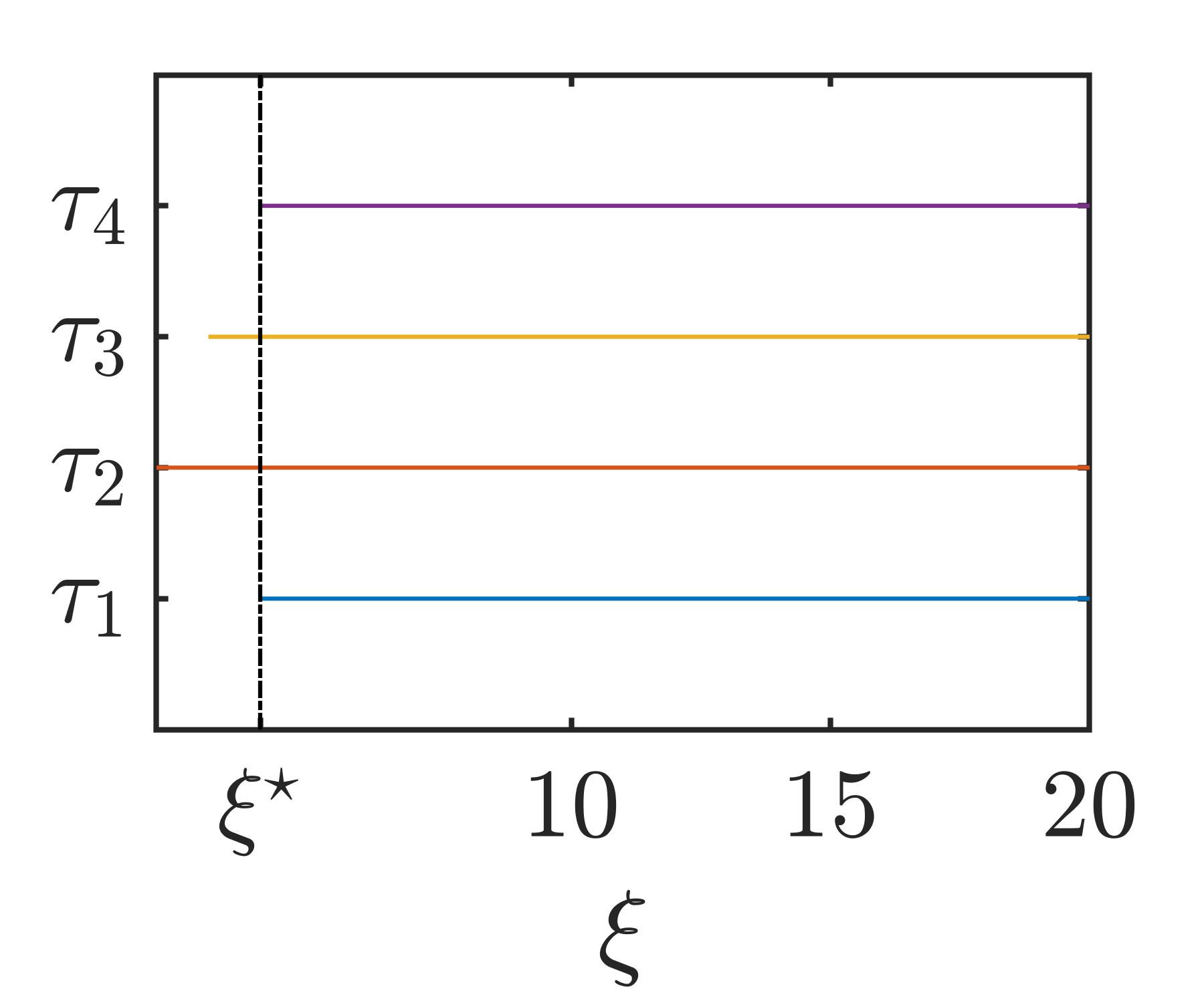}
  \caption{$\mathcal{S}_5+O^2S$}
  \label{f:tss5o2s}
\end{subfigure}%
\hfill
\begin{subfigure}{.32\textwidth}
  \centering
  \includegraphics[width=\textwidth]{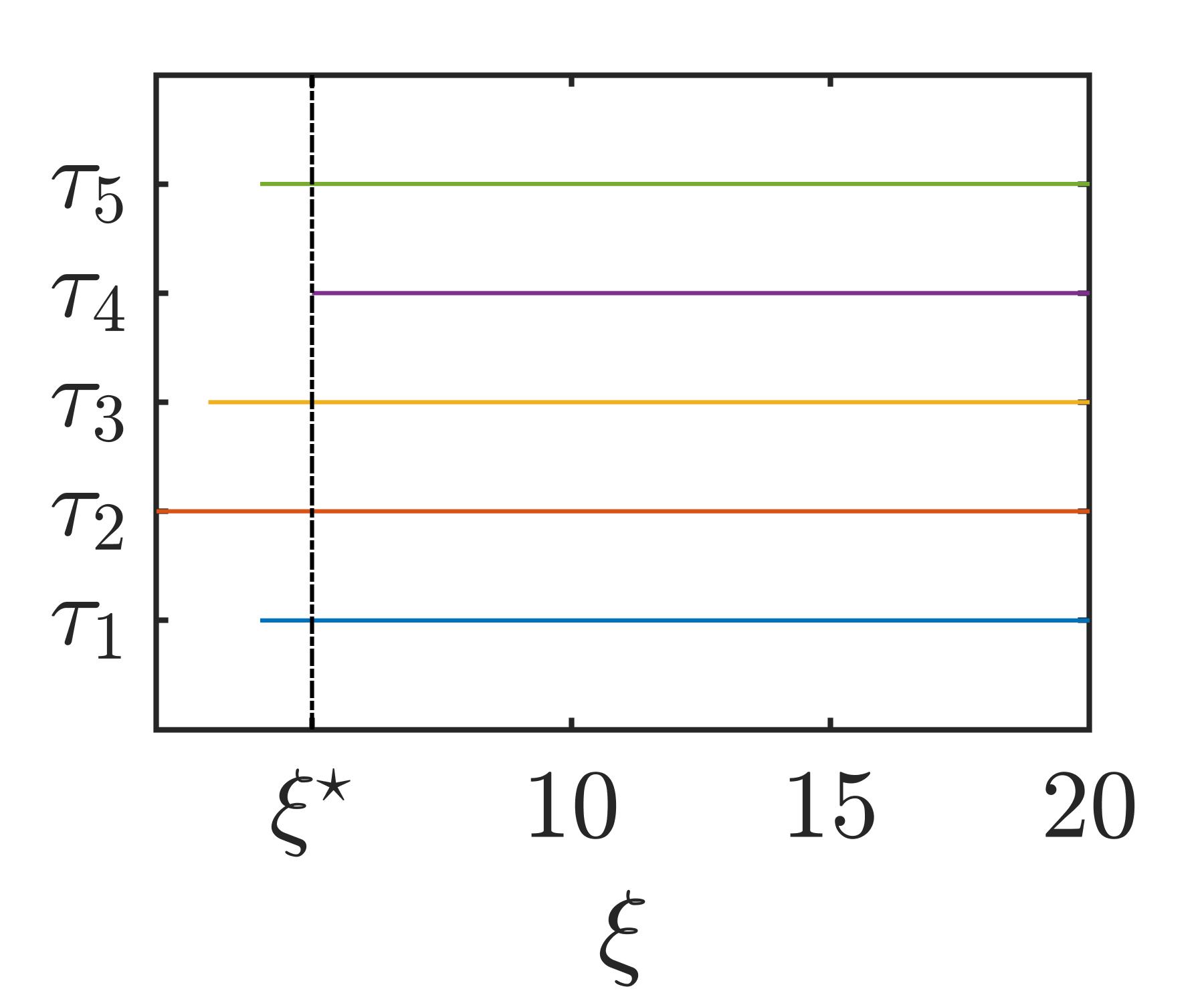}
  \caption{$\mathcal{S}_6+OSF$}
  \label{f:tss6osf}
\end{subfigure}%
\hfill
\begin{subfigure}{.32\textwidth}
  \centering
  \includegraphics[width=\textwidth]{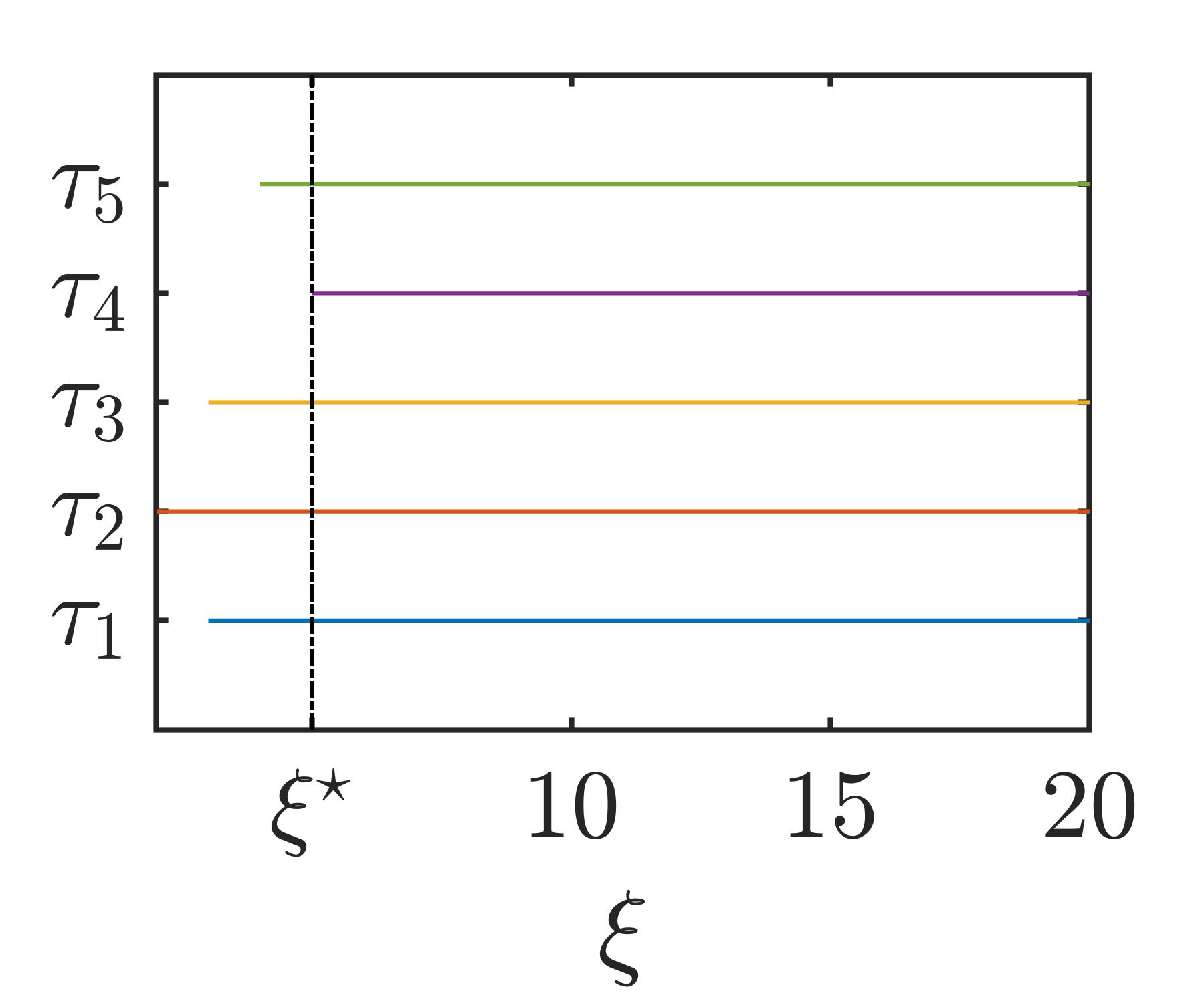}
  \caption{$\mathcal{S}_6+OIF$}
  \label{f:tss6oif}
\end{subfigure}
\hfill
\begin{subfigure}{.32\textwidth}
  \centering
  \includegraphics[width=\textwidth]{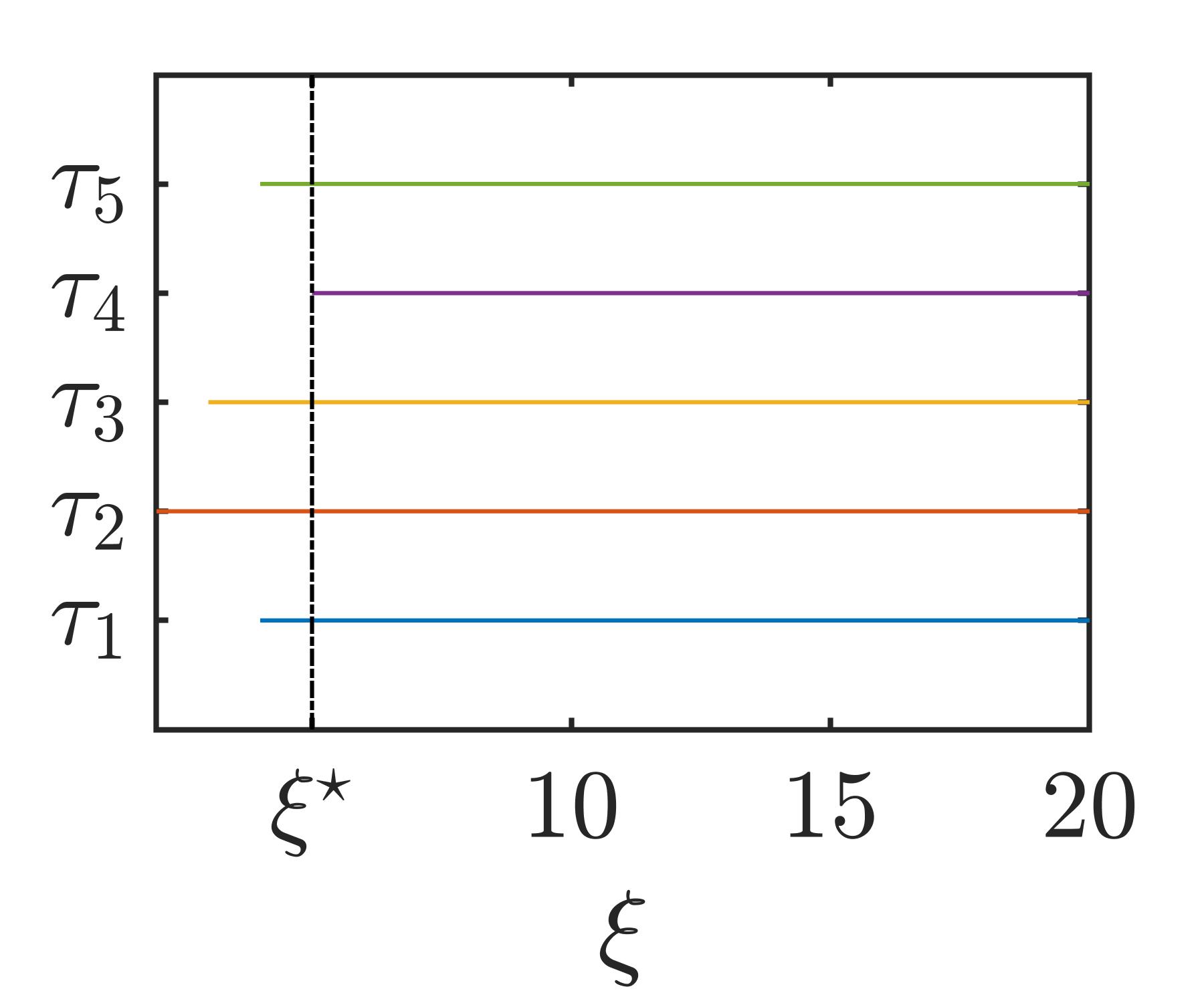}
  \caption{$\mathcal{S}_6+O^2S$}
  \label{f:tss6o2s}
\end{subfigure}%

\caption{Term Selection Behavior-III. $\xi^{\star}$ denotes actual cardinality of the system. The `dotted' vertical line indicate the value $\xi$ when all the system terms are identified.}
\label{f:TS3}
\end{figure*}
\subsection{Comparative Evaluation}
\label{s:compEval}

The objective of this part of the study is to compare the performance of orthogonal floating search algorithms. Following the procedure outlined Algorithm~\ref{al:genfloat}, the significant terms of the system shown in Table~\ref{t:sys} are identified. It is worth to note that while OSF and OIF do not have any control parameter, O$^2$S requires the selection of maximum search depth, $\mathcal{O}_{max}$. In this study, a maximum of $25\%$ terms is allowed to be exchanged, \textit{i.e.}, $\mathcal{O}_{max} = \ceil{0.25\times \min \{\xi, n -\xi \}}$.

The outcomes of model order selection are shown in Fig~\ref{f:mos}. Note that the `knee-point' in BIC coincides with the known cardinality (actual number of terms) of the systems. The selected cardinality (number of terms) for each system is shown in Table~\ref{t:SO} along with the qualitative search outcomes. It is interesting to see that the best possible search outcome, `\textit{Exact-Fitting}', is obtained for all combinations of system and search algorithm, except one. For the system $\mathcal{S}_1$, an over-fitted model (with one spurious term) is selected by OSF as seen in Fig.~\ref{f:S1} and Table~\ref{t:SO}. Note that this spurious term can easily be removed via a simple null-hypothesis test. Nevertheless, it is pertinent to identify the cause of this `over-fitting' which could be ascribed either to the information criterion or to the search algorithm. %


To investigate this further, the term selection behavior of the algorithms is determined. In particular, as the cardinality, $\xi$, is increased progressively from $\xi_{min}$ to $\xi_{max}$, the selection frequency of system terms is observed over the subsets identified by the algorithm, as follows: 
\begin{align}
    \label{eq:termFreq}
     \tau_{i,\xi} = \begin{cases}
                1, \ \textit{if} \ x_i \in \mathcal{X}_\xi\\
                0, \textit{otherwise}
               \end{cases}, \ \forall \ \mathcal{X}_\xi \in \Omega 
\end{align}
where, `$\tau_{i,\xi}$' encodes whether the $i^{th}$ term, `$x_i$', is included in the term subset $\mathcal{X}_\xi$ identified by the algorithm. Following this procedure, the term selection behavior is determined for all algorithms and shown here in Fig.~\ref{f:TS1}, \ref{f:TS2} and~\ref{f:TS3}. 


We first focus on the `\textit{over-fitting}' outcome by OSF on $\mathcal{S}_1$. The term selection by OSF for $\mathcal{S}_1$ is depicted in Fig.~\ref{f:tss1osf}. It is observed that all the system terms are first identified at $\xi = 6$ which is slightly higher than $\xi^{\star}$. This explains the `over-fitting' outcome. For the same system, OIF (Fig.~\ref{f:tss1oif}) and O$^2$S (Fig.~\ref{f:tss1o2s}) could identify all system terms from $\xi = \xi^{\star}$. This further underlines the efficacy of the `\textit{term swapping}' procedure introduced in OIF. 

Further, although the BIC could identify correct model order for all the systems in this study, the possibility of `over-fitting' cannot be ruled out. We therefore focus on the search behavior for $\xi > \xi^{\star}$. It is easier to follow that all the system terms must be included in the term subsets with cardinality $\xi\geq \xi^{\star}$. Thus it is desirable to have $\tau_{i,\xi}=1, \forall \ \xi\geq \xi^{\star}$. If this condition is satisfied then the correct system structure can still be obtained from the over-fitted structures through an appropriate secondary refinement procedure, \textit{e.g.}, \textit{null-hypothesis} test on the coefficients. The results show that this condition is satisfied for all systems by OIF and O$^2$S, as seen in Fig.~\ref{f:TS1}, \ref{f:TS2} and~\ref{f:TS3}. OSF could also satisfy this condition for all the systems except $\mathcal{S}_1$. 

\subsection{Comments on the Interval Selection}
\label{s:commentModelOrder}

The objective of this part of the study is to discuss the issues related to the selection of the cardinality interval. Given that the cardinality of the system under consideration is not known \textit{a priori}, a careful selection of the cardinality interval is crucial. In the previous subsections, a pragmatic approach is followed to select the cardinality interval, \textit{i.e.}, $[\xi_{min}, \xi_{max}]=[2,20]$; since in practice, over-parameterized models with higher cardinality are usually not desirable.

However, the selection of cardinality interval is system dependent. Therefore, the cardinality interval should be selected judiciously, especially for the higher order systems. In order to gain further insight into this issue, consider the following nonlinear system with $23$ terms, 
\begin{linenomath*}
\begin{small}
\begin{align}
	\label{eq:S4}
    S7 : y(k)  = & \ 0.8833u(k-1) + 0.0393u(k-2) + 0.8546u(k-3) + 0.8528u(k-1)^2 \\ \nonumber 
            & + 0.7582u(k-1)u(k-2) + 0.1750u(k-1)u(k-3) + 0.0864u(k-2)^2 \\ \nonumber
            & + 0.4916u(k-2)u(k-3)  + 0.0711u(k-3)^2 - 0.0375y(k-1) - 0.0598y(k-2)\\ \nonumber
            & - 0.0370y(k-3) - 0.0468y(k-4) - 0.0476y(k-1)^2 - 0.0781y(k-1)y(k-2)\\ \nonumber
            & - 0.0189y(k-1)y(k-3) - 0.0626y(k-1)y(k-4) - 0.0221y(k-2)^2\\ \nonumber 
            & - 0.0617y(k-2)y(k-3) - 0.0378y(k-2)y(k-4) - 0.0041y(k-3)^2\\ \nonumber
            & - 0.0543y(k-3)y(k-4) - 0.0603y(k-4)^2 + e(k)
\end{align}
\end{small}
\end{linenomath*}
For the identification purposes, $1000$ input-output data points are gathered with $u \sim WUN(0,1)$ and $e \sim WGN(0,0.01^2)$. A total of $286$ candidate terms are obtained with the following specification for the NARX model: $[n_u,n_y,n_l]=[5,5,3]$. 

\begin{figure*}[!t]
\centering
\small
\begin{subfigure}{.45\textwidth}
  \centering
  \includegraphics[width=\textwidth]{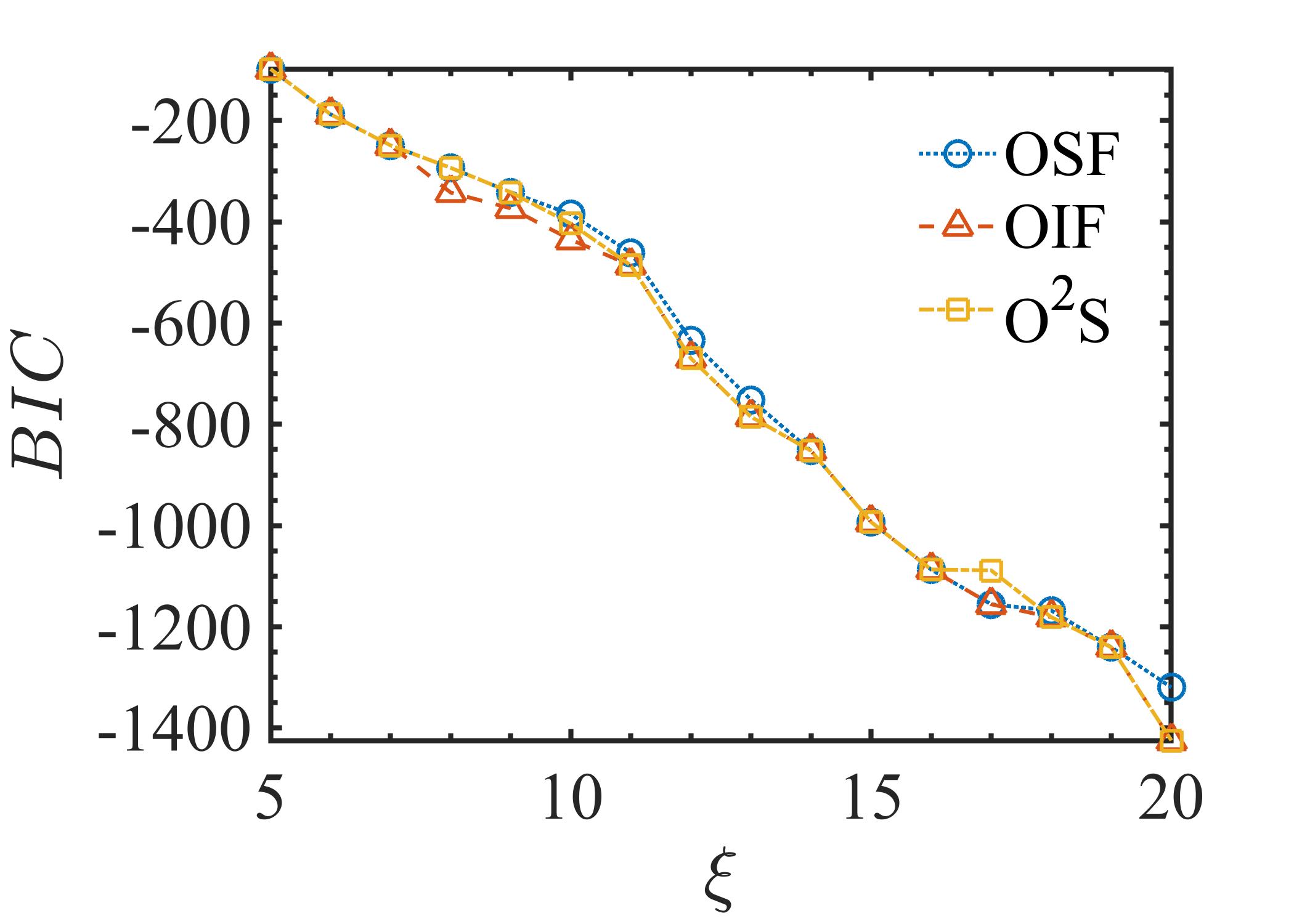}
  \caption{$\xi_{max} = 20$}
  \label{f:S4_1}
\end{subfigure}%
\hfill
\begin{subfigure}{.45\textwidth}
  \centering
  \includegraphics[width=\textwidth]{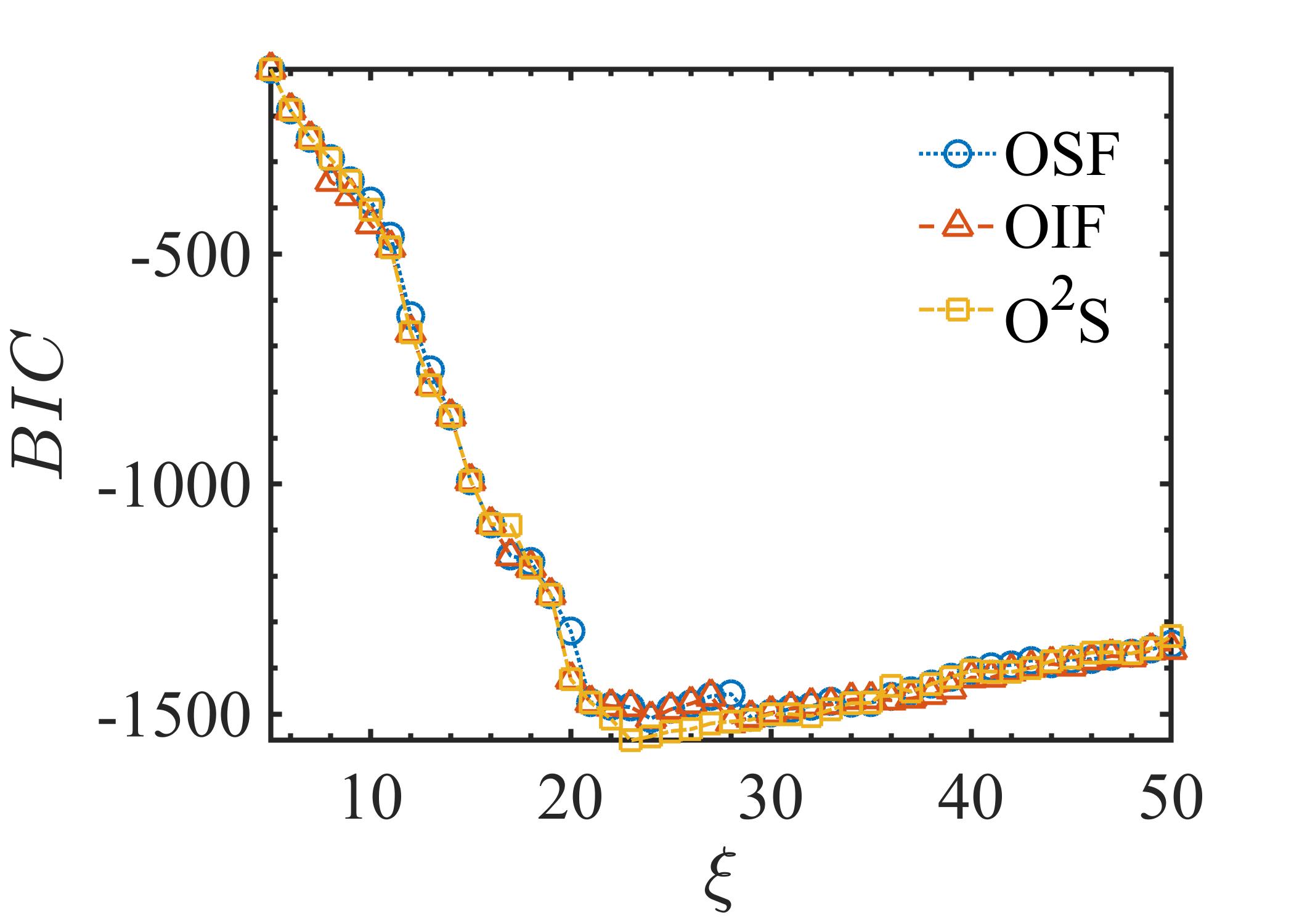}
  \caption{$ \xi_{max} = 50$}
  \label{f:S4_2}
\end{subfigure}
\caption{Determination of Cardinality Interval}
\label{f:S4Interval}
\end{figure*}

Initially, the system model is identified following the similar procedure, where the cardinality interval is specified as: $[\xi_{min}, \xi_{max}]=[2,20]$. The outcome of the cardinality selection for this interval is shown in Fig.~\ref{f:S4_1} which shows that the BIC is minimum at $\xi = 20$. However, it is not certain whether this is the desired `knee-point'. Therefore, the cardinality interval is increased to $\xi_{max}=50$. The consequent outcome is shown in Fig.~\ref{f:S4_2}. This result indicates a continual increase in BIC for $\xi \geq 30$ and thereby confirms that the plateau around $\xi \in [22,29]$ does contain the desired knee-point. 

As illustrated by this numerical example, the cardinality interval can easily be adjusted by observing the consequent `$\xi$ vs. BIC' relationship. We recommend the OSF to determine/adjust the cardinality interval due to its inexpensive computational requirements.  

\subsection{Orthogonal Floating Search Under Slow Varying Input}
\label{s:cERR}

As discussed in Section~\ref{s:termsig}, the present investigation uses the Error-Reduction-Ratio (ERR), instead of the other performance metrics, such as simulation error, ultra least square criterion. However, in recent years, the OFR-ERR has been criticized for leading to suboptimal search performance under certain scenarios~\cite{Piroddi:Spinelli:2003,Guo:Billings:2015,Falsone:Piroddi:2015,Baldacchino:Kadirkamanathan:2013}, where it is argued that this is due to the ERR criterion.

However, the results of this investigation on several benchmark nonlinear systems convincingly demonstrate that ERR could be an effective metric, provided it is paired with an appropriate search strategy. Thus, it can be argued that the suboptimal performance of OFR-ERR may be the consequence of the search strategy and not of the performance metric. Note that OFR-ERR is essentially a unidirectional sequential search approach which leads to the `\textit{nesting effect}', \textit{i.e.}, once a term is included it cannot be discarded. Thus the search performance may significantly be improved if the \textit{nesting effect} could be alleviated. The proposed orthogonal search framework effectively addresses this issue and therefore it can yield better search performance even with a relatively simple metric such as ERR. 

To demonstrate this, we consider the following numerical example, which has been used by various researchers for comparative investigation~\cite{Piroddi:Spinelli:2003,Guo:Billings:2015,Guo:Guo:2016}:
\begin{small}
\begin{align}
    \label{eq:SysColor}
    \mathcal{S}_8 : w(k) & = u(k-1) + 0.5u(k-2) + 0.25u(k-1)u(k-2) -0.3u(k-1)^3\\
                    y(k) & = w(k) + \frac{1}{1-0.8z^{-1}} e(k), \ \ e(\cdotp) \sim WGN(0.02) \nonumber\\
                \text{with} \ u(k) & = \frac{0.3}{1-1.6z^{-1} + 0.64z^{-2}} v(k), \ \ v(\cdotp) \sim WGN(0,1) \nonumber
\end{align}
\end{small}
where, `$y(\cdotp)$' is the observed value of `$w(\cdotp)$'. The system is excited with the slow-varying input $u(\cdotp)$ which is an AR process with a repeat pole at 0.8. A total of 165 candidate terms are obtained by the following NARX specifications: $[n_u, n_y, n_l]=[4, 4, 3]$. 

\begin{table}[!h]
  \centering
  \small
  \caption{OFR-ERR Search Dynamics for $\mathcal{S}_8^\dagger$}
  \label{t:slowofr}
 \begin{adjustbox}{max width=0.8\textwidth}
  \begin{threeparttable}
    \begin{tabular}{cccc}
    \toprule
    \bm{$\xi$} & \textbf{Subset$^\ddagger$} & \bm{$J(\cdotp)$} \\
    \midrule
    1     & $\{ \bm{y(k-1)} \}$ & 0.87426 \\ [1.2ex]
    2     & $\{ \bm{y(k-1)}, \ \bm{y(k-2)} \}$ & 0.92403 \\ [1.2ex]
    3     & $\{ \bm{y(k-1)}, \ \bm{y(k-2)}, \ u(k-1) \}$ & 0.94207 \\[1.2ex]
    4     & $\{ \bm{y(k-1)}, \ \bm{y(k-2)}, \ u(k-1), \ \bm{u(k-1)^2} \}$ & 0.98738 \\[1.2ex]
    5     & $\{ \bm{y(k-1)}, \ \bm{y(k-2)}, \ u(k-1), \ \bm{u(k-1)^2}, \ u(k-1)^3 \}$& 0.99383 \\[1.2ex]
    6     & $\{ \bm{y(k-1)}, \ \bm{y(k-2)}, \ u(k-1), \ \bm{u(k-1)^2}, \ u(k-1)^3, \ \bm{u(k-2)^3} \}$ & 0.99428 \\[1.2ex]
    7     & $\{ \bm{y(k-1)}, \ \bm{y(k-2)}, \ u(k-1), \ \bm{u(k-1)^2}, \ u(k-1)^3, \ \bm{u(k-2)^3}, \ \bm{u(k-3)^3} \}$ & 0.99871 \\[1.2ex]
    \bottomrule
    \end{tabular}%
 \begin{tablenotes}
      \small
      \item $\dagger$ \textit{spurious} terms are shown in \textit{bold} face
      \item $\ddagger$ The following system terms have not been included: $u(k-2)$, $u(k-1)u(k-2)$
    \end{tablenotes}
  \end{threeparttable}
 \end{adjustbox}
 \end{table}%
\begin{table}[!t]
  \centering
  \small
  \caption{OIF Search Dynamics for $\mathcal{S}_8^\dagger$}
  \label{t:slowoif}
 \begin{adjustbox}{max width=0.7\textwidth}
  \begin{threeparttable}
    
    \begin{tabular}{cccc}
    \toprule
    \textbf{Step} & \boldmath$\xi_{step}$ & \textbf{Subset} & \boldmath$J(\cdotp)$ \\
    \midrule
     1    & 1     & $\{ \bm{y(k-1)} \}$    & 0.87418 \\ [1ex]
    2     & 2     & $\{ \bm{y(k-1)}, \bm{y(k-2)} \}$ & 0.92385 \\[1ex]
    3     & 3     & $\{ \bm{y(k-1)}, \ u(k-1)^3, \ \bm{u(k-2)^3} \}$ & 0.98635 \\[1ex]
    4     & 3     & $\{ u(k-1), \ u(k-1)u(k-2), \ u(k-1)^3\}$ & 0.99632 \\[1ex]
    5     & 4     & $\{ u(k-1), \ u(k-2), \ u(k-1)u(k-2), \ u(k-1)^3 \}$ & 0.99757 \\[1ex]
    \bottomrule
    \end{tabular}%

     \begin{tablenotes}
      \small
      \item $\dagger$ \textit{spurious} terms are shown in \textit{bold} face
    \end{tablenotes}
  \end{threeparttable}
 \end{adjustbox}
 \end{table}%
\begin{table}[!t]
  \centering
  \small
  \caption{O$^2$S Search Dynamics for $\mathcal{S}_8^\dagger$}
  \label{t:slowo2s}
 \begin{adjustbox}{max width=0.95\textwidth}
  \begin{threeparttable}
    
    \begin{tabular}{cccccc}
    \toprule
    \textbf{Step} & \makecell{\textbf{Search} \\ \textbf{Depth} \\ `\bm{$o$}'} &\boldmath$\xi_{step}$ & \textbf{Subset} & \boldmath$J(\cdotp)$ & \textbf{Remarks} \\
    \midrule
    1     & -     & 4     & $\{\bm{y(k-1)}, \ \bm{y(k-2)}, \ u(k-1)^3, \ \bm{u(k-2)^3} \}$ & 0.98717 & \makecell{Initial\\ Model}\\[1.2ex]
    \midrule
    2     & \multirow{2}{*}{1} & 3     & $\{ u(k-1), \ u(k-1)u(k-2), \ u(k-1)^3  \}$ & 0.99632 & \multirow{2}{*}{\makecell{Down Swing\\ Improved}}\\[1.2ex]
    3     &       & 4     & $\{ u(k-1), \ u(k-2), \ u(k-1)u(k-2), \ u(k-1)^3  \}$ & \textbf{0.99757} \\[1.2ex]
    \midrule
    4     & \multirow{2}{*}{1} & 5 & $\{ \bm{y(k-1)}, \ u(k-1), \ u(k-2), \ u(k-1)u(k-2), \ u(k-1)^3  \}$ & 0.99758 & \multirow{2}{*}{\makecell{Up Swing\\ Did Not \\ Improve}} \\[1.2ex]
    5     &       & 4     & $\{ u(k-1), \ u(k-2), \ u(k-1)u(k-2), \ u(k-1)^3  \}$ & 0.99757 \\[1.2ex]
    \midrule
    6     & \multirow{2}{*}{1} & 3     & $\{ u(k-1), \ u(k-1)u(k-2), \ u(k-1)^3  \}$ & 0.99632 & \multirow{2}{*}{\makecell{Down Swing\\ Did Not \\ Improve}} \\[1.2ex]
    7     &       & 4     & $\{ u(k-1), \ u(k-2), \ u(k-1)u(k-2), \ u(k-1)^3  \}$ & 0.99757   & \\[1.2ex]

    \bottomrule
    \end{tabular}%

     \begin{tablenotes}
      \small
      \item $\dagger$ \textit{spurious} terms are shown in \textit{bold} face
    \end{tablenotes}
  \end{threeparttable}
 \end{adjustbox}
 \end{table}%

The system is identified by the following three algorithms: OFR-ERR, OIF and O$^2$S. The search dynamics of the compared algorithms are recorded and are shown in Table~\ref{t:slowofr} (OFR-ERR), Table~\ref{t:slowoif} (OIF) and Table~\ref{t:slowo2s} (O$^2$S). Note that, initially, all the algorithms selected the autoregressive terms `$y(k-1)$' and `$y(k-2)$'. As seen in Table~\ref{t:slowofr}, OFR-ERR could not remove the autoregressive terms due to the \textit{nesting effect}. Furthermore, several other spurious terms have been selected instead of two system terms: `$u(k-2)$' and `$u(k-1)u(k-2)$'. In contrast, OIF could easily remove the autoregressive terms through swapping procedure (see Step 3 and 4, Table~\ref{t:slowoif}). Similarly, the autoregressive terms are replaced by the system terms and the correct structure is identified in the first \textit{down-swing} procedure of O$^2$S (see Step 3, Table~\ref{t:slowo2s}).

It is obvious that the floating framework enabled OIF and O$^2$S to identify a better subset for the given cardinality. For example, all the system terms are identified by OIF and O$^2$S at $\xi=4$ with the criterion function $J(\cdotp)=0.99757$ (see Table~\ref{t:slowoif}-\ref{t:slowo2s}). In contrast, OFR-ERR identified only one system term at $\xi=4$. This is further reflected in the relatively lower value of the criterion function, $J(\cdotp)=0.98738$ (see Table~\ref{t:slowofr}). This example thus clearly demonstrate that ERR could be an effective performance metric when it is paired with a proper search strategy. 

It is worth to emphasize that, for identification purposes, the input signal to the system should be persistently exciting in nature. The example considered in (\ref{eq:SysColor}) is the exception to this fundamental fact, and it represents a worst-case scenario. In practice, such data should not be used for identification purposes. This example has been considered only for comparative analysis.

\subsection{Discrete models of Continuous-time Systems}
\label{s:Duff}

The effectiveness of the proposed framework is further demonstrated by fitting a polynomial NARX model to a continuous-time system. Since a discrete model of a continuous time system is not unique, the identified discrete-time model is validated by comparing Generalized Frequency Response Functions (GFRFs). Note that if the identified model has captured the system dynamics, the corresponding GFRFs should match.

\begin{figure*}[!t]
\centering
\small
\begin{subfigure}{.4\textwidth}
  \centering
  \includegraphics[width=\textwidth]{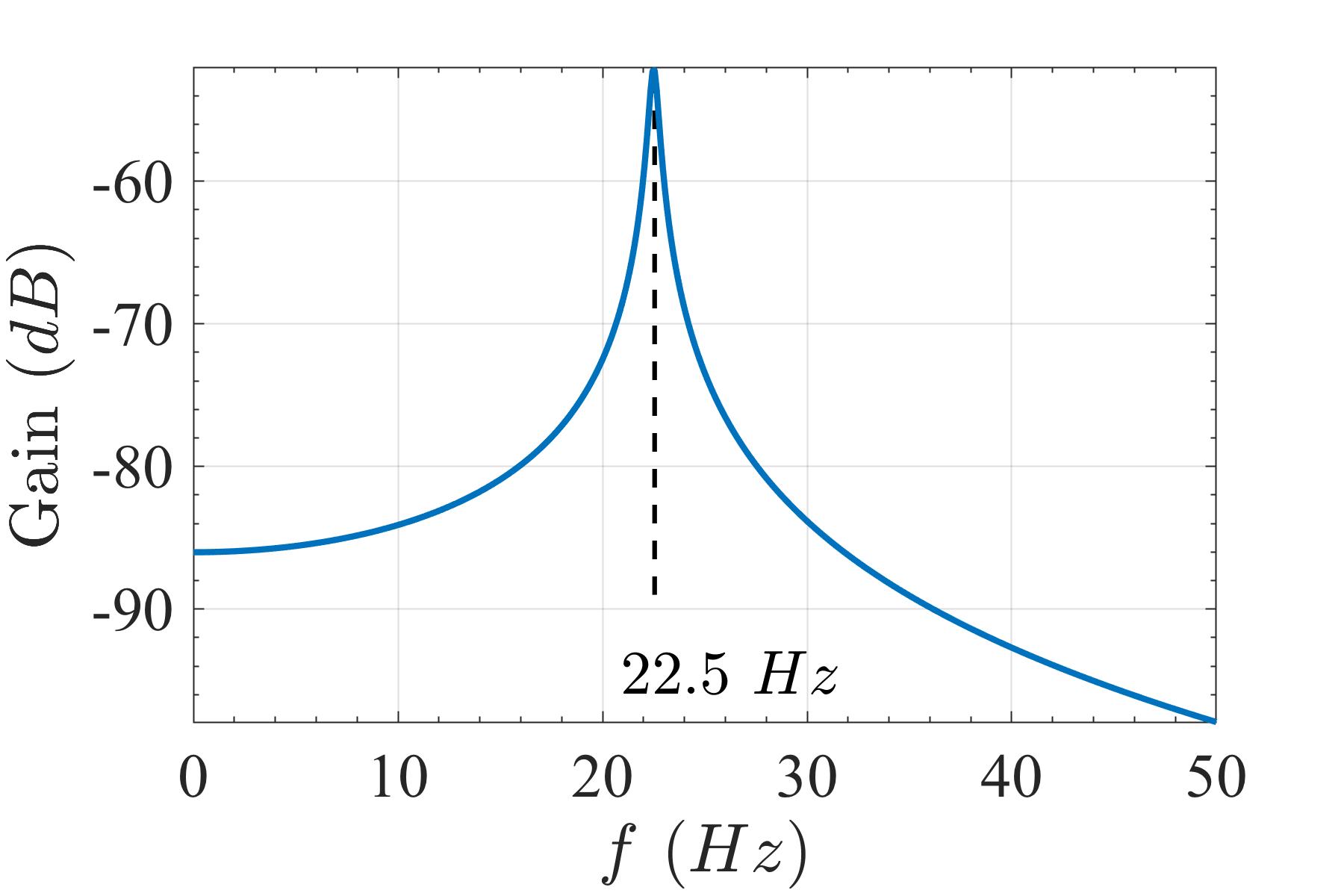}
  \caption{Linear GFRF (continuous)}
  \label{f:frfDuffC1}
\end{subfigure}%
\hfill
\begin{subfigure}{.4\textwidth}
  \centering
  \includegraphics[width=\textwidth]{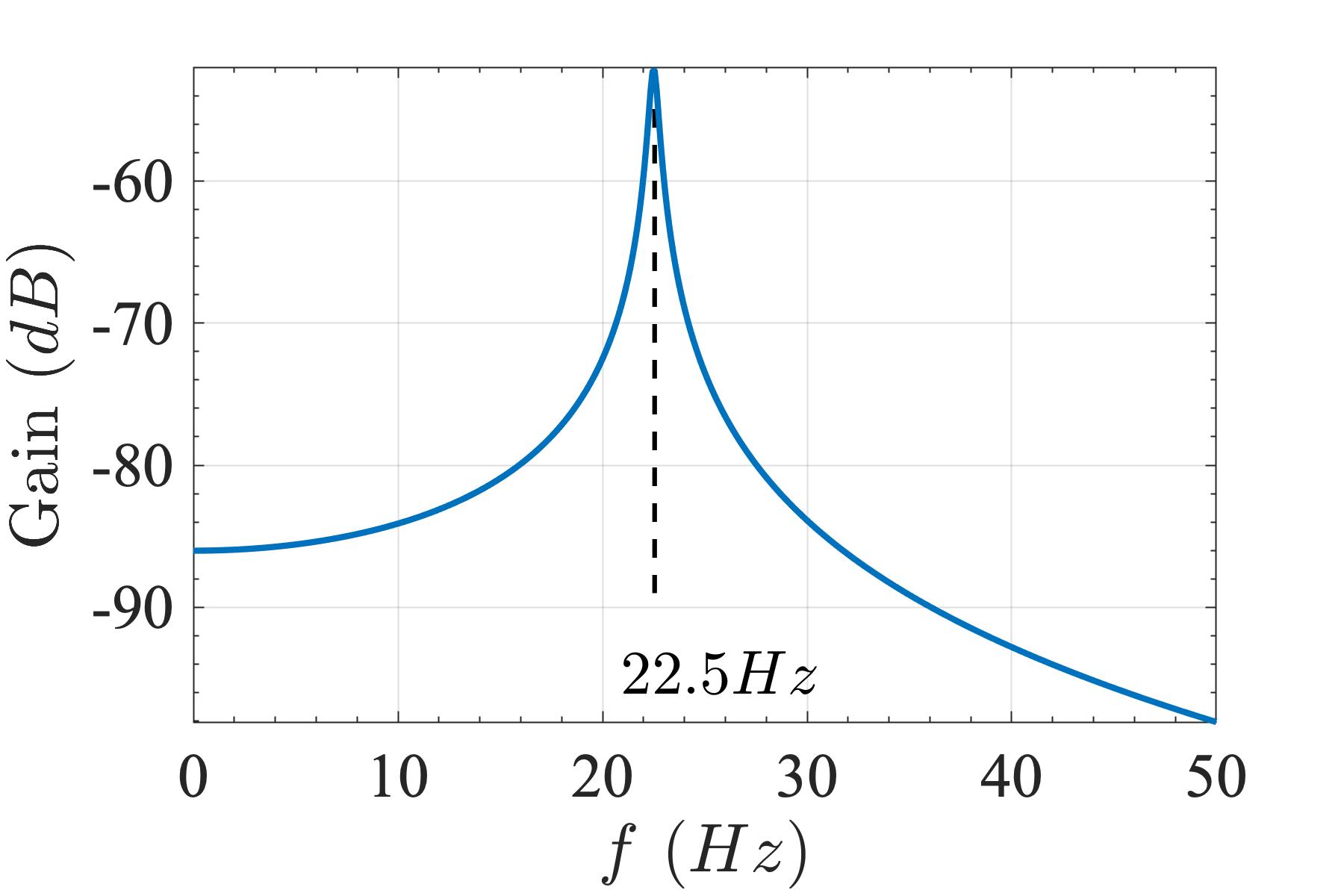}
  \caption{Linear GFRF (discrete)}
  \label{f:frfDuffD1}
\end{subfigure}
\medskip
\begin{subfigure}{.49\textwidth}
  \centering
  \includegraphics[width=\textwidth]{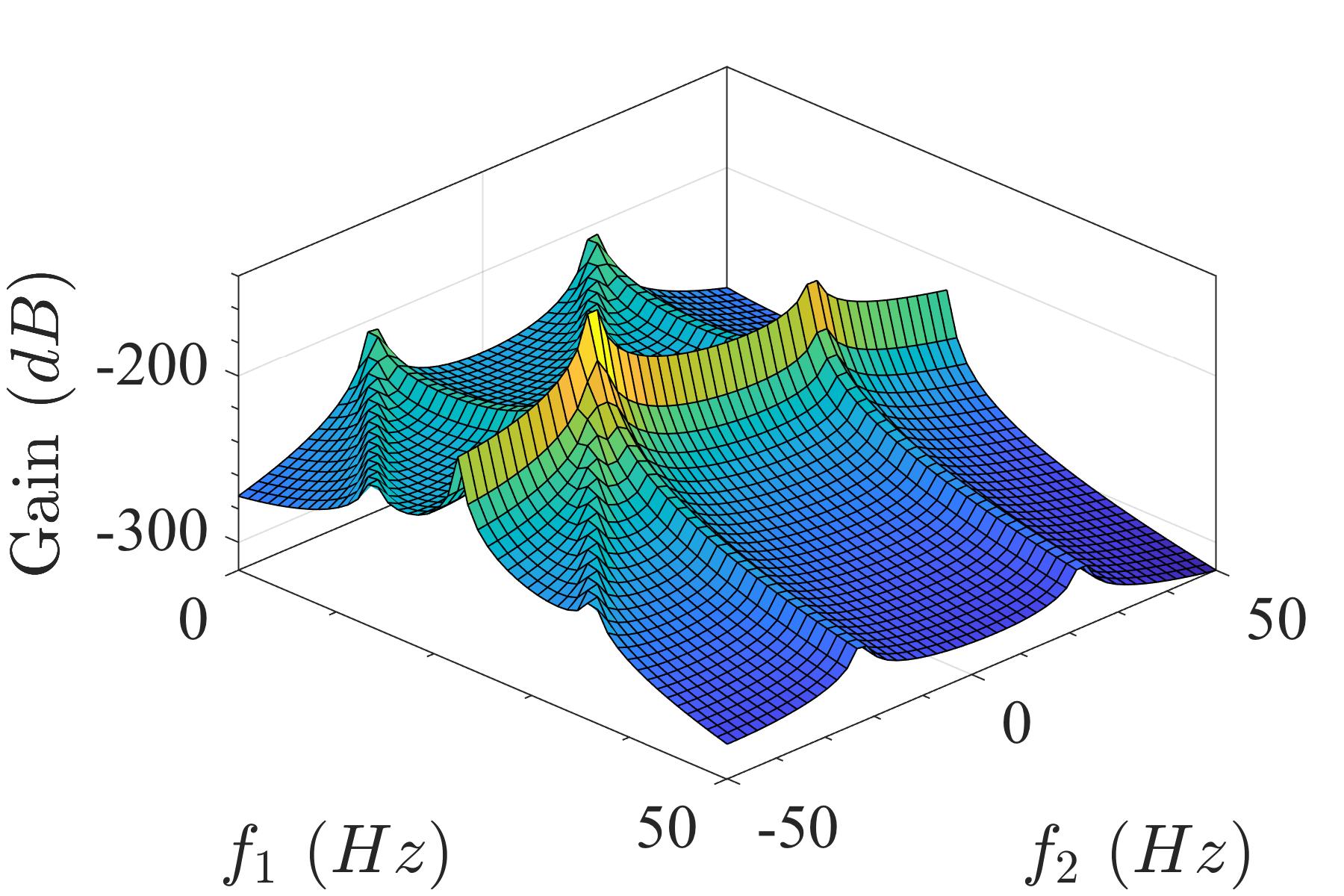}
  \caption{Third Order GFRF (continuous)}
  \label{f:frfDuffC3}
\end{subfigure}%
\hfill
\begin{subfigure}{.49\textwidth}
  \centering
  \includegraphics[width=\textwidth]{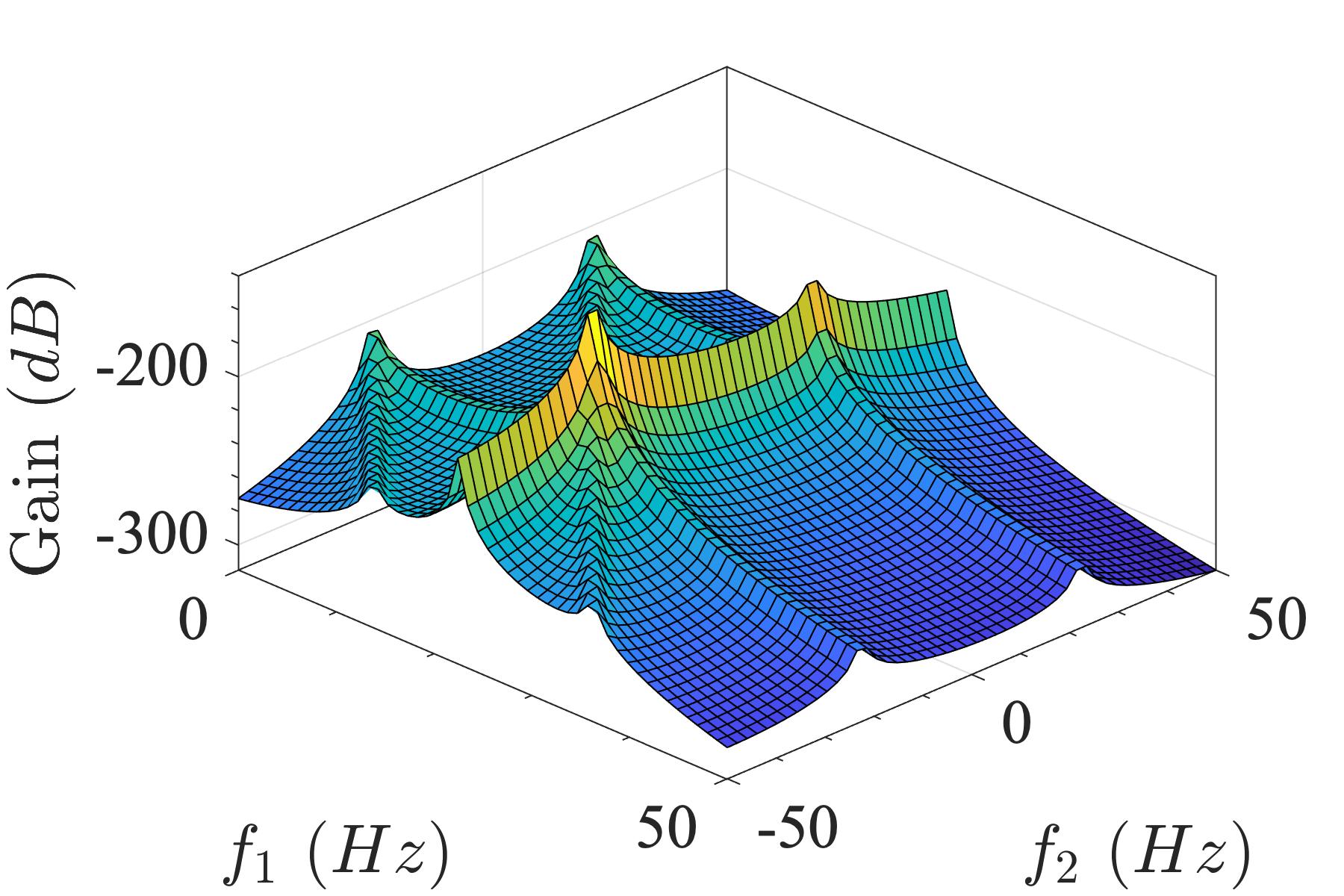}
  \caption{Third Order GFRF (discrete)}
  \label{f:frfDuffD3}
\end{subfigure}
\medskip
\begin{subfigure}{.46\textwidth}
  \centering
  \includegraphics[width=\textwidth]{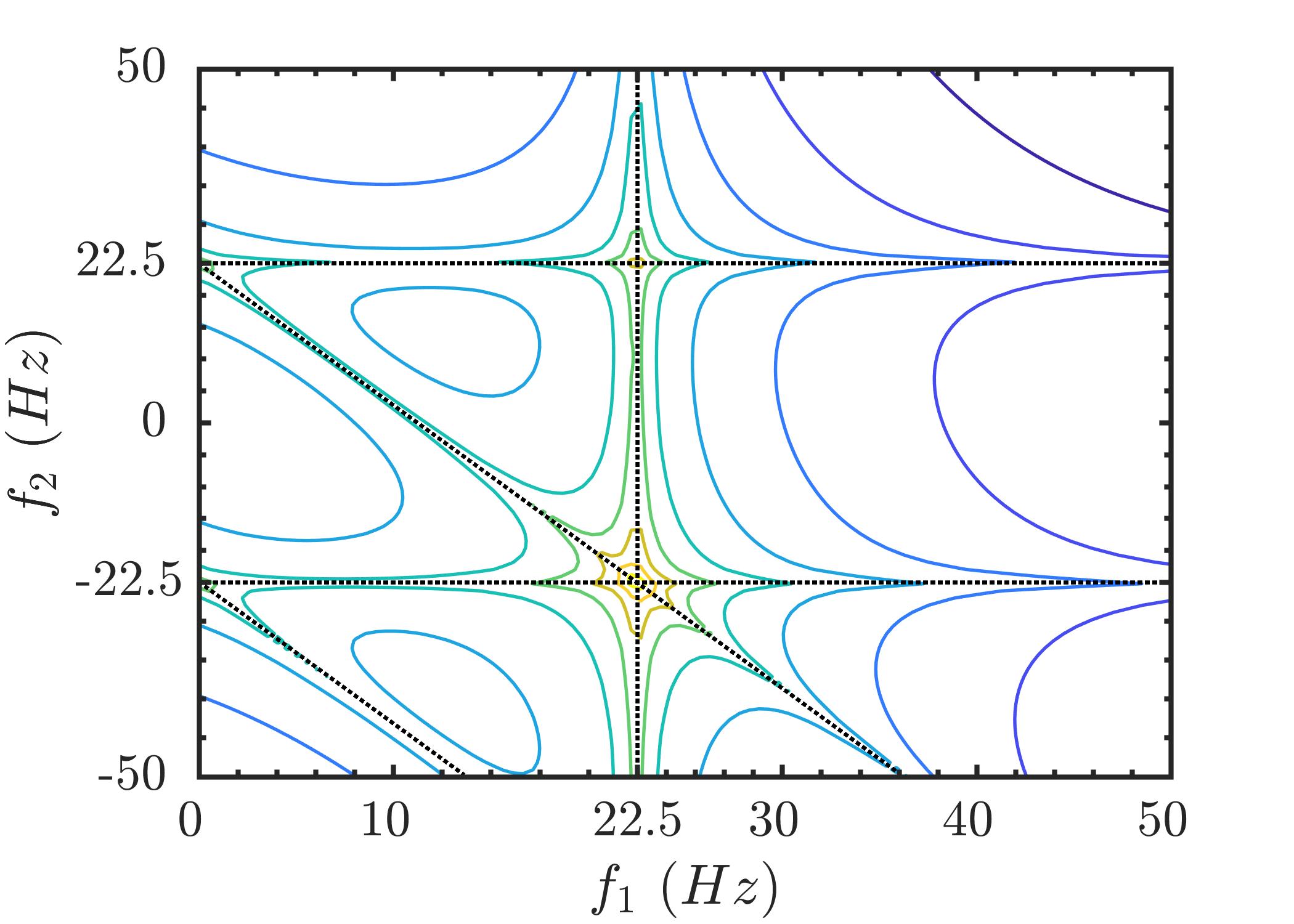}
  \caption{Third Order GFRF Contour (continuous)}
  \label{f:frfDuffC2}
\end{subfigure}%
\hfill
\begin{subfigure}{.46\textwidth}
  \centering
  \includegraphics[width=\textwidth]{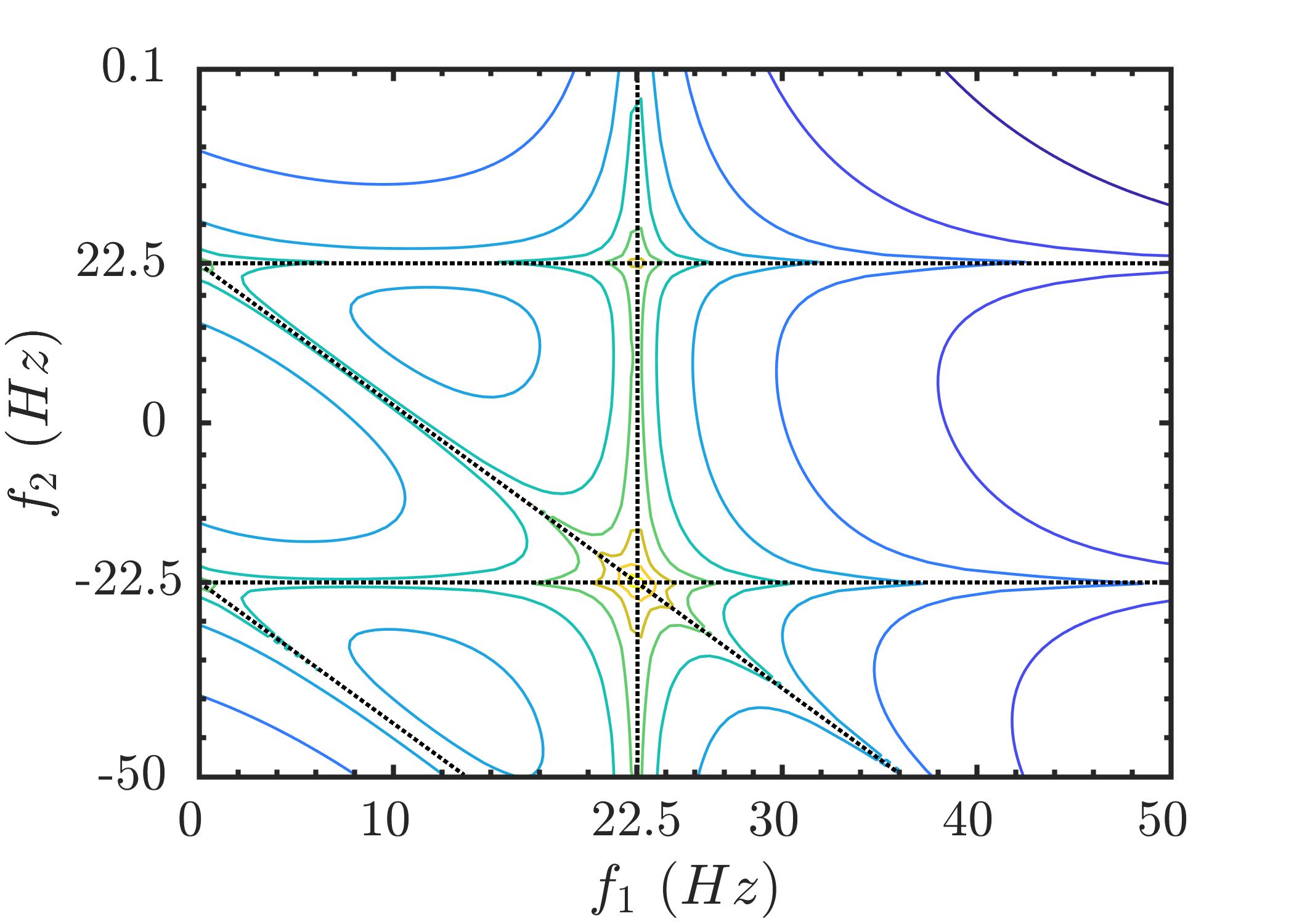}
  \caption{Third Order GFRF Contour (discrete)}
  \label{f:frfDuffD2}
\end{subfigure}
\caption{Linear and Third Order Frequency Response of Duffing's oscillator. For the third order GFRF $f_3=f_1$. Note that in the contour plots ridges align at $f_1+f_2+f_3= \pm 22.5$.}
\label{f:gfrfduff}
\end{figure*}

For this purpose, the \textit{Duffing}'s Oscillator is considered, which is given by,
\begin{linenomath*}
\begin{align}
	\label{eq:duffing}
	\ddot{y}(t) + 2\zeta\omega_n\dot{y}(t)+\omega_n^2y(t)+\omega_n^2\varepsilon y(t)^3-u(t)=0
\end{align}
\end{linenomath*}
where, $\omega_n = 45\pi$, $\zeta=0.01$ and $\varepsilon=3$. The input-output data are generated by exciting the system with $u(\cdotp)\sim WUN(0,1)$ and the output is sampled at $500 \ Hz$. A total of $1000$ data points are generated. The discrete NARX model is fitted using $700$ data points, and the remaining 300 data points are used for validation. The super-set containing a total of $286$ NARX terms is obtained by the following specifications: $[n_u, \ n_y, \ n_l] = [5, \ 5, \ 3]$.

The following model is identified by all the proposed algorithms OIS, OIF and O$^2$S: 
\begin{linenomath*}
    \begin{small}                                      
    \begin{align}
    \label{eq:eqsys}   
    y(k) = & \  1.9152 \, y(k-1) - 0.99436 \, y(k-2) \ + \ 1.983\times 10^{-6} \, u(k-1)\\ \nonumber
           & + \ 1.9792 \times 10^{-6} \, u(k-2) - \ 0.34289 \, y(k-1)^3 + 0.22813 \, y(k-2)y(k-1)^2 \\ \nonumber   
             & - \ 1.079 \times 10^{-7} \, u(k-1)y(k-1)^2 - \ 0.11847 \, y(k-1)y(k-2)^2    
    \end{align}                                   
    \end{small}                                        
\end{linenomath*}
The fitting error obtained over the validation data points is $3.96\times10^{-6}$.

The linear and the third order GFRFs of the continuous system are shown in Fig.~\ref{f:frfDuffC1},~\ref{f:frfDuffC3} and~\ref{f:frfDuffC2}. As indicated by these results, the frequency response of the Duffing's oscillator is characterized by the presence of several `ridges' and `peaks'. The ridges occur when the system is excited by the linear resonant frequency (\textit{i.e.}, $f_1$, $f_2$ or $f_3 = \pm 22.5$ Hz) or when the following condition is satisfied: $f_1+f_2+f_3=\pm 22.5$. These ridges are shown by `dotted' lines in the contour plot shown in Fig.~\ref{f:frfDuffC2}. Further, the peaks occur whenever all frequencies are equal to $\pm 22.5$. Note that these observations are in line with the earlier investigation~\cite{Billings:Peyton:1990}.

The GFRFs obtained with the discrete identified model are shown in Fig.~\ref{f:frfDuffD1},~\ref{f:frfDuffD3} and~\ref{f:frfDuffD2}. It is clear that the identified model could capture all the characteristics features of the Duffing's oscillator.  
\section{Discussion \& Conclusions}
\label{s:con}

A new Orthogonal Floating Search Framework has been proposed to identify significant terms of polynomial NARX models. The proposed framework integrates the classical ERR metric with floating search algorithms which retains the simplicity of OFR while eliminating its nesting effects. Following this framework, three well-known feature selection algorithms, such as SFFS, IFFS and OS, have been adapted to solve the structure selection problem. The effectiveness of this approach has been demonstrated considering several benchmark nonlinear systems including a system excited by a slowly varying input. The later system is often considered as a worst-case scenario. Further, a discrete-time model is fitted to Duffing's oscillator and it is validated by comparing generalized frequency response functions. These results convincingly demonstrate that the floating feature selection algorithms can easily be tailored for the purpose of structure selection in nonlinear system identification. For example, relatively simple procedures of the floating search, such as \textit{backtracking}, \textit{term swapping} and \textit{swing}, can effectively counter the \textit{nesting effect} and thereby yield significant improvement in the search performance (as shown in the results). The results of this investigation support the argument that the suboptimal performance of OFR-ERR is due in major part to its poor search strategy.

\linespread{1}

\appendix
\begin{small}

\section{Information Criteria}
\label{s:appIC}
Let $\mathcal{X}_\xi$, $\xi \in [\xi_{min}, \xi_{max}]$ denote a term subset identified the search algorithm. For this subset and the corresponding \textit{model-predicted output} ($\hat{y}$), the information criteria are given by,
\begin{itemize}
    \item Akaike Information Criterion (AIC):
\begin{linenomath*}
\begin{align}
\label{eq:aic}
AIC(\mathcal{X}_\xi) & = \mathcal{N}_v \ln(\mathcal{E}) + \varrho \xi, \ \text{with \ } \varrho=2
\end{align}
\end{linenomath*}
\item Bayesian Information Criterion (BIC):
\begin{linenomath*}
\begin{align}
\label{eq:bic}
BIC(\mathcal{X}_\xi) & = \mathcal{N}_v \ln(\mathcal{E}) + \ln(\mathcal{N}_v) \xi
\end{align}
\end{linenomath*}
\item Final Prediction Error (FPE):
\begin{linenomath*}
\begin{align}
\label{eq:fpe}
FPE(\mathcal{X}_\xi) & = \mathcal{N}_v \ln(\mathcal{E}) +  \mathcal{N}_v \ln (\frac{\mathcal{N}_v + \xi}{\mathcal{N}_v - \xi})
\end{align}
\end{linenomath*}
\item Khundrin's Law of Iterated Logarithm Criterion (LILC):
\begin{linenomath*}
\begin{align}
LILC(\mathcal{X}_\xi) & = \mathcal{N}_v \ln(\mathcal{E}) + 2\xi \ln{\ln{\mathcal{N}_v}}
\end{align}
\end{linenomath*}
\end{itemize}
where,`$\mathcal{N}_v$' denotes the length of the validation data; $\mathcal{E} = \frac{1}{\mathcal{N}_v} \sum \limits_{k=1}^{\mathcal{N}_v} [ y_k - \hat{y}_k ]^2$


\end{small}
\bibliographystyle{elsarticle-num}


\begin{thebibliography}{10}
\expandafter\ifx\csname url\endcsname\relax
  \def\url#1{\texttt{#1}}\fi
\expandafter\ifx\csname urlprefix\endcsname\relax\def\urlprefix{URL }\fi
\expandafter\ifx\csname href\endcsname\relax
  \def\href#1#2{#2} \def\path#1{#1}\fi

\bibitem{Billings:2013}
S.~A. Billings, Nonlinear system identification: NARMAX methods in the time,
  frequency, and spatio-temporal domains, John Wiley \& Sons, 2013.

\bibitem{Ljung:1999}
L.~Ljung, System Identification : Theory for the User, Prentice Hall PTR, Upper
  Saddle River, NJ, USA, 1999.

\bibitem{Hong:Mitchell:2008}
X.~Hong, R.~J. Mitchell, S.~Chen, C.~J. Harris, K.~Li, G.~W. Irwin, Model
  selection approaches for non-linear system identification: a review,
  International journal of systems science 39~(10) (2008) 925--946.

\bibitem{Billings:Chen:Korenberg:1989}
S.~A. Billings, S.~Chen, M.~J. Korenberg, Identification of {MIMO} non-linear
  systems using a forward-regression orthogonal estimator, International
  Journal of Control 49~(6) (1989) 2157--2189.

\bibitem{Baldacchino:Kadirkamanathan:2012}
T.~Baldacchino, S.~R. Anderson, V.~Kadirkamanathan, Structure detection and
  parameter estimation for {NARX} models in a unified {EM} framework,
  Automatica 48~(5) (2012) 857--865.

\bibitem{Baldacchino:Kadirkamanathan:2013}
T.~Baldacchino, S.~R. Anderson, V.~Kadirkamanathan, Computational system
  identification for bayesian {NARMAX} modelling, Automatica 49~(9) (2013)
  2641--2651.

\bibitem{Falsone:Piroddi:2015}
A.~Falsone, L.~Piroddi, M.~Prandini, A randomized algorithm for nonlinear model
  structure selection, Automatica 60 (2015) 227--238.

\bibitem{Tang:Long:2019}
X.~Tang, L.~Zhang, X.~Wang, Sparse augmented lagrangian algorithm for system
  identification, Neurocomputing.

\bibitem{Korenberg:Billings:1988}
M.~Korenberg, S.~Billings, Y.~Liu, P.~McIlroy, Orthogonal parameter estimation
  algorithm for non-linear stochastic systems, International Journal of Control
  48~(1) (1988) 193--210.

\bibitem{Chen:Billings:1989}
S.~Chen, S.~A. Billings, W.~Luo, Orthogonal least squares methods and their
  application to non-linear system identification, International Journal of
  control 50~(5) (1989) 1873--1896.

\bibitem{Chiras:Evans:2001}
N.~Chiras, C.~Evans, D.~Rees, Nonlinear gas turbine modeling using {NARMAX}
  structures, IEEE Transactions on Instrumentation and Measurement 50~(4)
  (2001) 893--898.

\bibitem{Mao:Billings:1997}
K.~Mao, S.~Billings, Algorithms for minimal model structure detection in
  nonlinear dynamic system identification, International journal of control
  68~(2) (1997) 311--330.

\bibitem{Piroddi:Spinelli:2003}
L.~Piroddi, W.~Spinelli, An identification algorithm for polynomial {NARX}
  models based on simulation error minimization, International Journal of
  Control 76~(17) (2003) 1767--1781.

\bibitem{Wei:Billings:2006}
H.~L. Wei, S.~A. Billings, Model structure selection using an integrated
  forward orthogonal search algorithm interfered with squared correlation and
  mutual information, Automatic Control and Systems Engineering, University of
  Sheffield, 2006.

\bibitem{Guo:Billings:2015}
Y.~Guo, L.~Guo, S.~A. Billings, H.-L. Wei, An iterative orthogonal forward
  regression algorithm, International Journal of Systems Science 46~(5) (2015)
  776--789.

\bibitem{Guo:Guo:2016}
Y.~Guo, L.~Guo, S.~Billings, H.-L. Wei, Ultra-orthogonal forward regression
  algorithms for the identification of non-linear dynamic systems,
  Neurocomputing 173 (2016) 715 -- 723.

\bibitem{Li:Peng:2006}
K.~Li, J.-X. Peng, E.-W. Bai, A two-stage algorithm for identification of
  nonlinear dynamic systems, Automatica 42~(7) (2006) 1189 -- 1197.

\bibitem{Hafiz:Swain:CEC:2018}
F.~Hafiz, A.~Swain, E.~M. Mendes, N.~Patel, Structure selection of polynomial
  {NARX} models using two dimensional {(2D)} particle swarms, in: 2018 IEEE
  Congress on Evolutionary Computation (CEC), 2018, pp. 1--8.

\bibitem{Michael:Lin:1973}
M.~Michael, W.~Lin, Experimental study of information measure and inter-intra
  class distance ratios on feature selection and orderings, IEEE Transactions
  on Systems, Man, and Cybernetics SMC-3~(2) (1973) 172--181.

\bibitem{Stearns:1976}
S.~D. Stearns, On selecting features for pattern classifiers, in: Proceedings
  of the 3rd International Joint Conference on Pattern Recognition, 1976, pp.
  71--75.

\bibitem{Kittler:1978}
J.~Kittler, Feature set search algorithms, Pattern recognition and signal
  processing (1978) 41--60.

\bibitem{Pudil:1994}
P.~Pudil, J.~Novovi{\v{c}}ov{\'a}, J.~Kittler, Floating search methods in
  feature selection, Pattern Recognition Letters 15~(11) (1994) 1119--1125.

\bibitem{Somol:Pudil:1999}
P.~Somol, P.~Pudil, J.~Novovi{\v{c}}ov{\'a}, P.~Pacl{\i}k, Adaptive floating
  search methods in feature selection, Pattern Recognition Letters 20~(11)
  (1999) 1157--1163.

\bibitem{Somol:Pudil:2000}
P.~Somol, P.~Pudil, Oscillating search algorithms for feature selection, in:
  Proceedings 15th International Conference on Pattern Recognition. ICPR-2000,
  Vol.~2, 2000, pp. 406--409 vol.2.

\bibitem{Nakariyakul:Casasent:2009}
S.~Nakariyakul, D.~P. Casasent, An improvement on floating search algorithms
  for feature subset selection, Pattern Recognition 42~(9) (2009) 1932--1940.

\bibitem{Stoica:Selen:2004a}
P.~Stoica, Y.~Selen, Model-order selection: a review of information criterion
  rules, IEEE Signal Processing Magazine 21~(4) (2004) 36--47.

\bibitem{Mendes:1995}
E.~M. A.~M. Mendes, Identification of nonlinear discrete systems with
  intelligent structure detection, Ph.D. thesis, University of Sheffield,
  England, UK (1995).

\bibitem{Bonin:Pirrodi:2010}
M.~Bonin, V.~Seghezza, L.~Piroddi, {NARX} model selection based on simulation
  error minimisation and {LASSO}, IET control theory \& applications 4~(7)
  (2010) 1157--1168.

\bibitem{Billings:Peyton:1990}
S.~A. Billings, J.~C.~P. Jones, Mapping non-linear integro-differential
  equations into the frequency domain, International Journal of Control 52~(4)
  (1990) 863--879.

\end{thebibliography}

\end{document}